\documentclass[aps,pra, twocolumn,floats,superscriptaddress]{revtex4}
\pdfoutput=1
\usepackage{epsfig,color}
\usepackage{bm}
\usepackage{amsmath}
\usepackage{subfigure}

\usepackage{graphicx}

\newcommand{\fig}[1]{figure \ref{fig:#1}}
\newcommand{\sect}[1]{section \ref{sec:#1}}

\begin{document}

\author{Sergio Boixo}
\affiliation{Information Sciences Institute and Department of Electrical Engineering, University of Southern California, Los Angeles, CA 90089, USA}
\author{Troels F. R{\o}nnow}
\affiliation{Theoretische Physik, ETH Zurich, 8093 Zurich, Switzerland}
\author{Sergei V. Isakov}
\affiliation{Theoretische Physik, ETH Zurich, 8093 Zurich, Switzerland}
\author{Zhihui Wang}
\affiliation{Department of Chemistry and Center for Quantum Information Science \& Technology,  University of Southern California, Los Angeles, California 90089, USA}
\author{David Wecker}
\affiliation{Quantum Architectures and Computation Group, Microsoft Research, Redmond, WA 98052, USA}
\author{Daniel A. Lidar}
\affiliation{
Departments of Electrical Engineering, Chemistry and Physics, and Center for Quantum Information Science \& Technology, University of Southern California, Los Angeles, California 90089, USA
}
\author{John M. Martinis}
\affiliation{Department of Physics, University of California, Santa Barbara, CA 93106-9530, USA}
\author{Matthias Troyer$^*$}
\affiliation{Theoretische Physik, ETH Zurich, 8093 Zurich, Switzerland}

\title{Quantum annealing with more than one hundred qubits}


\maketitle
 
{\bf
Quantum technology is maturing to the point where quantum devices, such as quantum communication systems, quantum random number generators and quantum simulators, may be built with capabilities exceeding classical computers. A quantum annealer, in particular, solves hard optimisation problems by evolving a known initial configuration at non-zero temperature towards the ground state of a Hamiltonian encoding a given problem. Here, we present results from experiments on a $108$ qubit D-Wave One device based on superconducting flux qubits. The strong correlations between the device and a simulated quantum annealer, in contrast with weak correlations between the device and classical annealing or classical spin dynamics, demonstrate that the device performs quantum annealing. We find additional evidence for quantum annealing in the form of small-gap avoided level crossings characterizing the hard problems. To assess the computational power of the device we compare it to optimised classical algorithms.}

Annealing a material by slow cooling  is an ancient  technique  to improve the properties of glasses, metals and steel that has been used for more than  seven millennia~\cite{Muhly1988}. Mimicking this process in computer simulations is the idea behind {\em simulated annealing} as an optimisation method~\cite{Kirkpatrick1983}, which views the cost function of an optimisation problem as the energy of a physical system. Its configurations are sampled in a Monte Carlo simulation using the Metropolis algorithm~\cite{Metropolis}, escaping from local minima by thermal fluctuations to find lower energy configurations. The goal is to find the global energy minimum (or at least a close approximation) by slowly lowering the temperature and thus obtain the solution to the optimisation problem. 

The phenomenon of quantum tunneling suggests that it can be more efficient to explore the state space {\em quantum mechanically} in a quantum annealer~\cite{Ray1989,Finnila1994,Kadowaki1998}. In {\em simulated quantum annealing}~\cite{sqa1,Santoro}, one makes use of this effect by adding quantum fluctuations, which are slowly reduced while keeping the temperature constant and positive -- ultimately ending up in a low energy configuration of the optimisation problem.  Simulated quantum annealing, using a quantum Monte Carlo algorithm, has been observed to be more efficient than thermal annealing for certain spin glass models~\cite{Santoro}, although the opposite has been observed for  $k$-satisfiability problems~\cite{ksat}. Further speedup may be expected in physical quantum annealing, either as an experimental technique for annealing a quantum spin glass~\cite{Brooke1999}, or -- and this is what we will focus on here --  as a computational technique in a programmable quantum device. 

In this work we report on computer simulations and experimental tests on a D-Wave One device~\cite{Johnson2011} in order to address central open questions about quantum annealers: is the device actually a quantum annealer, i.e., do the quantum effects observed on $8$ \cite{Johnson2011,Boixo2012} and $16$ qubits \cite{DWave-16q} persist when scaling problems up to more than $100$ qubits, or do short coherence times turn the device into a classical, thermal annealer? Which problems are easy and which problems are hard for a quantum annealer, for a simulated classical annealer, for classical spin dynamics, and for a simulated quantum annealer? How does the effort to find the ground state scale with problem size? Does the device have advantages over classical computers?

We consider the problem of finding the ground state of an Ising spin glass model with ``problem Hamiltonian"  
\begin{equation}
H_{\rm Ising} = -\sum_{i<j} J_{ij}\sigma_i^z\sigma_j^z -  \sum_i h_i\sigma_i^z,
\label{eq:H}
\end{equation}
with $N$ binary variables $\sigma_i^z = \pm 1$. This problem is non-deterministic polynomially (NP) hard~\cite{Barahona1982}, meaning that it is at least as hard as the hardest problems in NP, a class which includes notoriously difficult problems such as traveling salesman and satisfiability of logical formulas \cite{2013arXiv1302.5843L}.  It also implies that no efficient (polynomial time) algorithm to find these ground states is known and the computational effort of all existing algorithms scales  with problem size as ${\rm O}(\exp(cN^a))$. While quantum mechanics is not expected to turn the exponential scaling into a polynomial one, the constants $c$ and $a$ can be smaller on quantum devices, potentially giving substantial speedup over classical algorithms. 

To perform quantum annealing, we map each of the Ising variables $\sigma_i^z$ to the Pauli $z$-matrix (which defines the ``computational basis") and add a transverse magnetic field in the $x$-direction to induce quantum fluctuations,  obtaining the time-dependent $N$-qubit Hamiltonian 
\begin{equation}
\label{eq:Hquantum}
H(t) = -A(t) \sum_i \sigma_i^x +B(t) H_{\rm Ising}\ , \quad t\in[0,t_f]\ .
\end{equation}

Quantum annealing at positive temperature $T$ starts in the limit of a strong transverse field and weak problem Hamiltonian, $A(0) \gg \max(k_BT$, $B(0))$, 
with the system state close to the ground state of the transverse field term, the equal superposition state (in the computational basis) of all $N$ qubits. Monotonically decreasing $A(t)$ and increasing $B(t)$, the system evolves towards the ground state of the problem Hamiltonian, with $B(t_f) \gg A(t_f)$. 

Unlike adiabatic quantum computing \cite{farhi}, which has a similar schedule but assumes fully coherent adiabatic ground state evolution at zero temperature, quantum annealing \cite{Ray1989,Finnila1994,Kadowaki1998,Brooke1999} is a positive temperature method involving an open quantum system coupled to a thermal bath. Nevertheless, one expects that similar to adiabatic quantum computing, small gaps to excited states may thwart finding the ground state. In hard optimisation problems, the smallest gaps of avoided level crossings have been found to close exponentially fast with increasing problem size~\cite{Joerg2010,hen,Farhi2012}, suggesting an exponential scaling of the required annealing time $t_f$ with problem size $N$.\\

\noindent{\textbf{Experimental results}}\\
We performed our experiments on a D-Wave One chip, a device comprised of superconducting flux qubits with programmable couplings (see Methods). Of the $128$ qubits on the device, $108$ were fully functioning and were used in our experiments. The ``chimera" connectivity graph of these qubits is shown in figure 1 in the supplementary material. 
Instead of trying to map specific optimisation problem to the connectivity graph of the chip~\cite{ramsey,protein}, we chose random spin glass problems that can be directly realised. For each  coupler $J_{ij}$  in the device  we randomly assigned a value of either $+1$ or $-1$,  giving rise to a  very rough energy landscape. Local fields $h_i\ne0$ give a bias to  individual spins, tending to make the problem easier to solve for annealers. We thus set all  $h_i=0$ for most data shown here and provide data with local fields in the supplementary material.  We performed experiments for problems of sizes ranging from  $N=8$ to $N=108$. For each problem size $N$, we selected $1000$ different {\em instances} by choosing $1000$ sets of different random couplings $J_{ij}=\pm1$ (and for some of the data also random fields $h_i=\pm1$). For each of these instances, we performed $M=1000$ annealing {\em runs}
 and determined whether the system reached the ground state.

Our strategy is to discover the operating mechanism of the D-Wave device (DW) by comparing to three models: simulated classical annealing (SA), simulated quantum annealing (SQA), and classical spin dynamics (SD).  We study both the distribution of the success probabilities and the correlations between the D-Wave device and the models.

\begin{figure}
\centering
\includegraphics[width=\columnwidth]{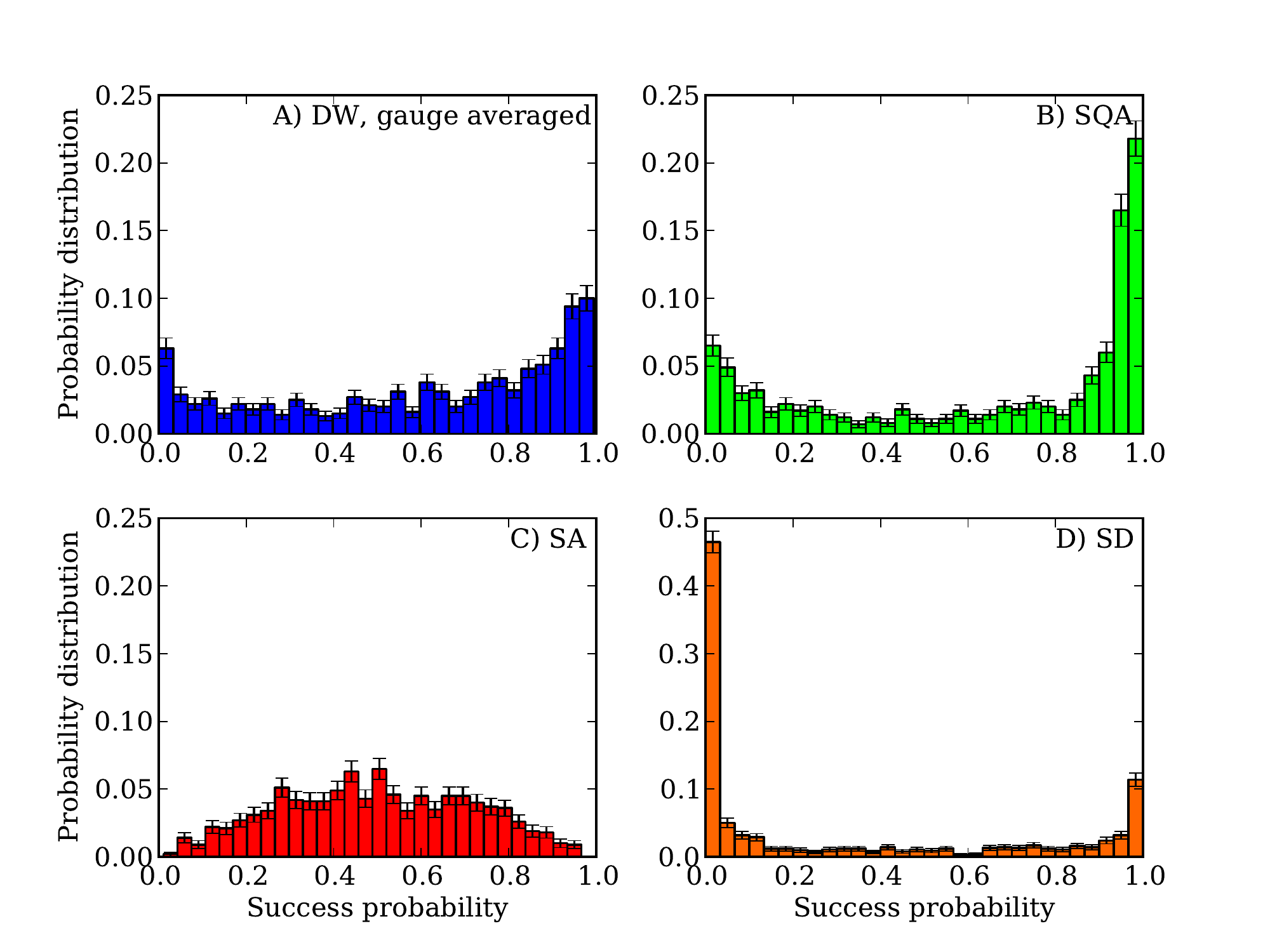}
\caption{{\bf Success probability distributions.}  
Shown are normalized histograms $p(s) = (\textrm{number of instances with probability }s)/K$ of the success probabilities of finding the ground states for $N=108$ qubits and $K=1000$ different spin glass instances.  We find similar bimodal distributions for the D-Wave results (DW, panel A) and the simulated quantum annealer (SQA, panel B), and somewhat similar distributions for spin dynamics (SD, panel D). The unimodal distribution for the simulated annealer (SA, panel C) clearly does not match.  The D-Wave data is taken with gauge averaging of $16$ sets. Note the different vertical axis scale for D).
}
\label{fig:histogram}
\end{figure}

For our first test, we counted for each instance the number of runs $M_{\rm GS}$  in which the ground state was reached, to determine the success probability as $s = M_{\rm GS}/M$.  Plots of the distribution of $s$ are shown in figure~\ref{fig:histogram}, where we see that the DW results match well with SQA, moderately with SD, and poorly with SA.  We find a unimodal distribution for the simulated annealer model for all schedules, temperatures and annealing times we tried, with a peak position that moves to the right as one increases $t_f$ (see supplementary material). In contrast, the D-Wave device, the simulated quantum annealer and the spin dynamics model exhibit a bimodal distribution, with a clear split into easy and hard instances.  Moderately increasing $t_f$ in the simulated quantum annealer makes the bimodal distribution more pronounced, as does lowering the temperature (see supplementary material).

\begin{figure}
\centering
\includegraphics[width=\columnwidth]{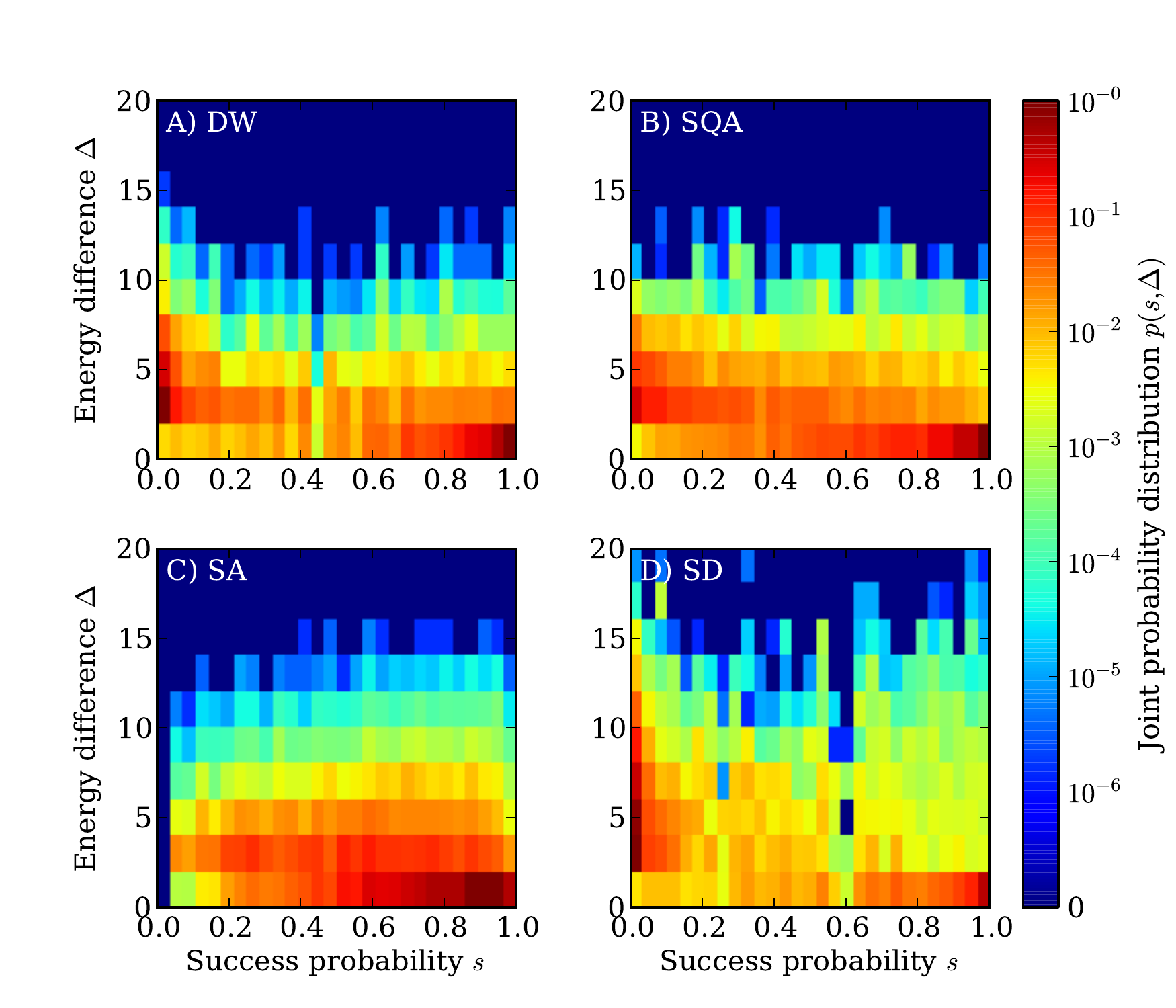}
\caption{{\bf Energy-success distributions.} 
Shown is the joint probability distribution $p(s, \Delta)$ (colour scale) of success probability $s$ and the final state energy $\Delta$ measured relative to the ground state.   We find very similar results for the D-Wave device (panel A) and the simulated quantum annealer (panel B).  The distribution for simulated classical annealing (panel C)  matches poorly and for spin dynamics (panel D) matches only moderately.  For the D-Wave device and SQA the hardest instances result predominantly in low-lying excited states, while easy instances result in ground states.  For SA most instances concentrate around intermediate success probabilities and the ground state as well as low-lying excited states. For classical spin dynamics there is a significantly higher incidence of relatively high excited states than for DW, as well as far fewer excited states for easy instances. The histograms of figure~\ref{fig:histogram}, representing $p(s)$, are recovered when summing these distributions over $\Delta$.  SA distributions for different numbers of sweeps are shown in the supplementary material.}
\label{fig:distribution}
\end{figure}

As a second test, we show in figure~\ref{fig:distribution} results for the joint probability distribution $p(s,\Delta)$, which also includes the probability distribution for the final state energy $\Delta$ measured relative to the ground state.  We find that the distribution for the D-Wave device (panel A) is very similar to that of the simulated quantum annealer (panel B), whereas it is quite different from that of a simulated classical annealer (panel C) and spin dynamics (panel D).

\begin{figure}
\centering
\includegraphics[width=\columnwidth]{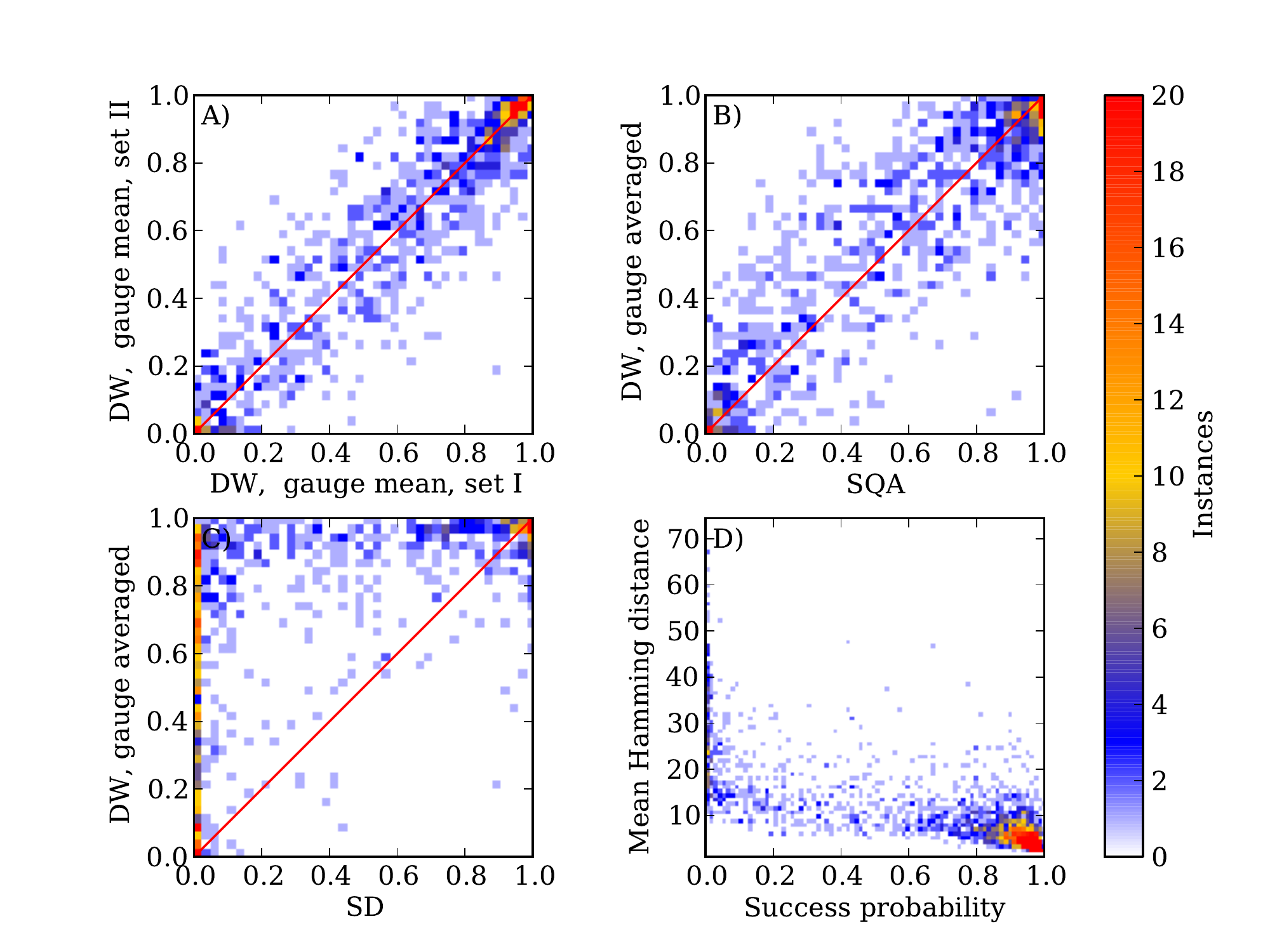}
\caption{{\bf Correlations.} 
Panels A-C show scatter plots of correlations of the success probabilities $p(s)$ obtained from different methods.  The red lines indicate perfect correlation.  Panel A is for the D-Wave device between two sets of eight different gauges.  This data shows the baseline imperfections in the correlations due to calibration errors in the D-Wave device.  Panel B is for the simulated quantum annealer (SQA) and the D-Wave device, with the latter averaged over $16$ random gauges.  This correlation is nearly as good as in panel A, indicating
good correlations between the two methods..  Panel C is for the classical spin dynamics and the D-Wave device, and shows poor correlation.  Panel D shows the correlation between success probability and the mean Hamming distance of excited states found at the end of the annealing for $N = 108$ spin instances with local random fields.  Easy
(hard) instances tend to have a small (large) Hamming distance.  The colour scale indicates how many of the instances are found in a pixel of the plots.
}
\label{fig:correlations}
\end{figure}

The third test, shown in figure~\ref{fig:correlations}, is perhaps the most enlightening, as it plots the correlation of the success probabilities between the DW data and the other models.  As a reference for the best correlations we may expect, we show in panel A) the correlations between two different sets of eight gauges (different embeddings of the same problem on the device, see Methods and supplementary material): no better correlations than the device with itself can be expected due to calibration errors. Panel B) shows a scatter plot of the hardness of instances for the simulated quantum annealer and the D-Wave device after gauge averaging.  The high density in the lower left corner (hard for both methods) and the upper right corner (easy for both methods) confirms the similarities between the D-Wave device and a simulated quantum annealer.  The two are also well correlated for instances of intermediate hardness. The similarity to panel A) suggests almost perfect correlation with SQA, to within calibration uncertainties.

In panel C) we show the correlation between the classical spin dynamics model and the device. 
Some instances are 
easily solved by the classical mean-field dynamics, simulated quantum annealing, and the device. However, as can be expected from inspection of their respective distributions in figure~\ref{fig:histogram}, there is no apparent correlation between the hard instances for the spin dynamics model and the success probability on the device, nor does there appear to be a correlation for instances of intermediate hardness, in contrast to the correlations seen in panel A). Similarly, there are poor correlations \cite{comment-SS} with a classical spin dynamics model of reference \cite{Smolin}.

The correlations between the simulated classical annealer and the D-Wave device, shown in the supplementary material, are significantly worse than between SQA and the device.

\begin{figure}[tb]
\centering
\includegraphics[width=0.48\columnwidth]{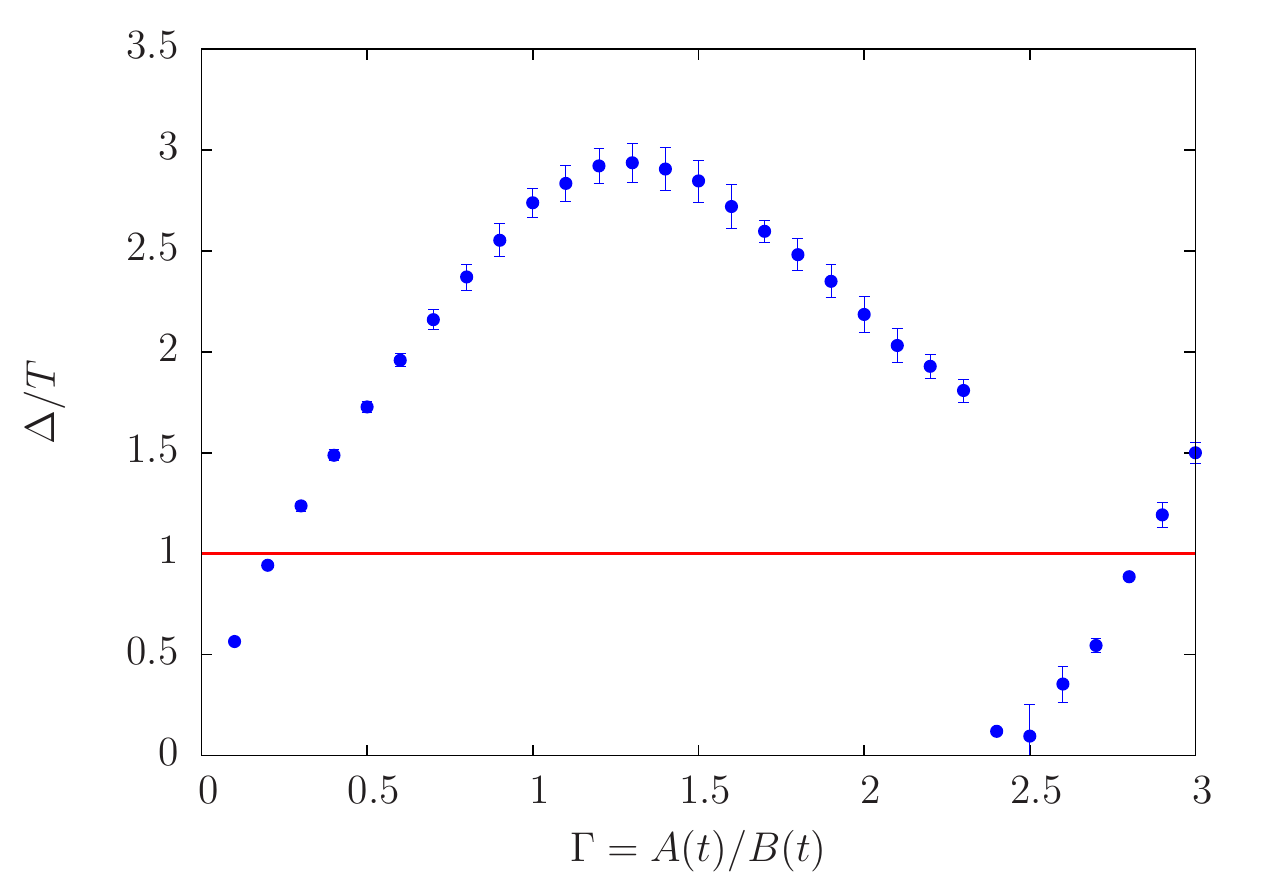}
\includegraphics[width=0.48\columnwidth]{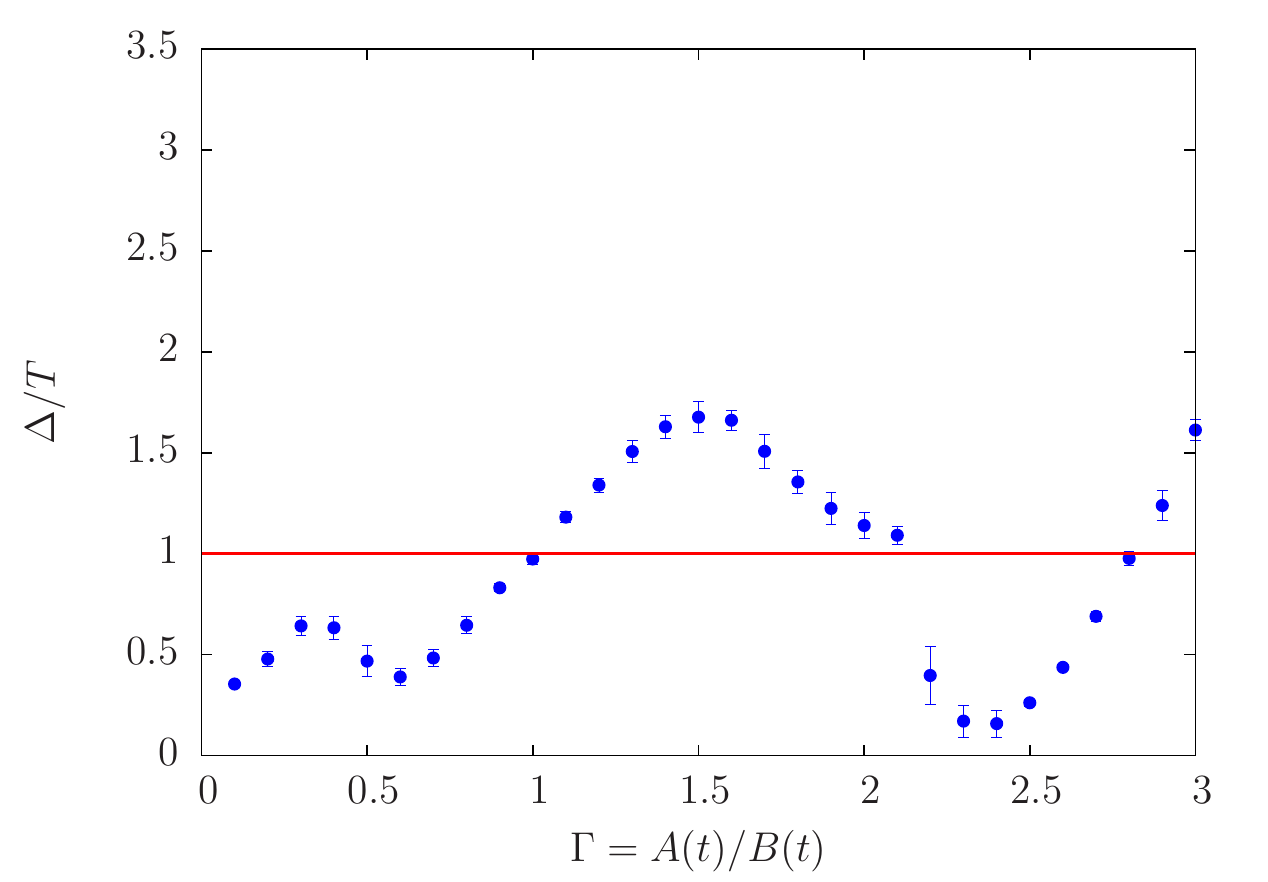}
\caption{{\bf Evolution of the lowest spectral gap.} Shown in blue are (upper bounds for) the gaps between ground state and the lowest excited state in units of the temperature for two typical spin glass instances of $N=108$ spins as a function of the ratio of transverse field to coupling, $\Gamma=A(t)/B(t)$. 
Note that $\Gamma$ depends on the annealing schedule details and decreases as a function of time. In the left panel we show the gap for an ``easy'' instance  with success probability 98\% and on the right for a ``hard'' instance with success probability 8\%. } 
\label{fig:gaps}
\end{figure}

We next provide evidence for the bimodality being due to quantum effects.  Our first evidence comes from the simulated quantum annealer. When lowering the temperature thermal updates are suppressed, quantum tunneling dominates thermal barrier crossing, and we observe a stronger bimodality; indeed a similar bimodal distribution arises also in an ensemble of (zero-temperature) Landau-Zener problems with a smooth distribution of gaps. In contrast, thermal effects become more important as we increase the temperature, and eventually the bimodality vanishes (see supplementary material). 
To provide further evidence we picked five hard and five easy instances and performed extensive QMC  simulations to estimate the spectral gap between the ground state and  the first excited state using a method similar to that of Refs. \cite{Kashurnikov1999,Young2008}. A representative result of one easy and one  hard instance is shown in figure~\ref{fig:gaps}; results for the other instances are shown in the supplementary material. For all instances, we found that the gap trivially closes around a ratio $\Gamma=A(t)/B(t)\approx2.3$ of  transverse field to Ising coupling, related to a global $Z_2$ spin inversion symmetry. The gap also closes towards the end of the schedule  as $\Gamma\rightarrow0$, when multiple states are expected to become degenerate ground  states. Neither of these small gaps has a detrimental effect on finding 
the ground state, since even after choosing the wrong branch at these avoided level crossings (either by thermalisation or diabatic transitions) the system still ends up in a ground state at the end of the annealing. The hard instances, however, typically 
have additional avoided level crossings with small gaps as is seen at 
$\Gamma\approx0.5$ in the right panel of figure~\ref{fig:gaps}.  These additional avoided level crossings cause failures of the annealing due to transitions to higher energy states, thus making the problem ``hard''. An explanation of the origin of small gap avoided level crossings for the hardest instances is presented in the supplementary material.

Investigating the excited states found by the device provides additional confirmation for the ``hard'' instances being due to avoided level crossings with small gaps. In panel D) of figure~\ref{fig:correlations}
we show a scatter plot of the mean Hamming distance of excited states versus success probability. The Hamming distance is the number of spins that need to be flipped to reach the closest ground state. We find that for the ``easy'' instances the Hamming distance is typically small. The associated spin flips are often due to thermal errors that can easily be corrected classically as discussed in the supplementary material. The ``hard'' instances on the other hand typically result in excited states with a large Hamming distance. This means that there the device typically finds local minima far away from ground states.  Many spins would need to be flipped to reach a ground state, which results in small tunneling matrix elements between the state found and the true ground state and thus small gap avoided level crossings \cite{Altshuler2010}.

Combining all these observations we have strong evidence for quantum behaviour in the device: unlike the classical annealer and classical spin dynamics, the simulated quantum annealer and the device split instances into hard and easy ones whose success probability is strongly and positively correlated. The same holds for the joint energy-success probability distributions. The bimodality of the success probability distribution in the case of the device can be understood as being due to quantum effects, with the hard instances being such due to small tunneling matrix elements (and corresponding small gaps) resulting between rarely found ground states and easily found excited states during the evolution.\\

\begin{figure}
\centering
\includegraphics[width=0.75\columnwidth]{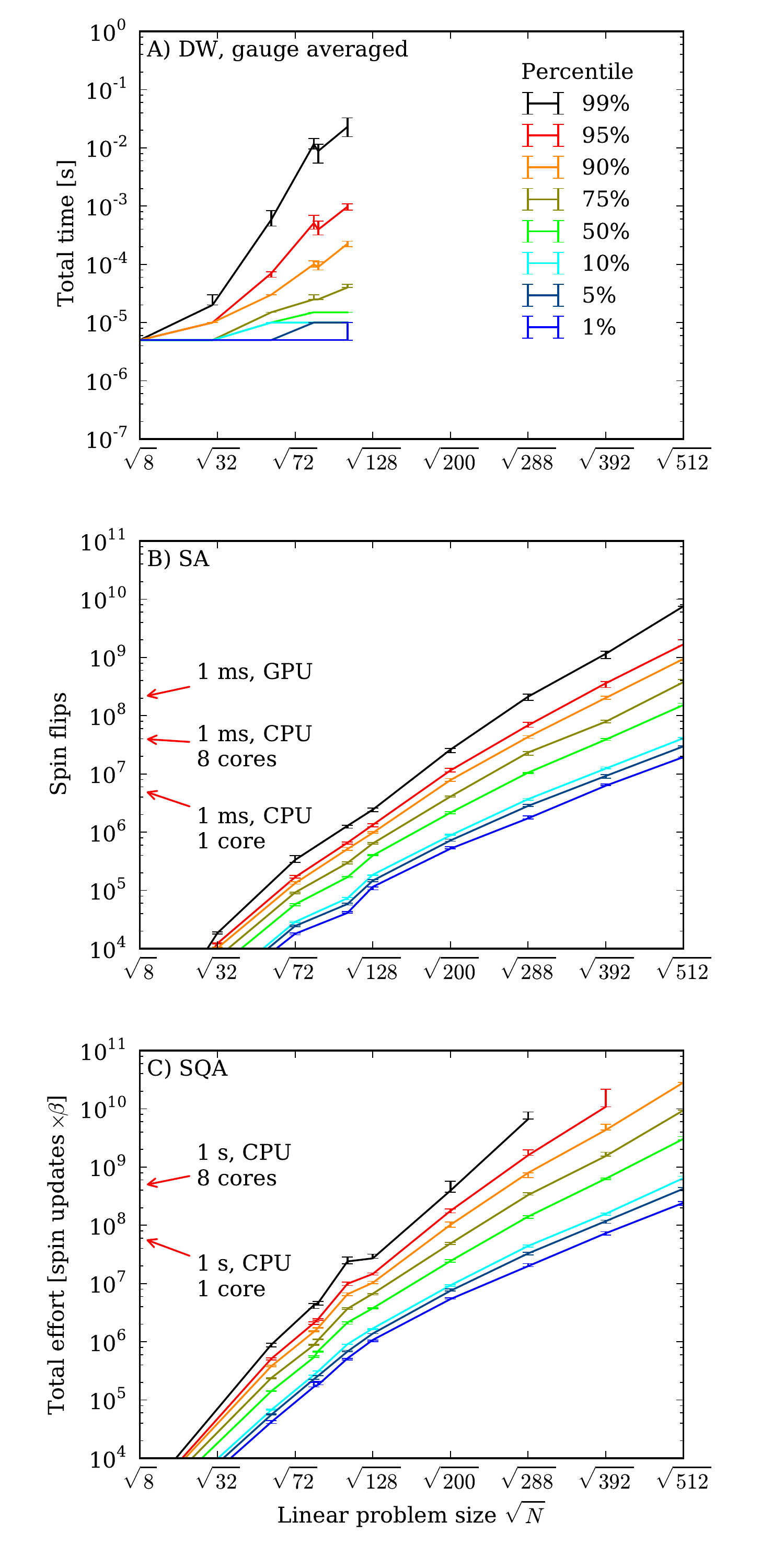}
\caption{{\bf Scaling with problem size.}  Shown is the scaling of the total effort to find the ground state with 99\% probability for A) the D-Wave device, B) the simulated annealer and C) the simulated quantum annealer.  The individual lines show the scaling of the various percentiles, from the 1\% easiest instances (0.01 percentile) to the 1\% hardest instances (0.99 percentile).  For the simulated annealers the vertical axis shows the total effort in number of spin updates for the simulated classical and quantum annealer. Arrows mark the number of spin updates that can be done in 1ms or 1s  on the reference machines.}
\label{fig:scaling}
\end{figure}

\noindent{\textbf{Scaling}}\\
We finally investigate the scaling of the annealing effort with problem size $N$. As a first reference we investigated four exact algorithms discussed in the supplementary material. An an exact version of belief propagation \cite{dechter1999bucket} performed fastest, requiring around $60\,\textrm{ms}$ for $N=128$ and $3$ minutes for $N=512$ on 16 cores of an Intel Xeon CPU, comparable to the timings reported in reference \cite{McGeoch}.

Since the tree width of the chimera graph scales as $\sqrt{N}$ \cite{Choi2}, exact solvers making use of the graph structure scale asymptotically no worse than $\exp(c \sqrt{N})$ and similar scaling is observed also for the simulated annealers discussed below.

For the D-Wave device (and the simulated annealers) we only take into account the intrinsic annealing time  and not overhead from programming the couplers and readout of the results. We calculate the total annealing 
time $Rt_f$, defined as the product of the annealing 
time $t_f$ of one annealing run multiplied by the number of repetitions $R$ needed to find the ground state at least once with $99\%$ probability.  From the success probability $s$ of a given percentile we calculate the required number of repetitions  $R_p=\log (1-p)/\log (1-s)$, with $p=0.99$. 

In figure~\ref{fig:scaling}A) we show the scaling of the typical (median) instance as well as various percentiles of hardness on the D-Wave device. The rapid increase of the higher percentiles is due to  calibration issues that cause problems for a fraction of problems.
 Focusing on the median we see only a modest increase in total annealing time from $5\,\mu$s to $15\,\mu$s, corresponding to three repetitions of the annealing. While an extrapolation of the observed experimental scaling is tempting, this will not yield the true asymptotic scaling. The reason is that the total annealing time depends sensitively on the choice of $t_f$ and 
for the device the minimal time of $t_f=5\, \mu s$ turns out to be suboptimal (see supplementary material).

For the simulated classical and quantum annealer, on the other hand, we can calculate the optimal annealing time and plot its scaling in figure  \ref{fig:scaling}B) and C). The effort here is measured in the number of spin updates, defined as $R NN_{\rm updates}$ where $N_{\rm updates}$ is the optimal number of updates per spin to minimise the total effort. Indicated by an arrow is the number of spin updates that can be performed in a millisecond on an  8-core Intel Xeon E5-2670 CPU and on an Nvidia K20X GPU.  At $N=108$ 
this yields a total annealing time of $4.3\,\mu s$ and $0.8\, \mu s$ respectively, slightly faster than the D-Wave device, while the simulated quantum annealer is substantially slower. Classical spin dynamics is not competitive (as we showed above it suffers from an abundance of hard instances) and was not considered.

Increasing the problem size up to $N=512$ spins our simulated quantum annealer shows that the fraction of easy instances drops rapidly (see supplementary material); perhaps surprisingly, $N=200$ is still a ``small" problem. As a consequence, both the annealing time and the number of repetitions need to be increased and the total effort grows exponentially both in the simulated quantum annealer and in the simulated classical annealer. We find an increase of the median effort by about three orders of magnitude when increasing the problem size from $N=108$ to $N=512$ spins. We note that even then the simulated annealer finds solutions in few milliseconds, which is faster than first benchmarks reported for a next-generation $512$ qubit D-Wave device \cite{McGeoch}.   We also observe that the simulated quantum annealer scales slightly worse than the simulated classical annealer for our problems. Investigating for which class of problems simulated quantum annealing is better than simulated classical annealing is an important open question to be addressed in future work.\\

\noindent{\textbf{Conclusions}}\\
Our  experiments have demonstrated that quantum annealing with more than one hundred qubits takes place in the D-Wave One device, despite limited qubit coherence times. The key evidence is the correlation between the success probabilities on the device and a simulated quantum annealer, where the hard instances are characterised  by avoided level crossings with small gaps. Sensitivity to these small gaps of the  quantum model demonstrates that  the device has sufficient ground state quantum coherence to realise quantum annealing of a transverse field Ising model. Considering the pure annealing time, the  performance for typical  (median) instances matches that of a  highly optimised classical  annealing code on a high-end Intel CPU.  

While for $108$ spins a majority of optimisation problems is still relatively easy, it should be possible to address the  open question of quantum speedup on future devices with more qubits, by comparing the scaling results of the simulated classical and quantum annealers to experiments (see supplementary material). Going to even larger problem sizes we soon approach the limits of classical computers. Optimistically extrapolating using the scaling observed in our simulations, the median time to find the best solution for our test problem will increase from milliseconds for $512$ variables to  minutes for $2048$ variables, and months for $4096$ variables.
A quantum annealer showing better scaling than classical algorithms for these problem sizes would be an exciting breakthrough, validating the potential of quantum information processing to outperform its classical counterpart. 

\section*{Methods}
{\small 
{\em Quantum annealing} was performed on the D-Wave One Rainer chip installed at the Information Sciences Institute of the University of Southern California. The device has been described in extensive detail elsewhere \cite{harris_experimental_2010_1,Harris2010,berkley2010scalable}. After programming the couplings, the device was cooled for $2.5\, s$, and then $1000$ annealing runs were performed using an annealing time of $t_f=5\,\mu s$. Annealing is performed at a temperature of $0.4\, \textrm{GHz}$, with an initial transverse field starting at $A(0)\approx 5\, \textrm{GHz}$, going to zero during the annealing, while the couplings and local fields are ramped up from near zero to about $B(t_f)\approx 5\, \textrm{GHz}$ at the end of the schedule. Details of the schedule and results 
for longer annealing times are provided in the supplementary material. 

 {\em Simulated annealing} was  performed using the Metropolis algorithm with local spin flips with codes optimised for the $\pm1$ couplings used as test problems. A total of $N_{\rm updates}$ flips per spin were attempted, increasing the inverse temperature $\beta=1/k_BT$ linearly in time from $0.1$ to $3$.  The {\em simulated quantum annealing} simulations were  performed in both discrete and continuous time path integral quantum Monte Carlo simulations with cluster updates along the imaginary time direction to account for the transverse field, combined with Metropolis rejection sampling for the Ising interactions (see the supplementary material for details).

{\em The classical spin dynamics model} replaces the quantum spins by $O(3)$ classical unit vectors $\vec{M}_i$, where the sign of the $z$-component of each spin is the value of the Ising variable. The spins are propagated via the equations of motion $\frac{\partial \vec{M}_i}{\partial t}  = \vec{H}_i(t) \times \vec{M}_i$,
where the time-dependent field $\vec{H}_i(t)$ acting on spin $i$ is a sum of a decaying transverse field (along the $x$ direction) and a growing coupling term (along $z$): $\vec{H}_i(t) \equiv  (1-t/t_f) h   \hat{e}_x- (t/t_f) \sum_j J_{i j} M^z_i \hat{e}_z$. The initial condition is to have all spins perturbed slightly from alignment along the $x$ direction:  $(- \sqrt{1 - \delta_i^2 - \eta_i^2}, \delta_i, \eta_i)$, where $|\delta_i|$, $|\eta_i|<0.1$.

{\em Gauge averaging} was performed on the device by using gauge symmetries to obtain a new model with the same spectrum. This was achieved by picking a gauge factor $a_i=\pm 1$ for each qubit, and transforming the couplings as $J_{ij}\rightarrow a_ia_jJ_{ij}$ and $h_i\rightarrow a_ih_i$. Success probabilities $s_g$ obtained from $G$ gauge choices were arithmetically averaged for the correlation plots and  as $\overline{s} = \prod_{g=1}^{G}(1-s_g)^{1/G}$ for the scaling of total effort (see supplementary material for a derivation).

{\em The ground state energies} were obtained using exact optimisation algorithms, an exact version of belief propagation using Bucket Sort \cite{dechter1999bucket} and a related optimised divide-and-conquer algorithm described in the supplementary material.
}

\medskip

{\small \noindent {\bf Acknowledgements} We acknowledge useful discussions with M.H.~Amin, M.H.~Freedman, H.G.~Katzgraber, C. Marcus, B.~Smith and K.~Svore. We thank Lei Wang for providing  data of spin dynamics simulations, G. Wagenbreth for help optimizing the belief propagation code and P.  Messmer for help with optimizing the GPU codes. Simulations were performed on the Brutus cluster at ETH Zurich and on computing resources of Microsoft Research with the help of J. Jernigan. This work was supported by the Swiss National Science Foundation through the NCCR QSIT, by ARO grant number W911NF-12-1-0523, by ARO MURI grant number W911NF-11-1-0268, by the Lockheed Martin Corporation, by DARPA grant number FA8750-13-2-0035, and by NSF grant number CHE-1037992. MT acknowledges hospitality of the Aspen Center for Physics, supported by NSF grant PHY-1066293. The initial planning of the experiments by MT was funded by Microsoft Research.}

\medskip

{\small \noindent {\bf Author Contributions}  MT, JM and DL designed the experiments and wrote the manuscript, with input from all other authors. SB and ZW performed the experiments on D-Wave One. SI, TR and MT wrote the simulated classical and quantum annealing codes and TR, SI, MT and DW performed the simulations. SB and TR wrote the bucket sort code and divide and conquer codes.  TR, SI, MT, SB, ZW and DL evaluated the data. All authors contributed to the discussion and the presentation of the results.}

\newpage 
\vphantom{10cm}
\onecolumngrid

{\large\bfseries\centering{$\phantom{\hbox{\tt X}}$\\$\phantom{\hbox{\tt X}}$\\Supplementary material for ``Quantum annealing with more than one hundred qubits''\\$\phantom{\hbox{\tt X}}$\\$\phantom{\hbox{\tt X}}$\\$\phantom{\hbox{\tt X}}$\\}}
\twocolumngrid
\setcounter{figure}{0}
\clearpage
\setcounter{page}{1}
\setcounter{equation}{0}

\section{Overview}
Here we provide additional details in support of the main text. Section \ref{sec:graph} shows details of the chimera graph used in our study and the choice of graphs for our simulations. Section \ref{sec:algo} expands upon the algorithms employed in our study.  Section \ref{sec:histograms} presents additional success probability histograms for different numbers of qubits and for instances with magnetic fields, explains the origin of easy and hard instances, and explains how the final state can be improved via a simple error reduction scheme. Section \ref{sec:correlations} presents further correlation plots and provide more details on gauge averaging. Section \ref{sec:scaling} gives details on how we determined the scaling plots and how quantum speedup can be detected on future devices. Finally, section \ref{sec:gap} explains how the spectral gaps were calculated by quantum Monte Carlo (QMC) simulations.

\section{The chimera graph of the D-Wave device.}
\label{sec:graph}

 \begin{figure}[t]
\centering
\includegraphics[width=0.6\columnwidth]{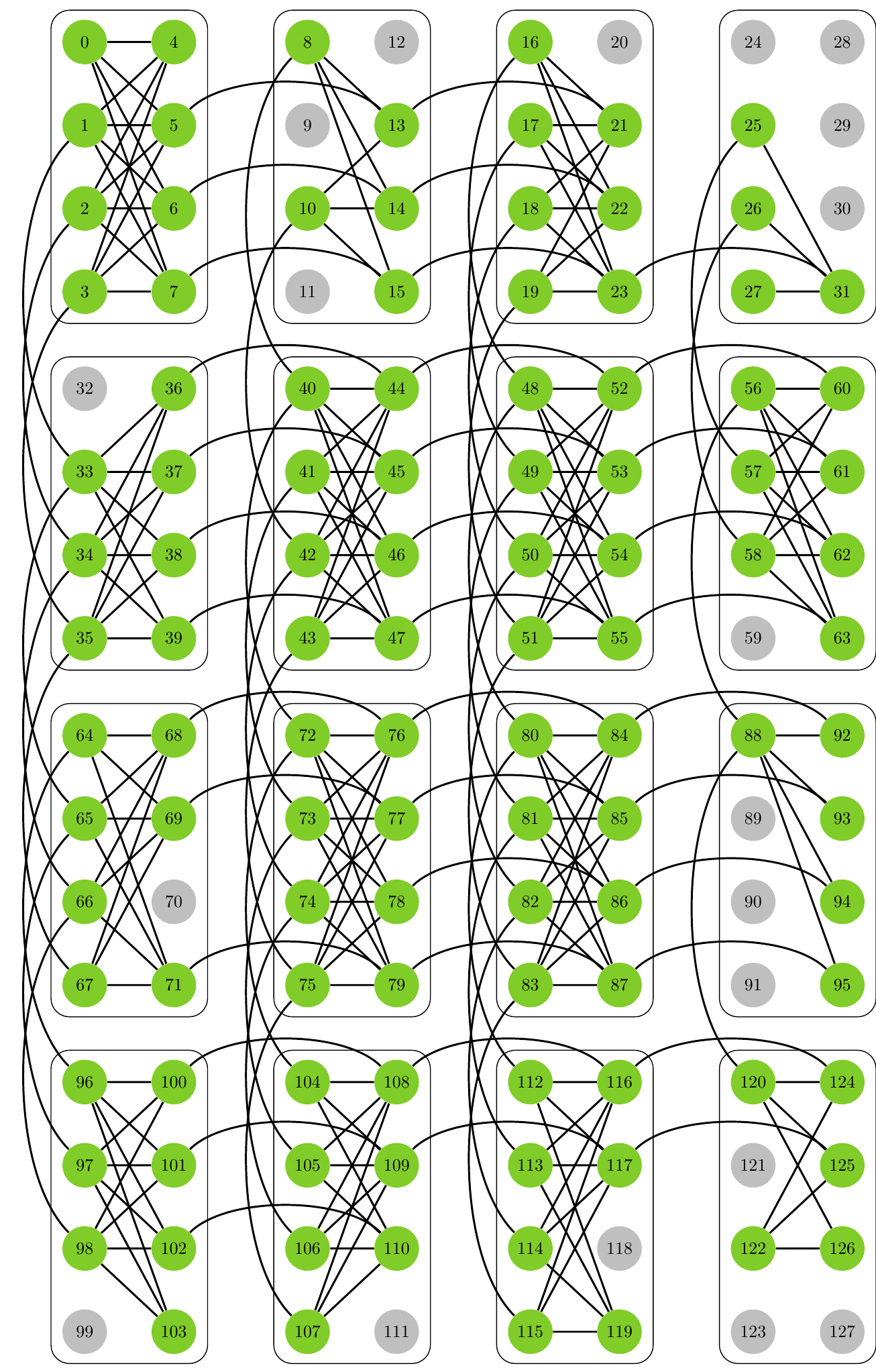}
\caption{{\bf Qubits and couplers in the D-Wave device.} The D-Wave
One Rainer chip consists of $4\times4$ unit cells of eight qubits, connected by programmable inductive couplers as shown by lines.}
\label{fig:qubits}
\end{figure}

The qubits and couplers in the D-Wave device can be thought of as the vertices and edges, respectively, of a bipartite graph, called the ``chimera graph'', as shown in figure~\ref{fig:qubits}. This graph is built from unit cells containing eight qubits each. Within each unit cell the qubits and couplers realise a  complete bipartite graph $K_{4,4}$ where each of the four qubits on the left is coupled to all of the four on the right and vice versa. Each qubit on the left is furthermore coupled to the corresponding qubit in the unit cell above and below, while each of the ones on the right is horizontally coupled to the corresponding qubits in the unit cells to the left and right (with appropriate modifications for the boundary qubits). Of the 128 qubits in the device, the 108 working qubits used in the experiments are shown in green, and the couplers between them are marked as black lines. 

For our scaling analysis we follow the standard procedure for scaling of finite dimensional models by considering the chimera graph as an $L\times  L$ square lattice with an eight-site unit cell and open boundary conditions. The sizes we typically used in our numerical simulations are $L=1,\ldots,8$ corresponding to $N= 8L^2 = 8, 32, 72, 128, 200, 288, 392$ or $512$ spins. For the simulated annealers and exact solvers on sizes of $128$ and above we used a perfect chimera graph. For sizes below $128$ where we compare to the device we use the working qubits within selections of $L\times  L$ eight-site unit cells from the graph shown in figure~\ref{fig:qubits}.

In references \cite{Choi1,Choi2} it was shown that an optimisation problem on a complete graph with $\sqrt{N}$ vertices can be mapped to an equivalent problem on a chimera graph with $N$ vertices through minor-embedding. The tree width of $\sqrt{N}$ mentioned in the main text arises from this mapping. See Section~\ref{sec:exactscaling} for additional details about the tree width and tree decomposition of a graph.

\section{Classical algorithms}
\label{sec:algo}

\subsection{Simulated annealing}
Simulated annealing (SA) is performed by using the Metropolis algorithm to sequentially update one spin after the other. One pass through all spins is called one {\em sweep}, and the number of sweeps is our measure of the annealing time for SA. Our highly optimised simulated annealing code, based on a variant of the algorithm in Ref.  \cite{J.Stat.Phys.44.985,Comput.Phys.Commun.59.387}, uses multi-spin coding to simultaneously perform 64 simulations in parallel on a single CPU core: each bit of a 64-bit integer represents the state of a spin in one of the 64 simulations and all 64 spins are updated at once. A similar code for GPUs uses 32-bit integers and additionally performs many independent annealing runs and  updates many spins in parallel in multiple threads.

\begin{table}[b]
\center{
\begin{tabular}{ | l | c | c |}
\hline
   & spin flips per ns & relative speed \\
\hline
Intel Xeon E5-2670, 1 core  & 5     &  1  \\
Intel Xeon E5-2670, 8 cores & 40    &  8  \\
Nvidia Tesla K20X GPU       & 210   &  42 \\
\hline
\end{tabular}
}
\caption{Performance of  the classical annealer  on our reference CPU and GPU.}
\label{table:sqperf}
\end{table}

The performance of our codes on the classical reference hardware is shown in Table~\ref{table:sqperf}. We  use  high-end chips at the time of writing, an 8-core Intel Xeon E5-2670 ``Sandy Bridge'' CPU and an Nvidia Tesla K20X  ``Kepler'' GPU. To find a ground state of our hardest 108-spin instances with a probability of 99\%,  this translates to a median annealing $32\mu$s on a single core of the CPU, $4\mu$s on eight cores, and 0.8$\mu$s on the GPU, which should be compared to $15\mu$s pure annealing time on the D-Wave device for the same problems.

\begin{figure}[t]
  \centering
  \includegraphics[width=\columnwidth]{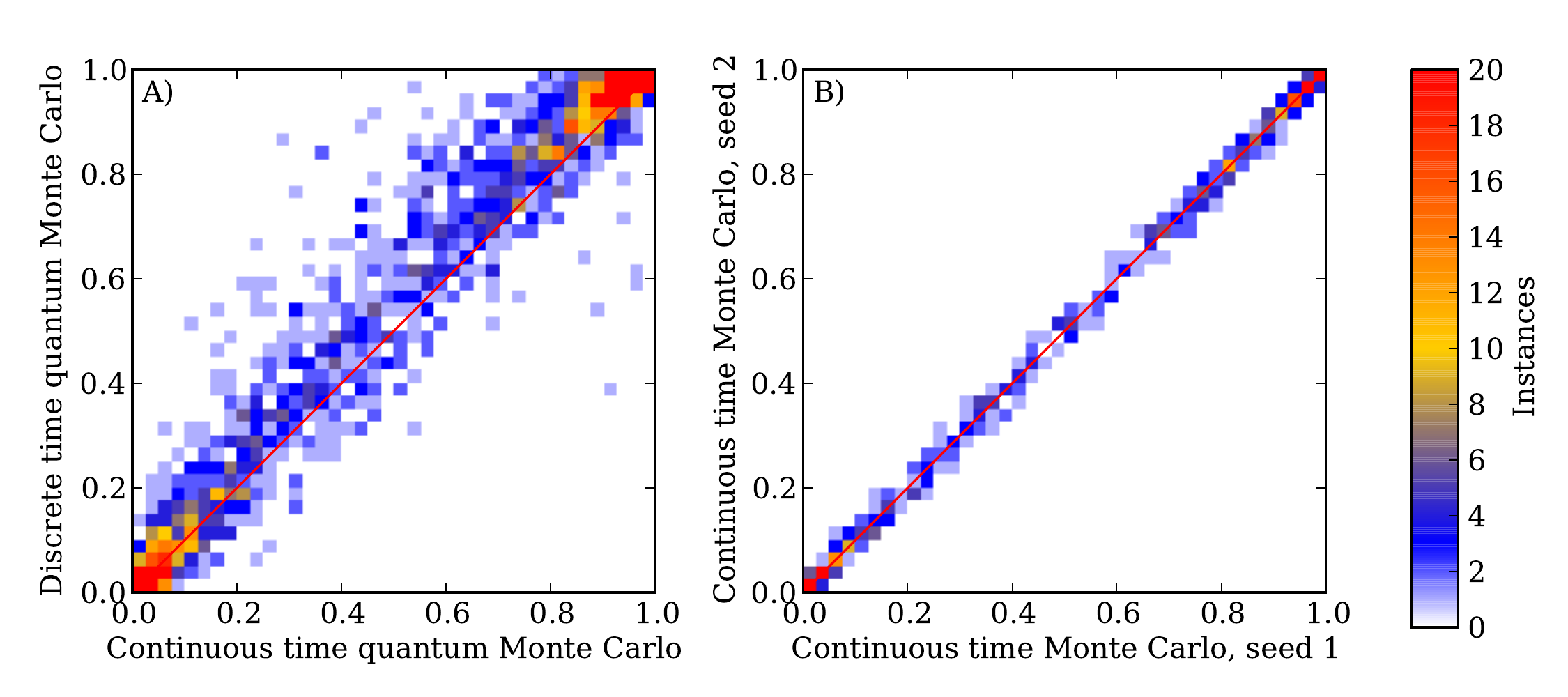}  
  \caption{{\bf Correlation between simulated quantum annealers.} Axes corresponds to success probabilities and pixels are colour-coded according to the number of instances. A) correlations between continuous- and discrete time Monte
    Carlo simulations. The scatter observed here is a measure for the dependence of success probabilities on details of the simulated quantum annealing implementation, for instances with $N=108$ spins performing 10,000 sweeps.
	B) Correlations between two independent sets of 1000 simulations with different initial starting points. Schedule II and 10,000 sweeps are used, see figure~\ref{fig:schedules}. Both simulations were performed at $T=0.1$.}
  \label{fig:ctq_dtq}
\end{figure}

\subsection{Simulated quantum annealing}
\label{sec:qannealing}
For simulated quantum annealing we use both a continuous time algorithm and a discrete time algorithm.

The continuous time algorithm \cite{Eur.Phys.J.B.9.233} constructs segments of a world line in the (imaginary) time direction and flips them using the Metropolis algorithm. Specifically, we pick a random site and introduce new cuts in
the time direction via a Poisson process. Then we calculate the overlaps with
the neighbouring sites and use these overlaps to calculate the Metropolis
acceptance ratios $P_\text{Met}$ and flip a segment with probability
$P_\text{Met}/2$. We cannot grow the cluster along the space directions as in Ref.~\onlinecite{Eur.Phys.J.B.9.233},
which connects segments into larger clusters, since such cluster updates are inefficient in frustrated models like our spin glass.

In order to implement the fastest possible simulated quantum annealing code we 
also implemented a discrete time algorithm similar to that outlined in Ref. \onlinecite{PhysRevB.66.094203}. 
However, unlike Ref. \onlinecite{PhysRevB.66.094203}.  we again used cluster updates along the imaginary time direction, typically with 64 time slices.

To verify that the discrete time code produces results similar to the continuous time 
algorithm -- {\it i.e.}, that the error in the imaginary time direction is small -- we show
a correlation plot between the discrete- and continuous time in \fig{ctq_dtq}A). The scatter observed in this plot is a measure for the dependence of success probabilities on details of the simulated quantum annealing implementation.  We also show correlations for a continuous time Monte Carlo using two different sets of random seeds for initial configurations and updates in \fig{ctq_dtq}B). The scatter of points is within the 3\% ($1/\sqrt{1000}$)  error expected for the success probabilities when performing 1000 annealing runs per instance.

\begin{figure}[t]
  \centering
  \includegraphics[width=\columnwidth]{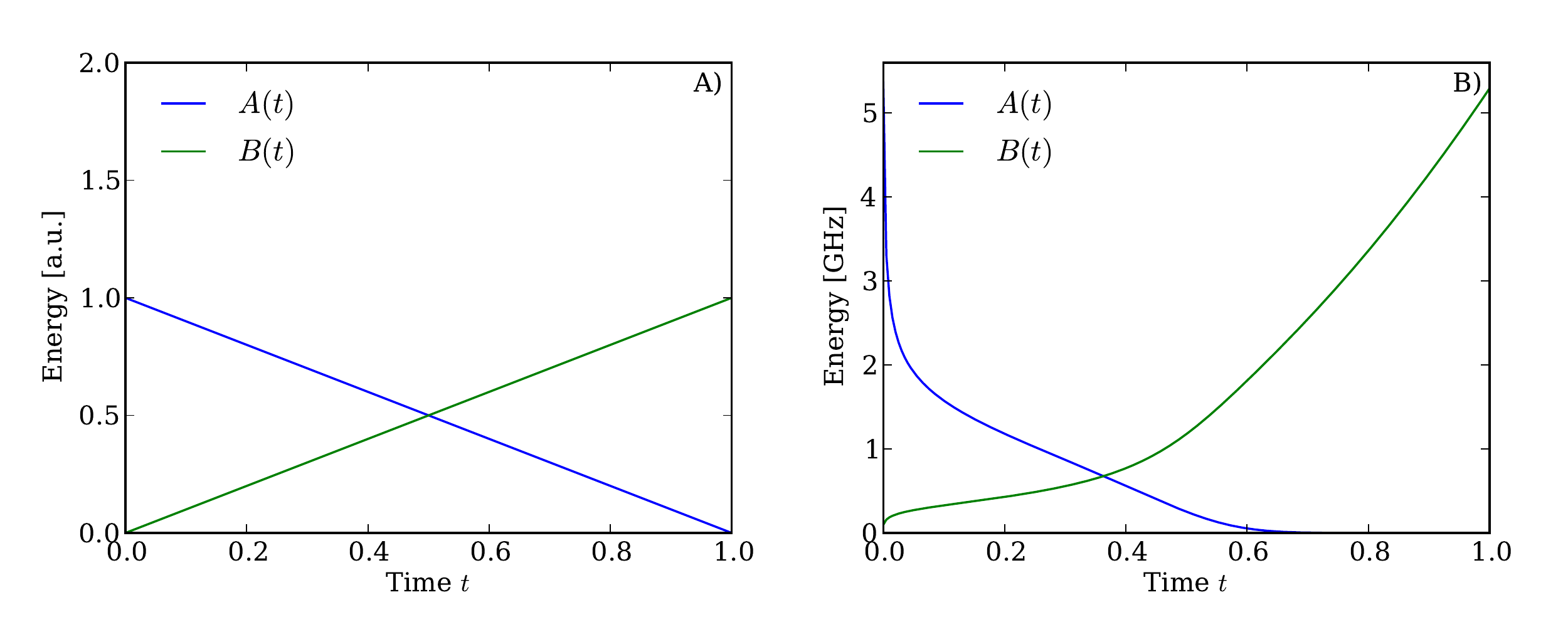}  
  \caption{{\bf Annealing schedules used for Monte Carlo codes} A) schedule I, the linear schedule. B) schedule II, the schedule of the D-Wave device.}
  \label{fig:schedules}
\end{figure}

In Table~\ref{table:qmcperf}, we summarise the performance of these two codes for typical cases using the linear schedule of \fig{schedules}A).

We have performed simulated quantum annealing with two different schedules shown in figure~\ref{fig:schedules}: A) a linear schedule (referred to as schedule I) where the transverse field is ramped down linearly in time and the Ising couplings are ramped up linearly, and B) the schedule used in the D-Wave device (referred to as schedule II). The performance was similar in both cases. For the scaling plots we used the linear schedule I at an optimised temperature ranging from $T=0.33$ to $T=1$  depending on system size, which gives slightly better performance than schedule II. 

For the correlation plots we use the continuous time code  (CTQ) with schedule II --  the average over slightly different schedules for the individual qubits on the device -- but at up to ten times lower temperature than the device temperature of 20mK (0.4 GHz). The simulated quantum annealer requires a temperature about three times lower than the nominal temperature of the device to exhibit a clear bimodal distribution. This can be explained and motivated as follows. When the transverse field is strong the quantum Monte Carlo updates mimic the quantum tunneling taking place in the device, however when the transverse field becomes smaller these Monte Carlo updates turn into local spin flips of a classical (thermal) annealer. The device in this regime, on the other hand, has high tunneling barriers between the two states of the qubits of the device that suppress thermal tunneling over the barrier. To achieve a similar suppression of thermal excitations we need to lower the temperature in the quantum Monte Carlo simulations by at least a factor two. Lowering by more than a factor of ten does not significantly change histograms or correlations, indicating that at the chosen temperature the simulated quantum annealer is dominated by quantum tunneling and not thermal effects.

\begin{table}[t]
\center{
\begin{tabular}{ | l | c | c |}
\hline
 & spin updates per $\mu$s & relative speed \\
\hline
CTQ, 1 core  & 1.3      &  1    \\
CTQ, 8 cores & 3.8      &  8    \\
DTQ, 1 core  & 5.8      &  4.5  \\
DTQ, 8 cores & 46       &  35   \\
\hline
\end{tabular}
}
\caption{{\bf Performance of simulated quantum annealers.} We show performance figures of both our continuous time  (CTQ) and discrete time (DTQ) implementations on an Intel Xeon E5-2670 CPU using schedule II at inverse temperature $\beta = 10$. }
\label{table:qmcperf}
\end{table}

\subsection{Exact solvers}
\label{sec:bucket}

We investigated four exact solvers, akmaxsat \cite{akmaxsat},  the biqmac algorithm \cite{biqmac} used in the spin glass server \cite{sgserver}, exact belief propagation using bucket sort \cite{dechter1999bucket} and a related divide-and-conquer algorithm. The latter is specifically designed for the 
chimera graph, generalising divide-and-conquer for the square lattice. We consider a chimera graph of $M \times M$ 8-spin unit cells ($8M^2$ spins). For each possible configuration of the $4M$ spins on the left side of the unit cells in the first row (which couple vertically) we find and store the optimal configuration of the remaining $4M$ spins with effort $4M$, giving a total effort of $4M\,2^{4 M}$. Finding the $2^{4 M}$ optimal configurations of the next
row of unit cells, one builds up the solution row by row, scaling as ${\cal
  O}\left(M^2\,2^{4 M} \right)$. Since $M\propto \sqrt{N}$, where $N$ is the number of spins, this scaling demonstrates explicitly the claim made in the main text that exact solutions scale no worse than ${\cal O}(\exp(c\sqrt{N}))$, as for the bucket sort algorithm. 
We present the scaling of these algorithms in section \ref{sec:exactscaling}.

\section{Success probability histograms}
\label{sec:histograms}

In this section we provide additional experimental and simulation data, complementing the data shown in the main text.

\subsection{Experimental annealing histograms}

\begin{figure}[t]
  \centering 
  \includegraphics[width=\columnwidth]{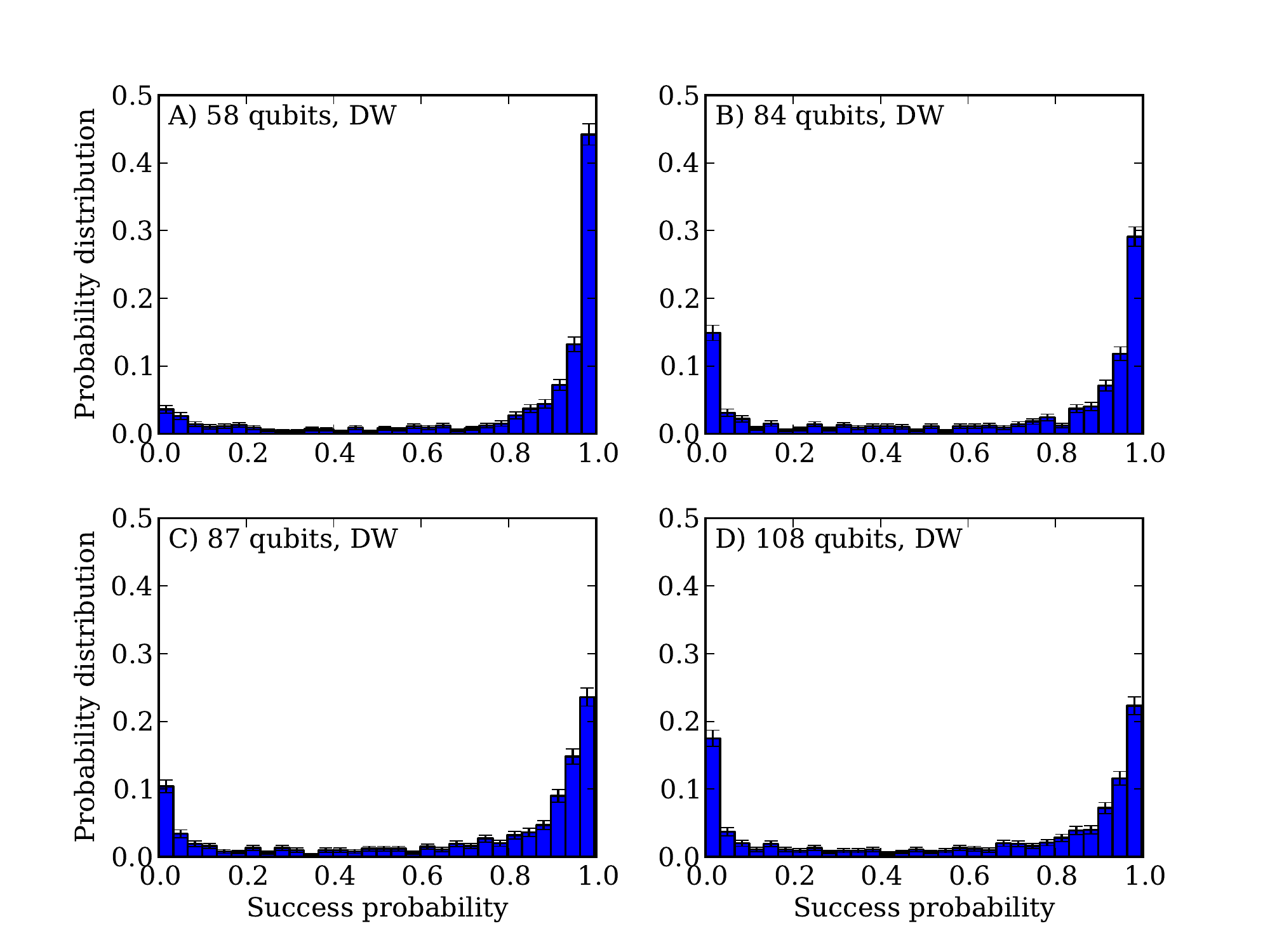}  
  \caption{{\bf Success probability histograms for the D-Wave device for instances without local fields.}  A) using 58 qubits, B) 84 qubits, C) 87 qubits and D) 108 qubits. The peak at low success probability grows with the number of qubits, reflecting the increasing hardness of the corresponding problem instances.}
  \label{fig:dwavehist1}
\end{figure}

\begin{figure}[t]
  \centering 
  \includegraphics[width=\columnwidth]{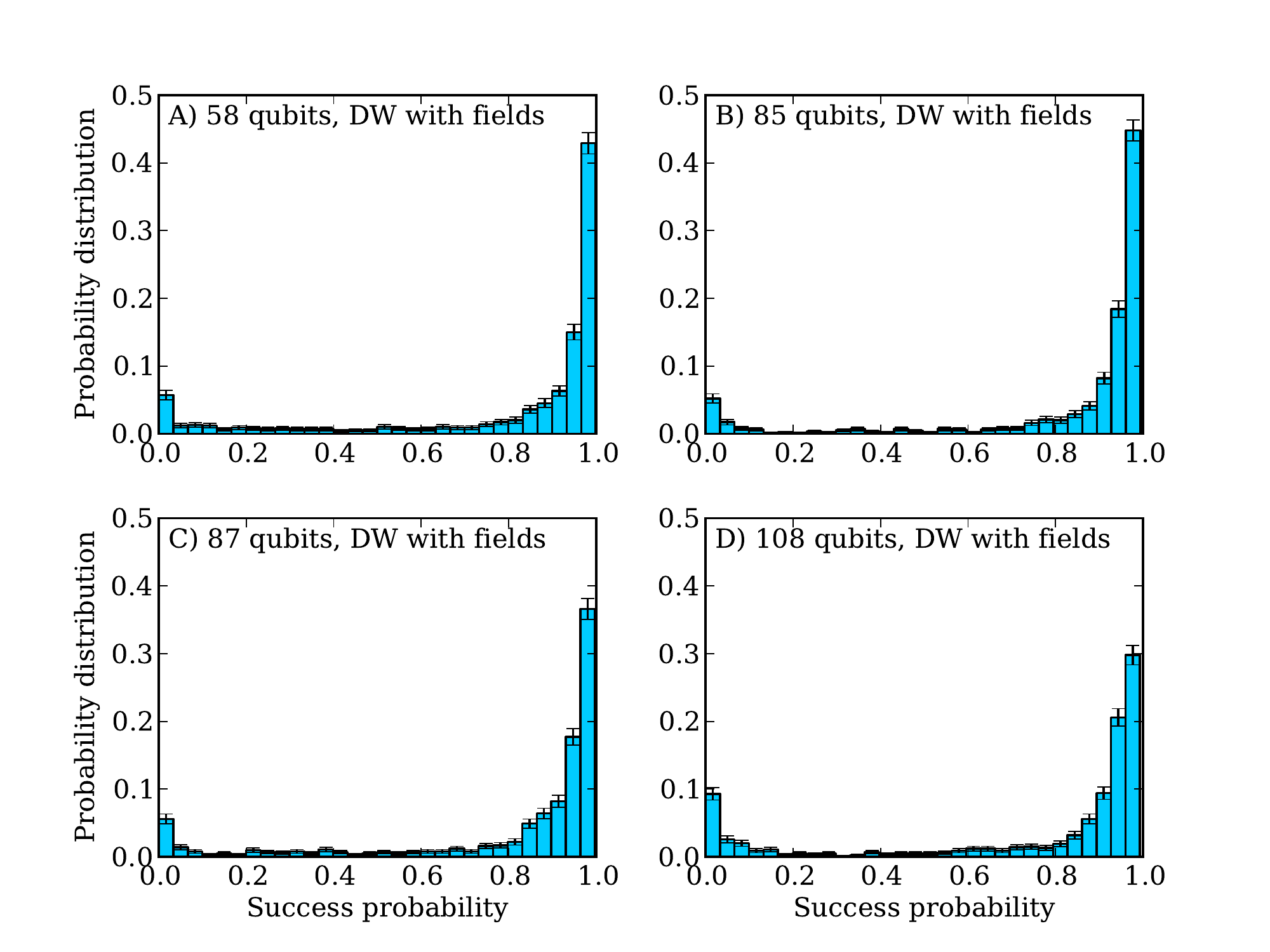}  
  \caption{{\bf Success probability histograms for the D-Wave device for instances with local fields.}  A) using 58 qubits, B) 85 qubits, C) 87 qubits and D) 108 qubits.}
  \label{fig:dwavehist2}
\end{figure}

\begin{figure}[t]
  \centering
  \includegraphics[width=\columnwidth]{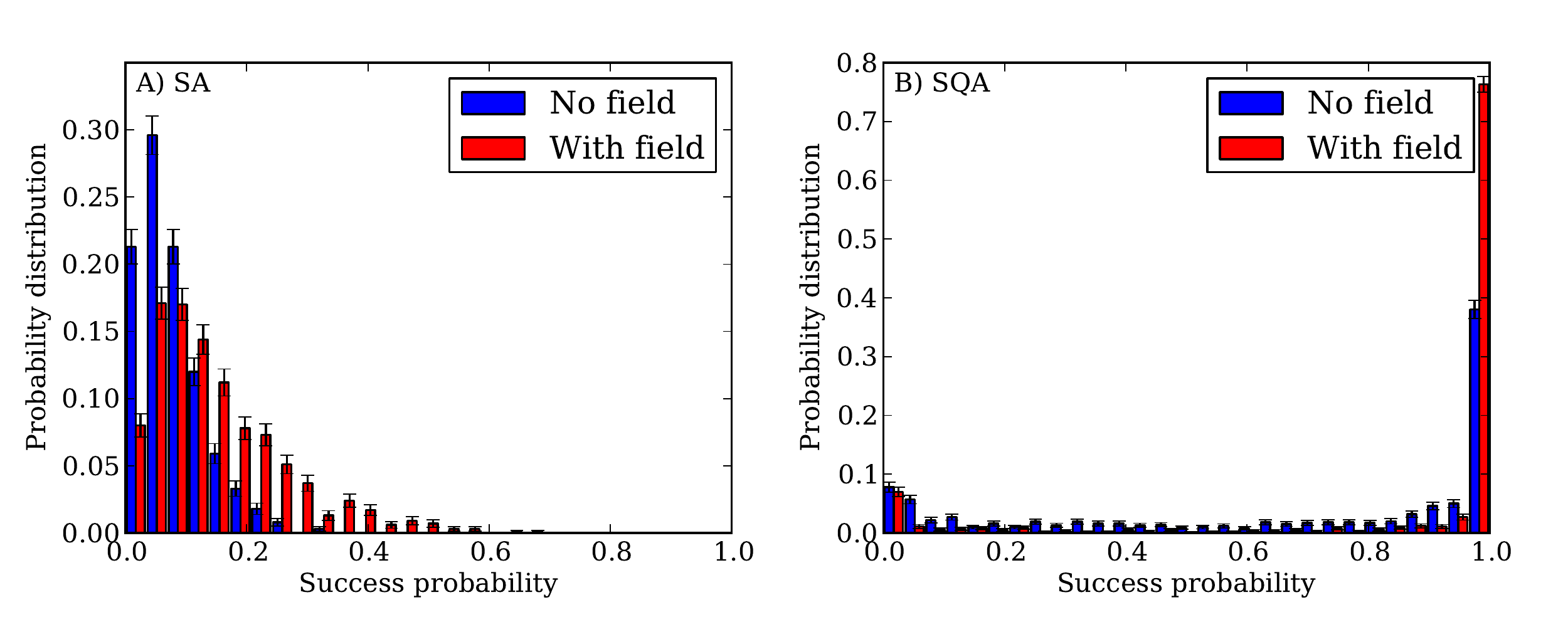}  
  \caption{{\bf Comparison of instances with and without local fields}. A) success probability histograms for the simulated classical annealer with 50 sweeps (updates per spin). B) success probability histograms for the simulated quantum annealer using 10,000 sweeps.  }
  \label{fig:with_without_field}
\end{figure}

In addition to the histogram for $108$ qubits shown in the main text, we
also show histograms for $58, 84, 87$ and $108$ qubits without local fields in
figure~\ref{fig:dwavehist1} and with local fields in figure~\ref{fig:dwavehist2}. 
By comparing these two figures, one
notes that the cases with local fields are in general easier for the D-Wave device. We also verified that this is the case for the simulated
annealers - see figure~\ref{fig:with_without_field}.

To check that the bimodality is not due to faulty couplers or qubits we performed the following analysis for the hardest instances of $N=108$ spin problems where the D-Wave device did not find the ground state with one gauge choice. For each of these instances we looked at the lowest energy configurations reached and determined which spins differ compared to the closest ground state configuration. Closeness is measured in terms of the Hamming distance, which is the number of spins that have to be flipped to reach a ground state configuration. We did not observe a strongly peaked distribution which would have indicated a singly faulty qubit or coupler.

\subsection{Simulated annealing and simulated quantum annealing}

\begin{figure}[b]
  \centering 
  \includegraphics[width=\columnwidth]{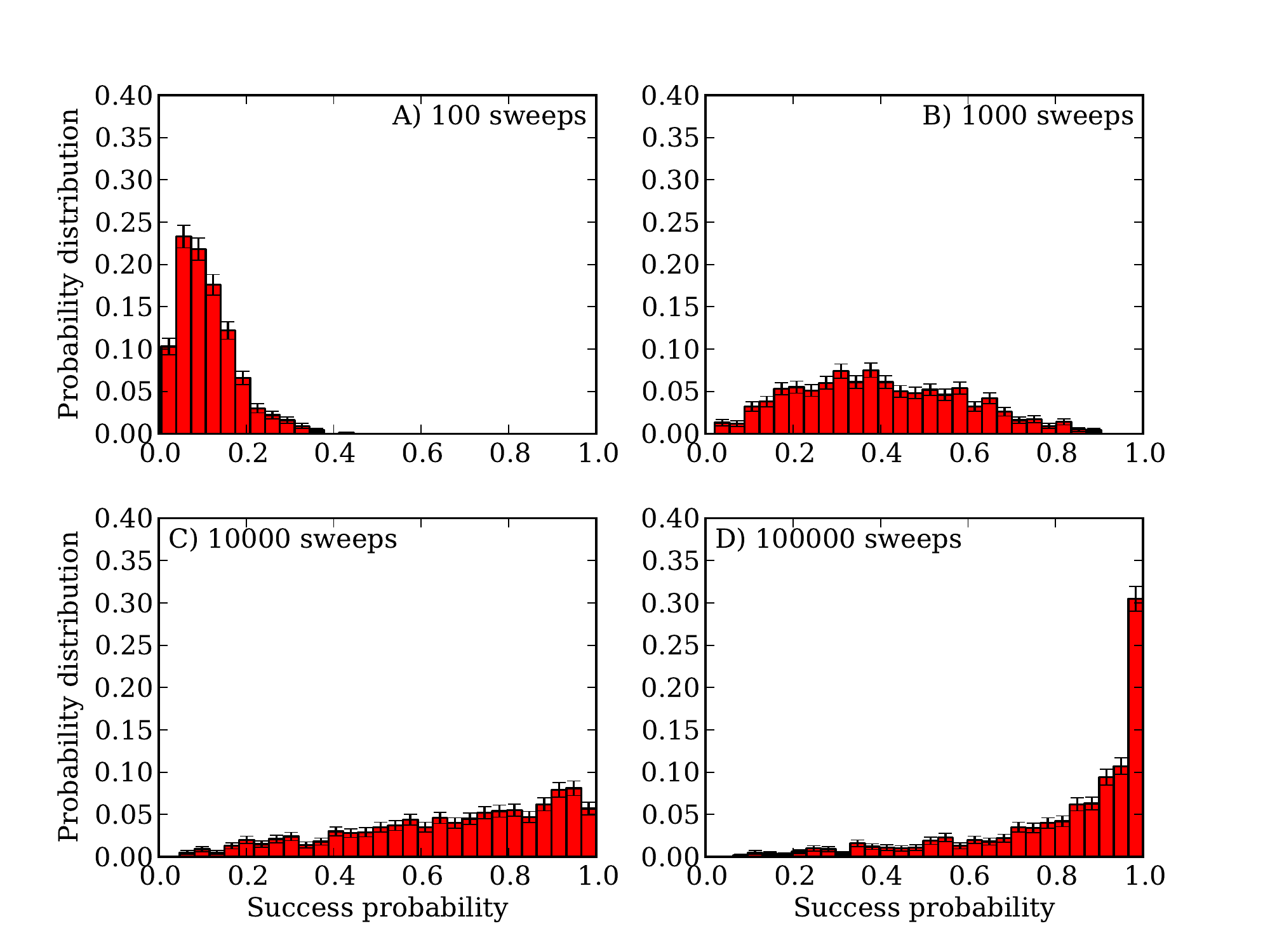}  
  \caption{{\bf Success probability histogram for the simulated classical annealer.}  Annealing times are A)  $100$ sweeps, B) $1,000$ sweeps, D) $10,000$ sweeps and D) $100,000$ sweeps for instances without local fields. }
  \label{fig:classical_histo}
\end{figure}

\begin{figure}[t]
  \centering
  \includegraphics[width=\columnwidth]{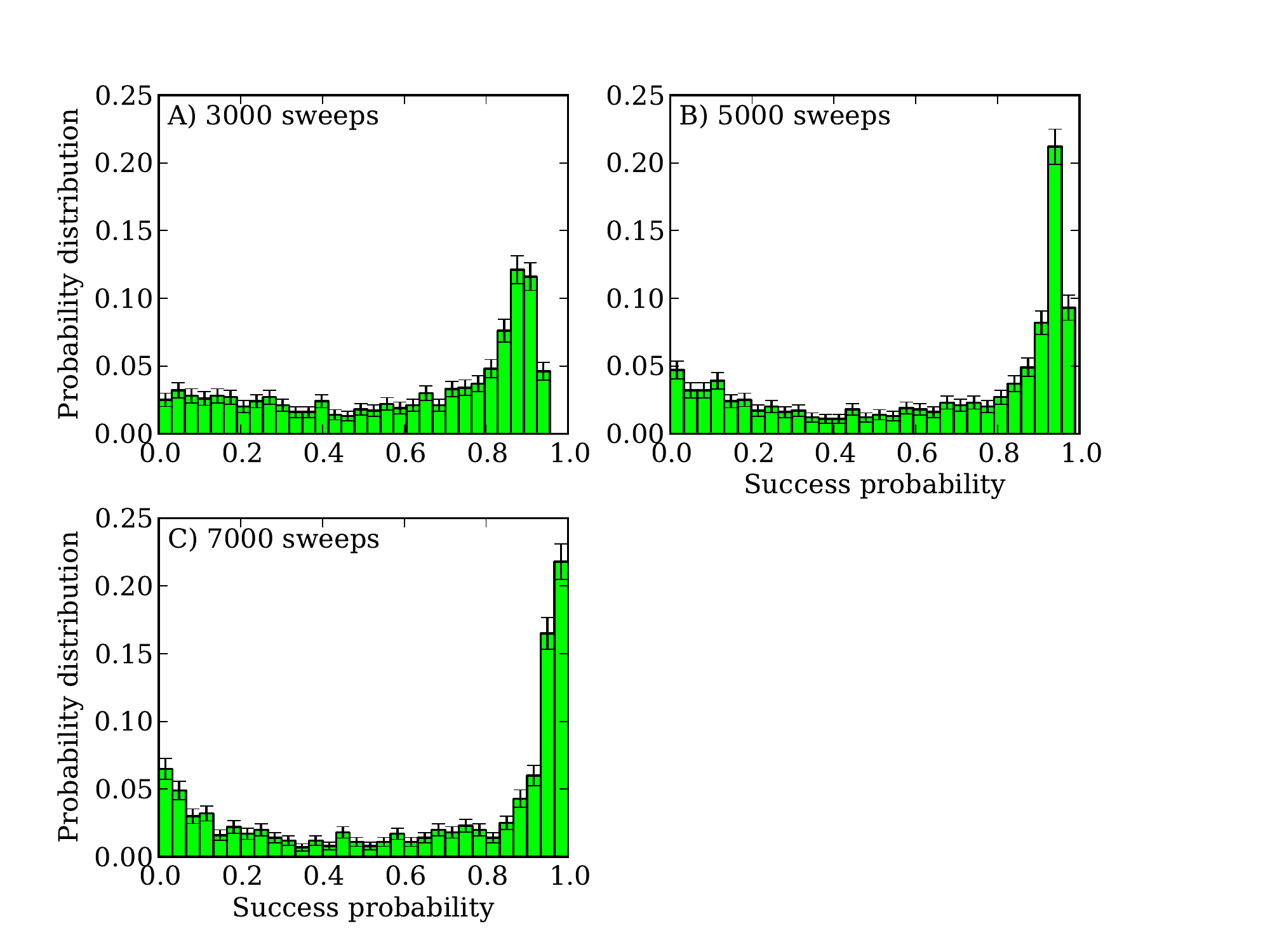}  
  \caption{{\bf Success histograms for simulated quantum annealing.} The bimodal structure becomes more pronounced upon increasing the annealing time from A) $3,000$ sweeps to B) $5,000$ and C) $7,000$ sweeps. All three histograms were obtained for instances without local fields using schedule II and a temperature $T=0.1$.}
  \label{fig:qa_histo_sweeps}
\end{figure}

\begin{figure}[t]
  \centering
  \includegraphics[width=\columnwidth]{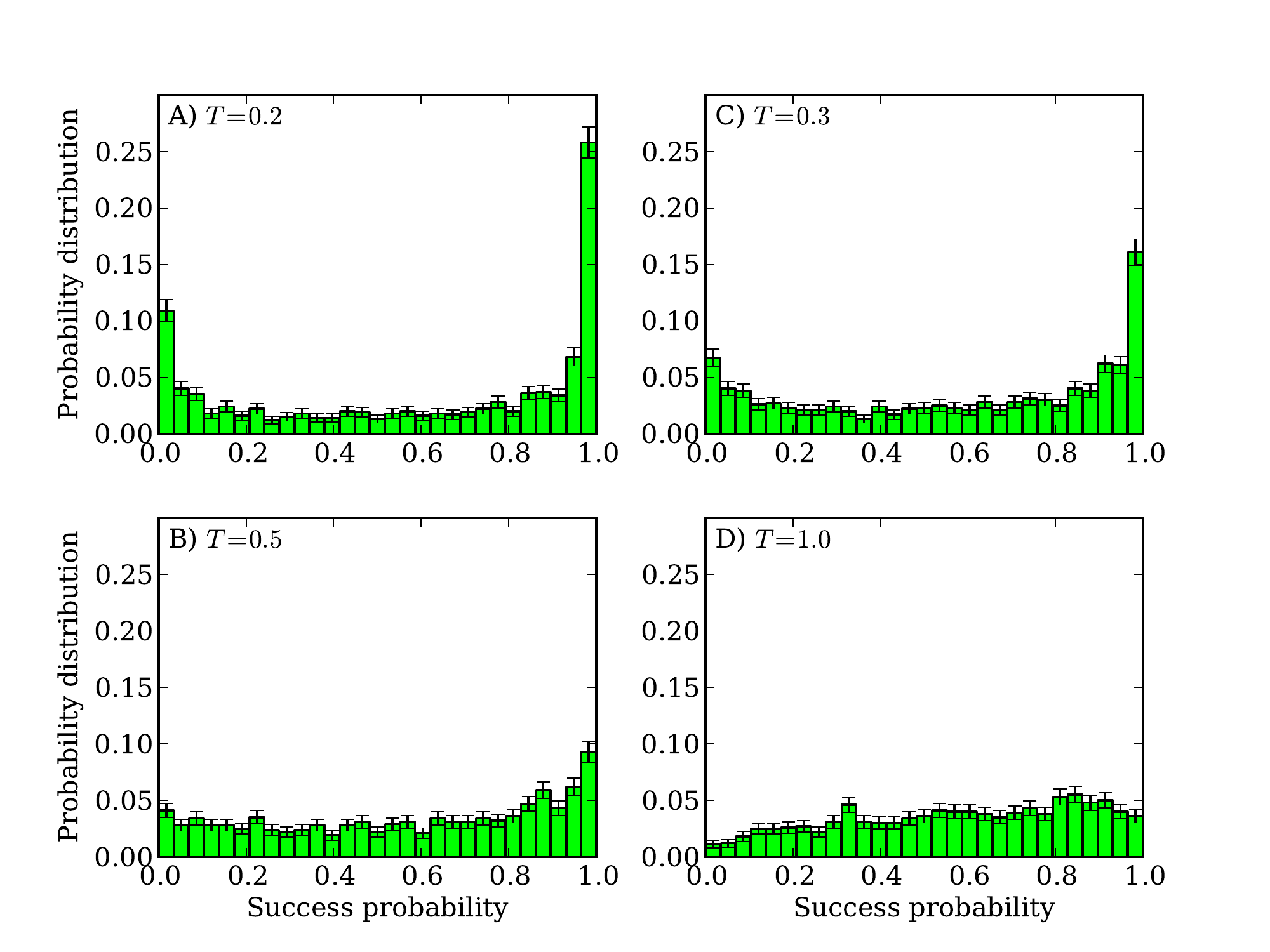}  
  \caption{{\bf Success histograms for simulated quantum annealing as a function of temperature.} The bimodal structure becomes more pronounced at lower temperature and vanishes when increasing the temperature. Shown are results at  A) $T=0.2$ B) $T=0.3$, C) $T=0.5$ and D) $T=1.0$. All histograms were obtained for instances without local fields using schedule II with 7000 sweeps. }
  \label{fig:qa_histo_temperature}
\end{figure}

Figures   \ref{fig:classical_histo}, \ref{fig:qa_histo_sweeps} and \ref{fig:qa_histo_temperature} show the success probability histograms for $N=108$ spin problems for simulated annealing, simulated quantum annealing with different number of annealing sweeps (updates per spin), and simulated quantum annealing at different temperatures. While in the simulated classical annealer the distribution is always unimodal and shifts towards larger success probabilities upon increasing the annealing time, the simulated quantum annealer becomes more strongly bimodal when increasing annealing times.

\begin{figure}[t]
\centering
\includegraphics[width=\columnwidth]{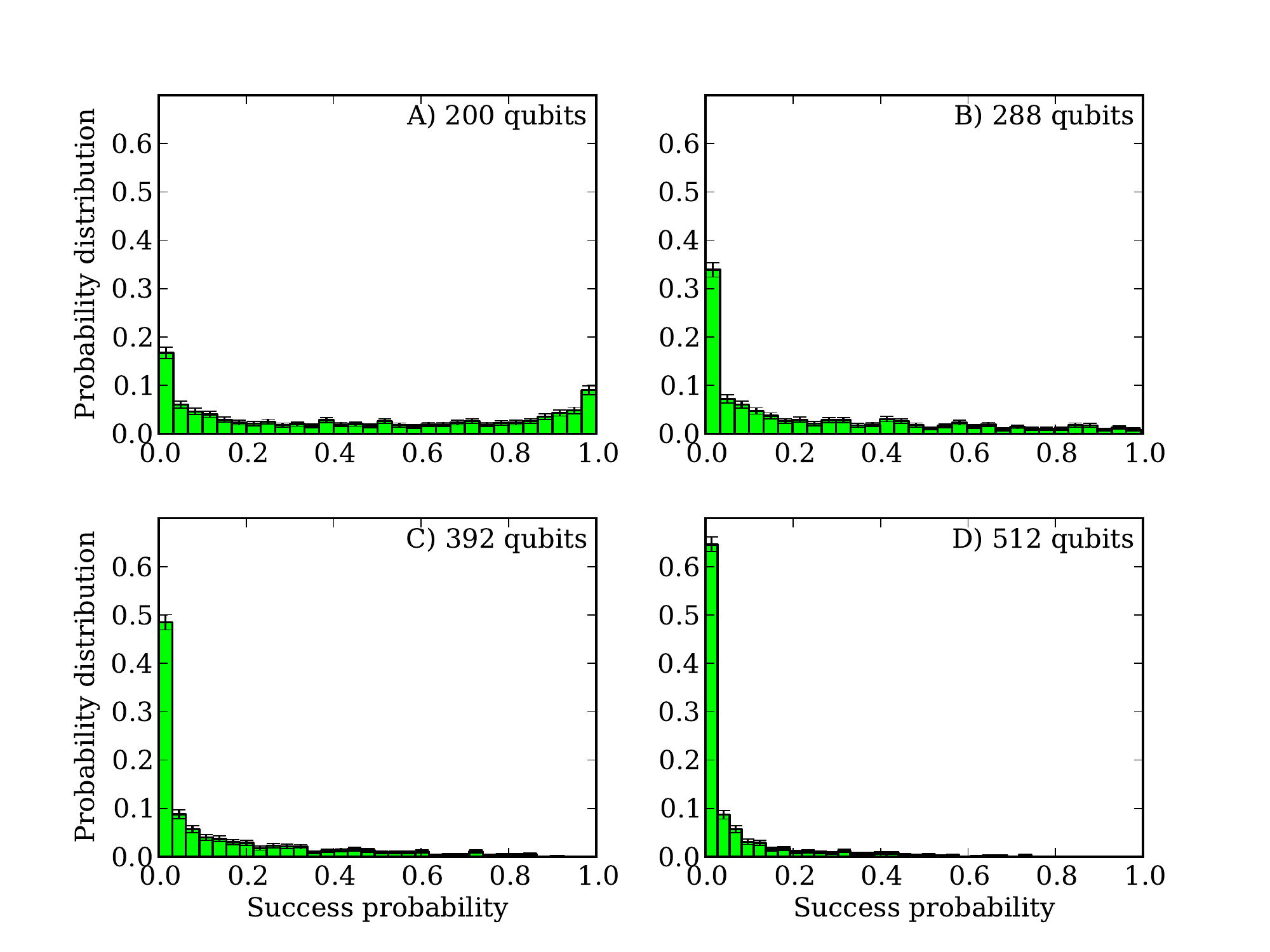}
\caption{{\bf Scaling to larger problem sizes for the simulated quantum annealer for instances without local fields.} Histograms of the hardness distribution obtained by simulated quantum annealing with $N_{\rm updates}=10000$ Monte Carlo updates per spin using schedule II at $T=0.1$. The fraction of easy problems rapidly decreases when increasing the problem size beyond $200$ spins. }
\label{fig:scalingup}
\end{figure}

\begin{figure}[b]
\centering
  \includegraphics[width=\columnwidth]{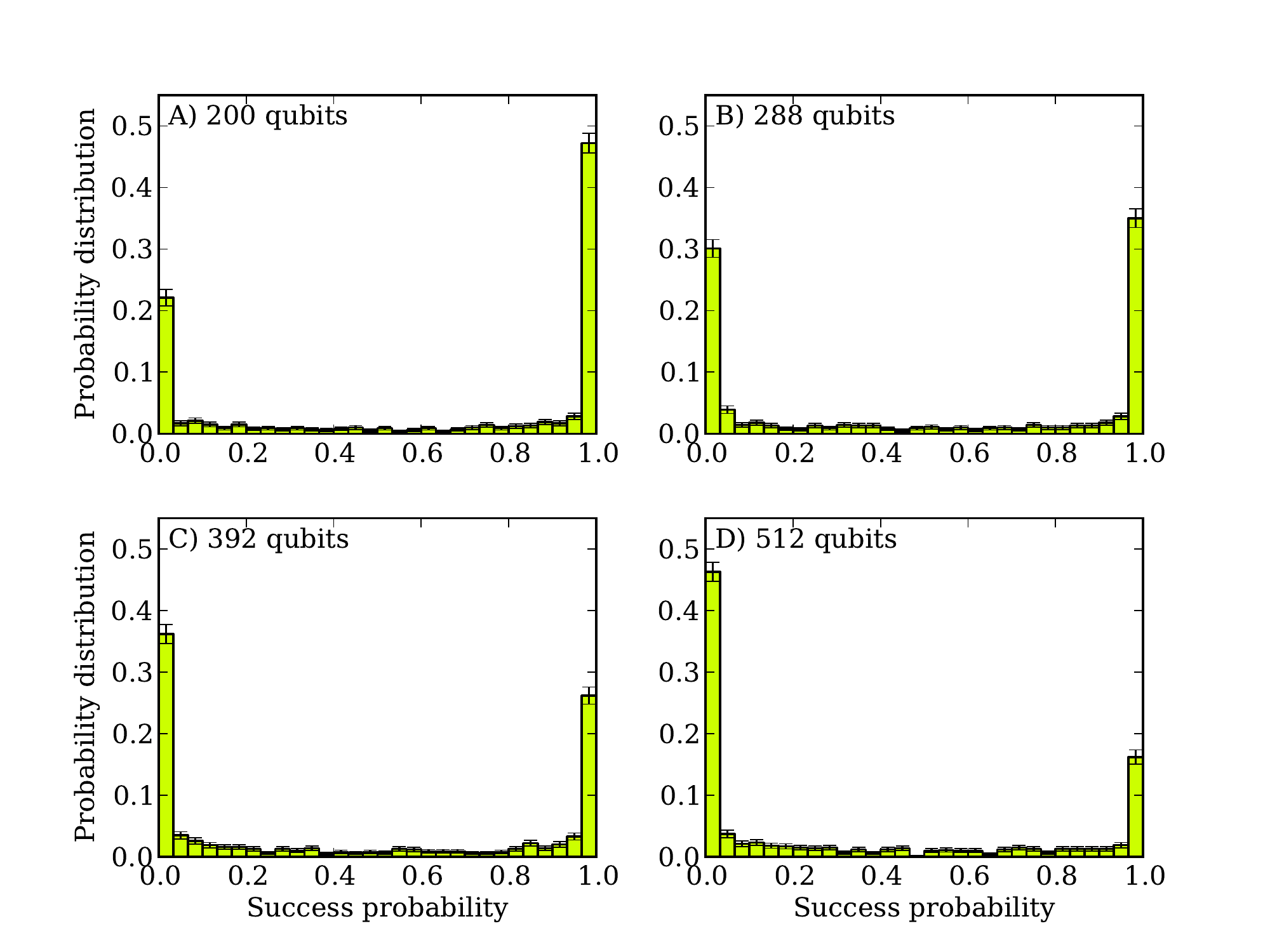}  
\caption{{\bf Scaling to larger problem sizes for the simulated quantum annealer for instances with local fields.} Histograms of the hardness distribution obtained by simulated quantum annealing with $N_{\rm updates}=10000$ Monte Carlo updates per spin using schedule II with $T=0.1$.}
\label{fig:scalingupfields}
\end{figure}

Figures \ref{fig:scalingup}  and \ref{fig:scalingupfields} show the success probability histograms for simulated quantum annealing for instances without and with local fields when increasing the problem size.  As the problem size is increased the weight in the peak at low success probability increases. While for the instances without local fields the bimodality of easy and hard instances vanishes at about $N=288$ spins, it remains up to about $512$ spins for instances with local fields. This indicates that there are more easy instances here than in the case without fields, and hence that instances with fields can be used to test for quantumness in a device scaled up to more qubits. However, once the ``easy'' problems disappear beyond  $512$ spins, and with them the bimodality, alternative analysis methods will be necessary to test for quantumness.

\begin{figure}[t]
\centering
  \includegraphics[width=\columnwidth]{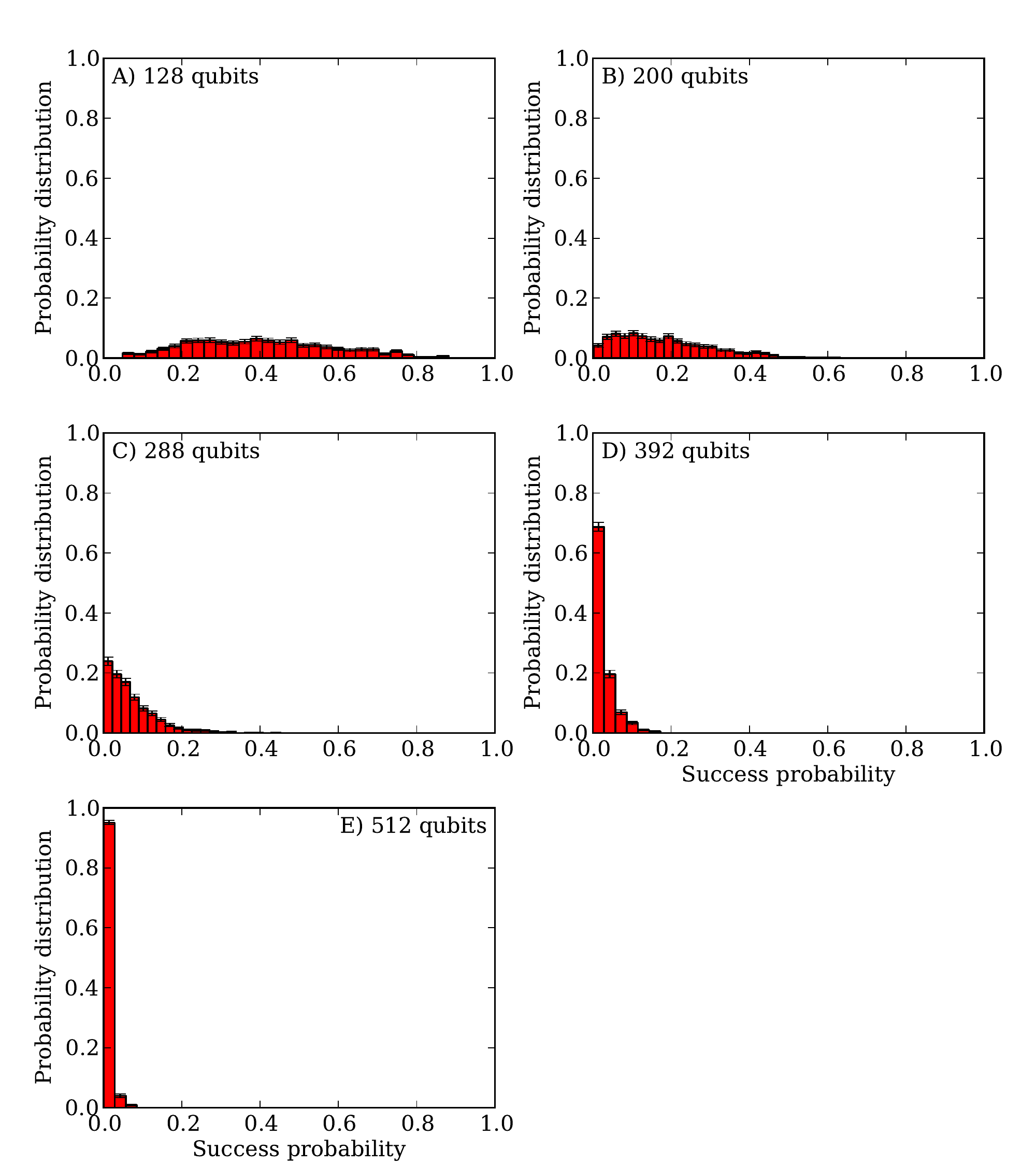}  
\caption{{\bf Scaling to larger problem sizes for the simulated classical annealer.} Histograms of the hardness distribution obtained by simulated  annealing with $N_{\rm updates}=10000$ Monte Carlo updates per spin for instances without local fields. }
\label{fig:scalingup_sm}
\end{figure}

Supplementing the last two figures, we show the evolution of the success probability histograms for different numbers of spins for a simulated classical annealer in figure~\ref{fig:scalingup_sm}. It is seen that, in contrast to simulated quantum annealing where the weight in the two peaks of the bimodal distribution changes, here at fixed annealing time the unimodal peak gradually shifts to smaller success probabilities.

\subsection{Hardness, gaps and free qubits}

Figures 3 and 4 of the main text show that the ``hard'' instances typically exhibit small gaps during the evolution and often get trapped in excited states with a large Hamming distance to the closest ground state. The correlation between small gaps and large Hamming distance can be understood with a simple perturbative argument in the regime of small transverse fields $- \Gamma \sum \sigma^x$, where most of the small gap avoided level crossings appear. At an avoided level crossing between two states with  Hamming distance $d$, $d$ spins need to be flipped to adiabatically follow the ground state. In perturbation theory, the tunneling matrix element between the two states  is of order $\Gamma^d$, exponentially small in the Hamming distance. The small matrix element not only poses problems for adiabatic evolution but also suppresses quantum Monte Carlo updates that connect the two states. This common origin of the hardness in both quantum annealing and simulated quantum annealing explains the observed correlations despite the different underlying dynamics (deterministic \textit{vs} stochastic).

\begin{figure}[t]
  \centering 
  \includegraphics[width=\columnwidth]{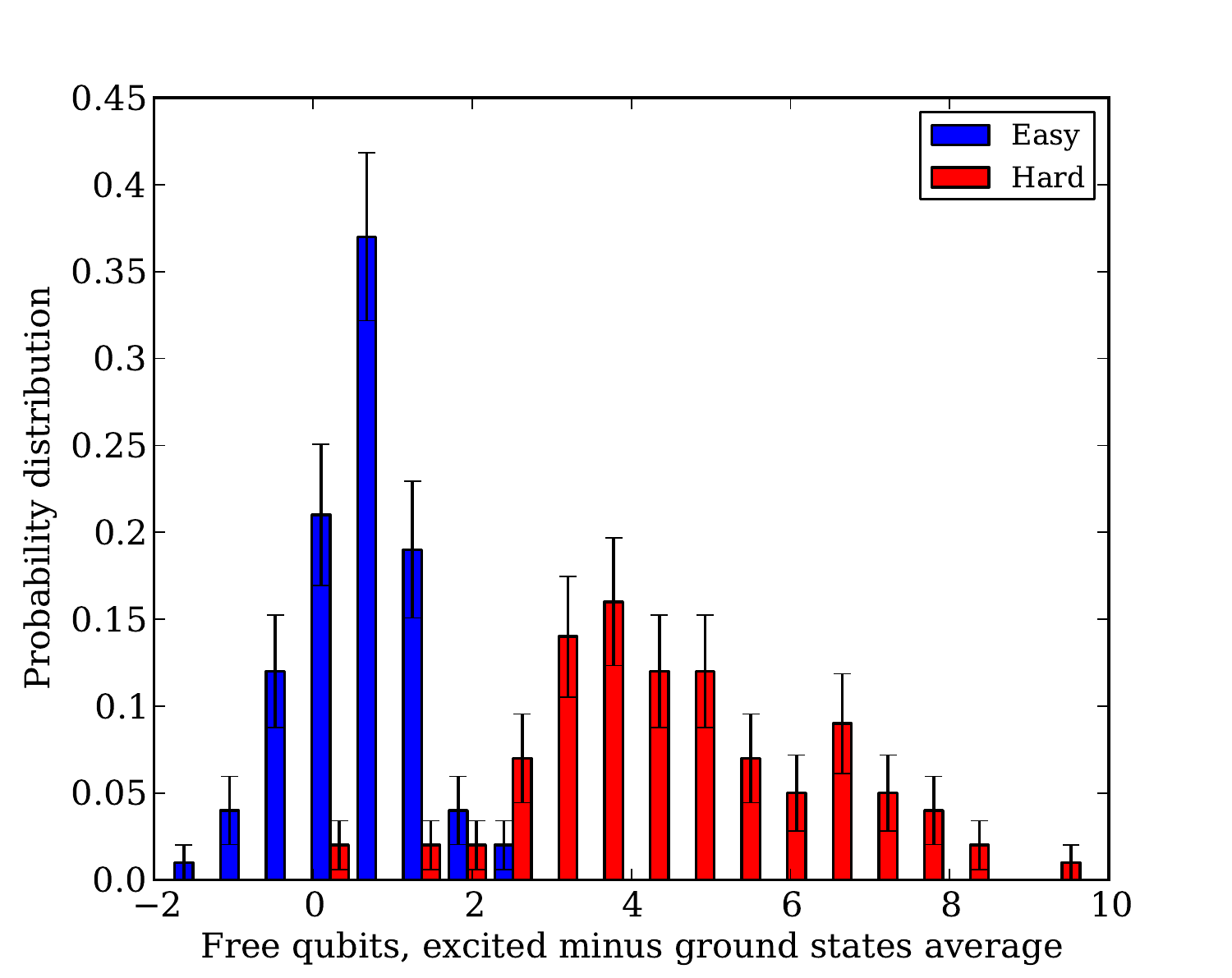}  
  \caption{{\bf Degeneracies and hardness.} To illustrate correlations between hardness and degeneracies we calculate the average number of free qubits in excited states and ground states found by the D-Wave device. The histogram shows $\langle$free qubits in excited states$\rangle$ - $\langle$free qubits in ground states$\rangle$. Hard instances (here the hardest 10\% from $1000$ instances) are typically trapped in excited states with many free qubits, which lead to avoided level crossings. The easiest decile of instances, in contrast, typically does not have more free qubits in excited states than in ground states. Results from the simulated quantum annealer are very similar.}  \label{fig:measured_floppy_qubits} \end{figure}

An intuitive understanding of how such small gap avoided level crossings can arise can be obtained by considering degenerate states that are connected by single spin flips.  The {\em free} qubits of a state are the qubits that can be flipped without changing the energy. From perturbation theory, a small transverse field $\Gamma$ breaks the degeneracy of each free qubit~\cite{PhysRevB.63.224401,amin2009,Boixo2012}. In the simplest case, degenerate states form a hypercube and have the same free qubits. If the unperturbed energy was $E_0$, the  lowest energy state of a hypercube with $F$ free qubits will then have energy $E_0-  \Gamma F$ to first order in perturbation theory. If the low energy excited states have more free qubits than the ground states, their energy will be lowered more, resulting in avoided level crossings and small gaps. As seen in figure~\ref{fig:measured_floppy_qubits}, hard instances tend to have more free qubits in low energy excited states. This phenomenon has been previously observed in the random subcubes model, a specially constructed toy model \cite{Bapst2013}. It can also be seen that on the device the experimental average of free qubits for first excited states is higher than the unweighted average over all excited states. This is not the case in simulated annealing. 

In the spin glasses with $\pm 1$ couplings chosen here, we can expect to find a significant number of free qubits per state. This is because many qubits have six couplings, that can cancel each other. We also expect more low energy excited states than ground states, and consequently some low energy excited states will have more free qubits than most ground states. As argued above, both physical and simulated quantum annealing are sensitive to this problem: many spins updates are needed to move between the states at both sides of the avoided crossing. On the other hand, classical simulated annealers, not having a transverse field, do not suffer from these avoided crossings.  This might explain why simulated quantum annealing scales slightly worse than classical annealing for the $\pm1$ spin glasses considered in this work.

\subsection{Improving the final state by fixing single spin errors}

\begin{figure}[t]
  \centering
  \includegraphics[width=\columnwidth]{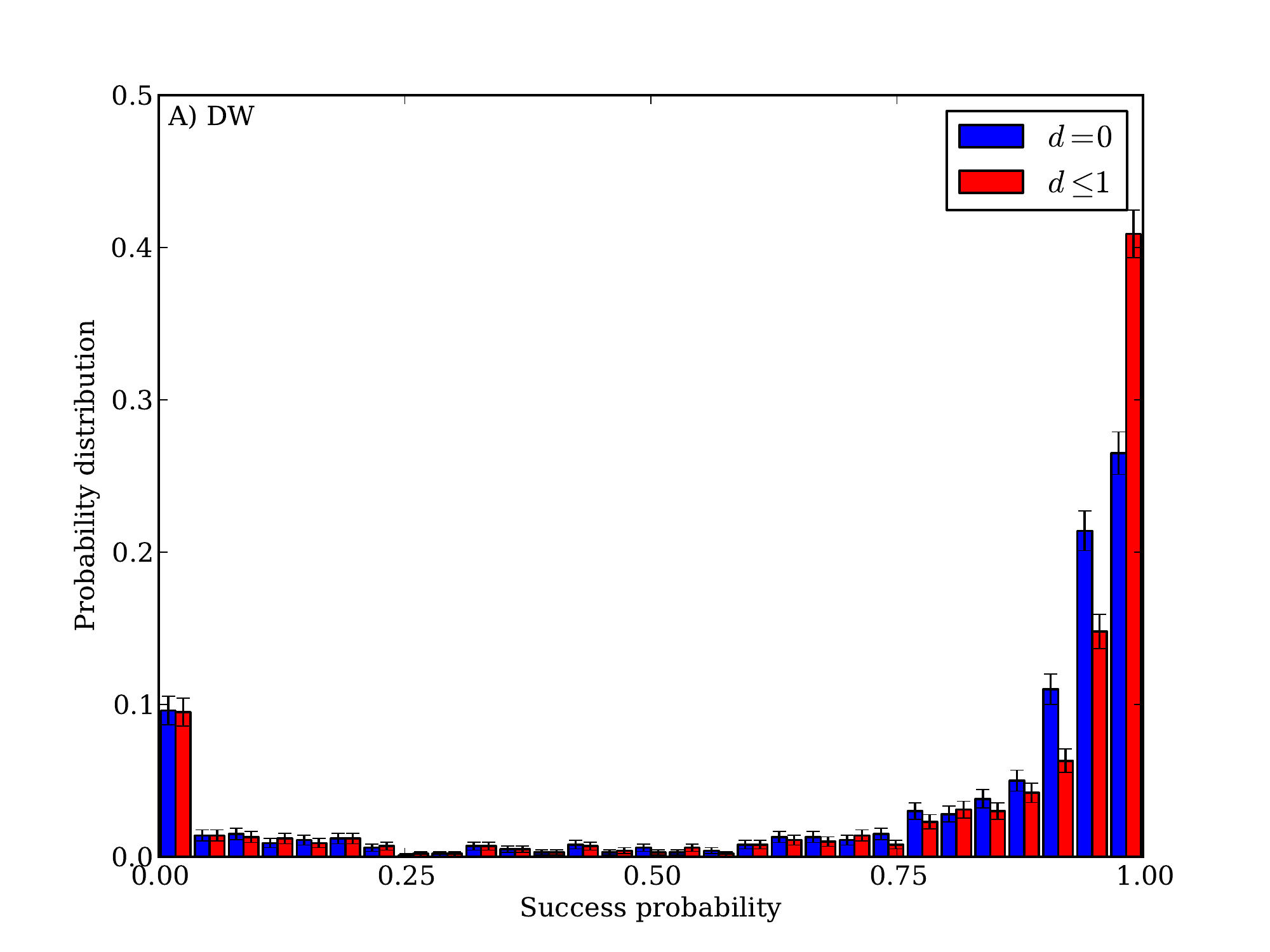}  
  \includegraphics[width=\columnwidth]{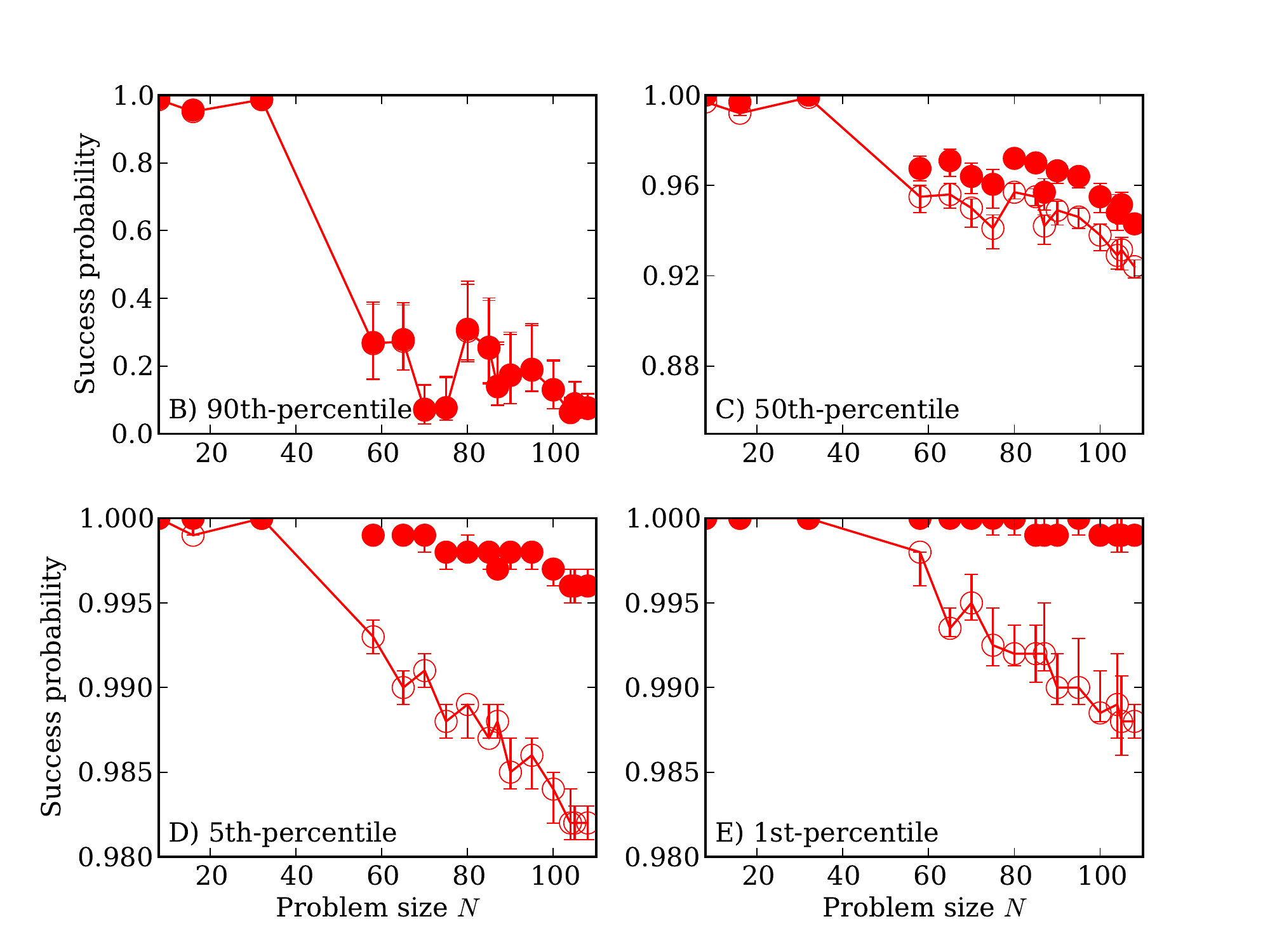}  
  \caption{{\bf Effect of fixing single spin flip errors.} Shown are results for $N=108$ spin instances with local fields: A) Histogram of the success probability $s$ with ($d\leq 1$) and without ($d=0$) single spin flips. The dependence of the effect of such error reduction on problem size $N$ can be seen by following the percentiles of success probability as a function of system size for B) the 90\% percentile, C) 50\% percentile D) 5\% percentile and E) 1\% percentile.  Empty (filled) symbols are the success probabilities without (with) single spin flips.}
  \label{fig:errorcorrect}
\end{figure}

Single spin flip errors can  be fixed with a linearly scaling effort by checking once for each spin whether the total energy can be lowered by flipping it, and then flipping the spin that lowers the energy the most. To illustrate the effect of this procedure we show in figure~\ref{fig:errorcorrect}A), the success probability histogram for $N=108$ spin instances with local fields, with and without error reduction. Without error reduction only ground states (Hamming distance $d=0$) count as a successful annealing run, while with error reduction also the runs giving states a Hamming distance $d=1$ away from the ground state will, after error reduction, end up in the ground state. It is clear from figure~\ref{fig:errorcorrect}A) that while ``hard" instances do not benefit from this error reduction scheme, such single spin flip errors are a dominant error source for the ``easy'' instances and their failure rate is substantially reduced using error reduction. The most likely source for such errors are thermal excitations or readout errors. 

To understand the dependence on problem size $N$ and hardness, we consider the percentiles of the success probability distribution as a function of $N$ in figure~\ref{fig:errorcorrect}B) to E). We see that the success probability decreases upon increasing $N$, i.e., the probability of such errors increases for larger problems. Independently of $N$, single-bit error reduction is ineffective in increasing the success probability for the ``hard'' instances (high percentiles). It becomes more and more effective as problems become easier (low percentiles), improving the success probability to very nearly 1, independently of $N$, at the $1\%$ percentile. 

Variants of this simple error reduction scheme can be derived that, also with linear scaling, fix not only one single spin errors, but also ``disconnected'' single spin errors ({\it i.e.}, flips of spins that are not connected by one of the couplings). One such example is zero temperature Glauber dynamics, flipping randomly selected spins as long as they lower the energy. However, we saw only minimal improvements using such alternative schemes and sometimes worse results when the ``wrong'' spins are flipped and one reaches a different local minimum. The reason may be that disconnected single spin errors are not common in the problem sizes we studied, but that may change in larger problems.

\section{Correlations}

\label{sec:correlations}

\subsection{Copulae}
To better understand the correlations we here show copulae in addition to the correlations shown in the main text. Copulae
\begin{equation}
  c(x_1,x_2) = \frac{f(x_1,x_2)}{f_1(x_1) f_2(x_2)}.
\end{equation}
 factor out the marginal distributions $f_1(x_1)$ and $f_2(x_2)$ ({\it e.g.}, the strong bimodal distribution) from the joint probability function $f(x_1,x_2)$ describing the correlation density. For independent sets the copula density is $c(x_1,x_2)=1$ while for perfect correlations it is a delta function $c(x_1,x_2)=\delta(x_1-x_2)/f_1(x_1)$.
 
 \begin{figure}[t]
  \centering
  \includegraphics[width=\columnwidth]{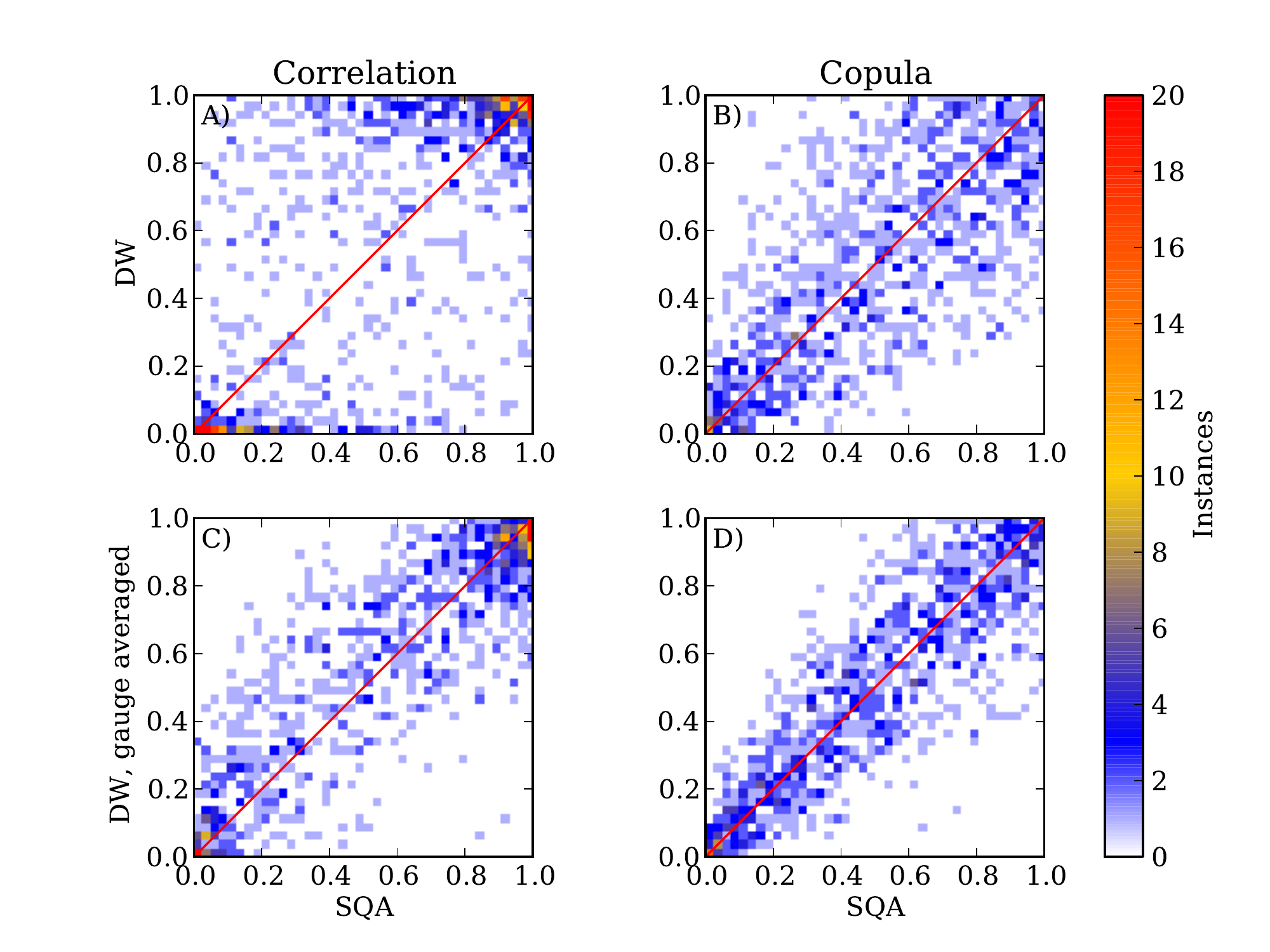}  
  \caption{{\bf Correlations and copulae} between the simulated quantum annealer and the D-Wave device. Axes corresponds to success probabilities and pixels are colour-coded according to the number of instances. The simulation was carried out using schedule II at $T=0.3$ using 7000 sweeps.}
  \label{fig:classical_corr2}
\end{figure}

To plot copula densities of success probabilities we replace each success probability by its rank after sorting the success probabilities divided by the number of values, and then plot the correlation densities of these normalised ranks.  In figure~\ref{fig:classical_corr2} we plot copulae corresponding to the correlations shown in the main text, enhancing the visibility of the correlations especially in the corners. 

\begin{figure}[t]
  \includegraphics[width=\columnwidth]{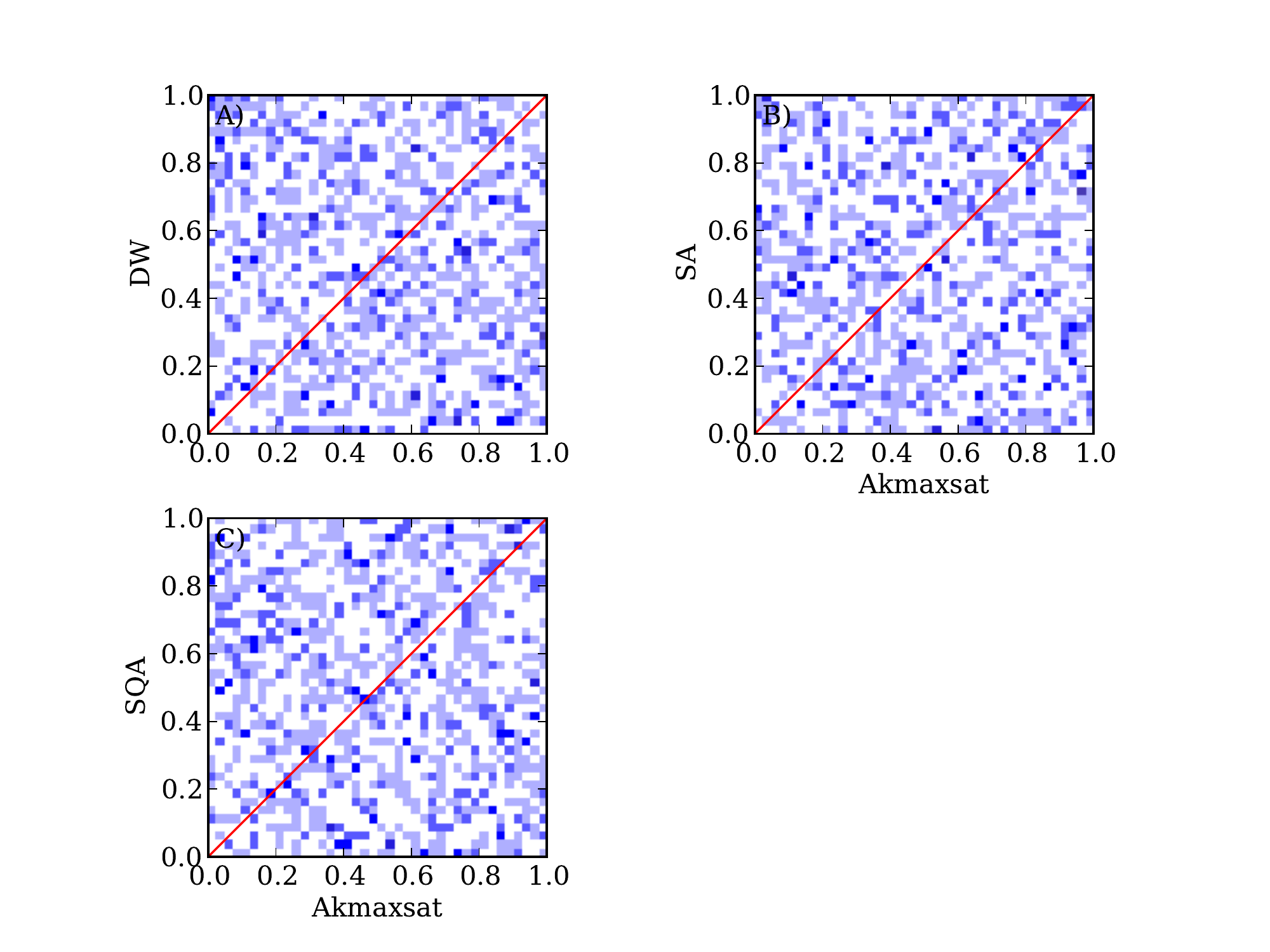}  
  \caption{{\bf Copulae of the hardness between exact optimisers akmaxsat and  the annealers.}  A) for the D-Wave device, B) the simulated annealer and C) the simulated quantum annealer, showing the absence of correlations.}
  \label{fig:exact_corr}
\end{figure}

In figure~\ref{fig:exact_corr} we show the copulae between the hardness for akmaxsat and the D-Wave device and the simulated classical and quantum annealers. We do not observe any correlations.

\subsection{Reproducibility}

\begin{figure}[t]
  \centering
  \includegraphics[width=\columnwidth]{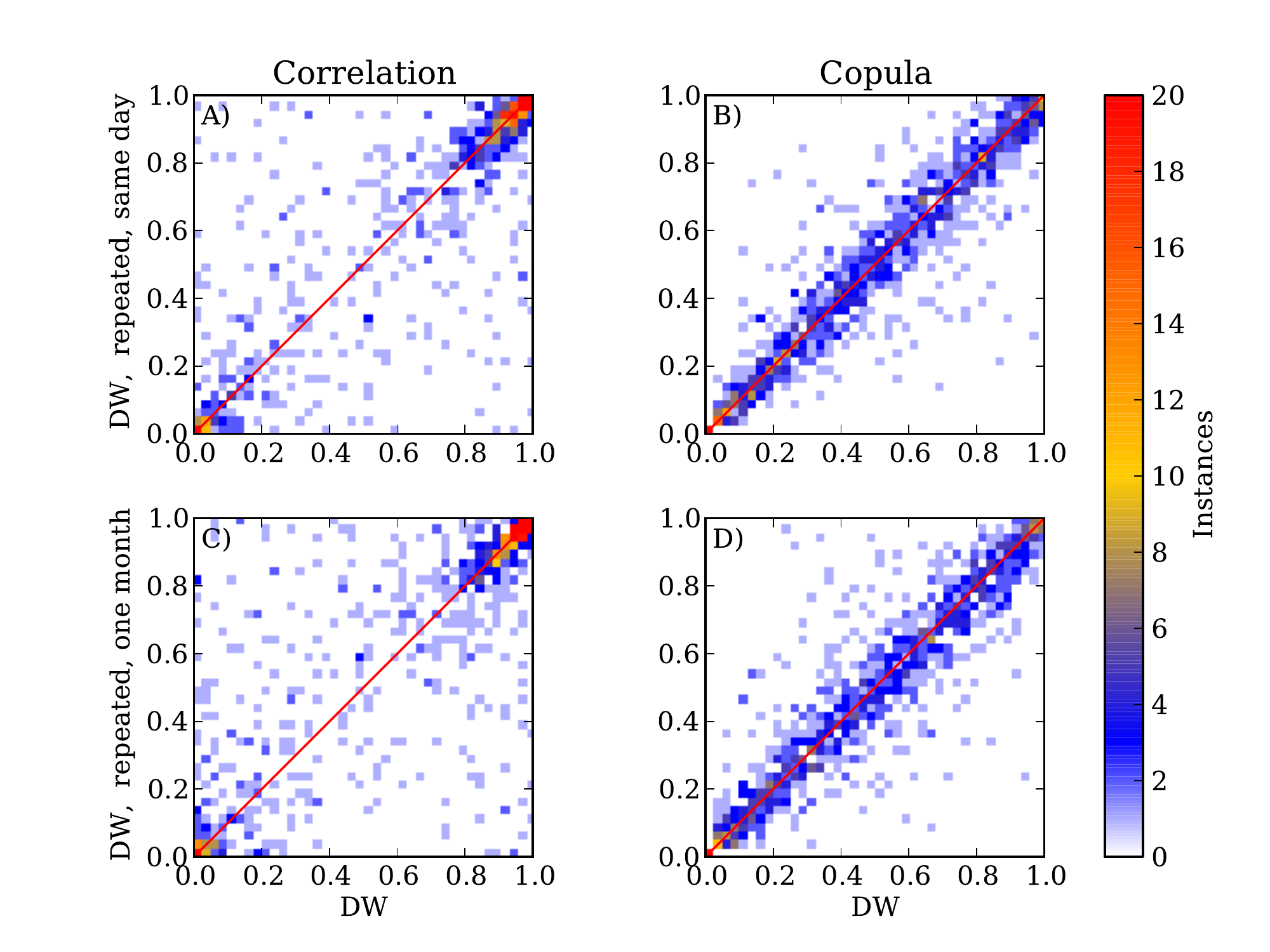}  
  \caption{{\bf Reproducibility of the experiments}. We show correlations and copulae for repetitions of experiments: A) and B) on the same day, and, C) and D) one
    month apart. A) and C) show correlations between success probabilities,
    while B) and D) show the corresponding copulae. }
  \label{fig:dwavereproducability}
\end{figure}

To verify reproducibility of the QA data we performed experiments on $N=108$ spin instances three times, with a month between the first and the second two repetitions. The correlations (see figure~\ref{fig:dwavereproducability}) show that the device is stable over the time of a month.

Strong deviations are seen for a small fraction of the instances. These are most likely due to 1/f noise or ``programming errors'' when flux quanta are loaded into the programmable magnetic memories to program a specific set of couplings into the device. Since these programming errors will limit the correlations between any model and the device, no better correlations than shown in this figure can be expected between our simulations and the device.

\subsection{Gauge averaging} 
\label{sec:gaugeaverage}

The spectrum of an Ising spin glass is invariant under a gauge transformation that changes spins $\sigma^z_i\rightarrow  a_i\sigma^z_i$, with $a_i=\pm1$, when at the same time changing the couplings $J_{ij} \rightarrow a_ia_jJ_{ij}$ and the local fields as $h_i\rightarrow a_ih_i$. While the simulated annealers are invariant under such a gauge transformation, the calibration of the D-Wave device is not perfect and breaks the gauge symmetry. Different gauges hence realise slightly different physical systems with different success probabilities. We average over gauges to reduce calibration uncertainties.

\begin{figure}
\centering
\includegraphics[width=\columnwidth]{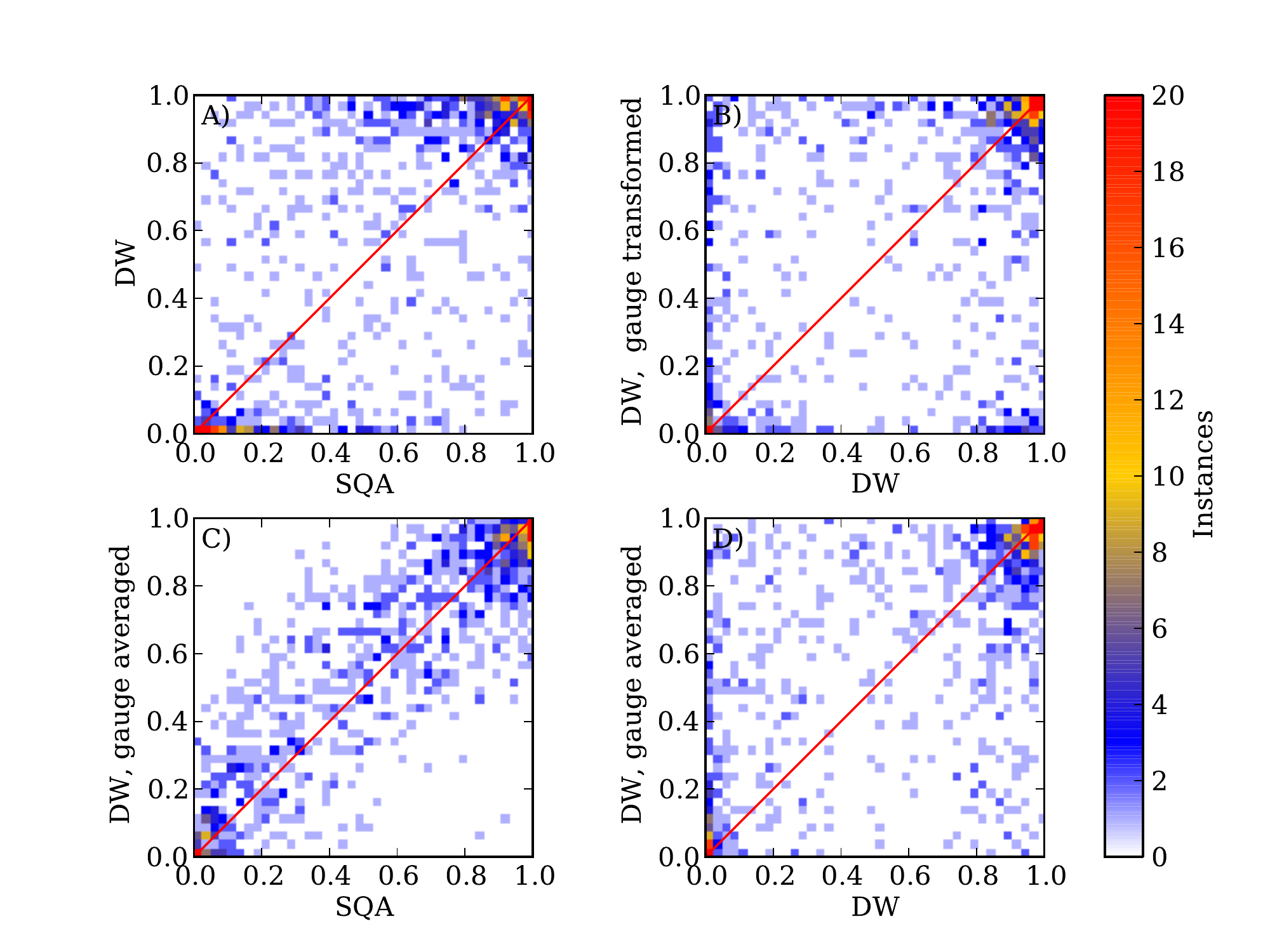}
\caption{{\bf Correlations.} Shown are scatter plots of correlations of
  the success probabilities of 1000 instances of $N=108$ spins. The colour scale
  indicates how many of the instances are found in a pixel of the
  plots. A) between a simulated quantum annealer (SQA) and the D-Wave
  device (DW) B) the quantum device and a gauge-transformed encoding of
  the same instance on the device where the sign of all
  couplings is changed. C) the SQA and an average over 16 random gauges
  on the device. D) a single gauge choice on the device against the
  average over 16 random gauges. The simulations were done at $T=0.3$ with schedule II using 7000 sweeps.
}
\label{fig:correlations-supp}
\end{figure}

Panel A of figure~\ref{fig:correlations-supp} 
shows a scatter plot of the hardness of instances in the simulated 
quantum annealer (SQA) and the D-Wave device  (DW). The high 
density in the lower left corner (hard for both methods) and the upper 
right corner (easy for both methods) confirms the similarities between 
the quantum device and a simulated quantum annealer.   
However, a small percentage of instances appears in the other corners 
(hard for one method but easy for the other). We attribute these to 
calibration errors of the couplers in the device which vary by about 10\%~\cite{Harris2010}. 
To test for calibration errors we performed a gauge transformation
on each instance  to realise a different 
encoding of the same spin glass instance on the device. The correlations 
between two different encodings (gauge choices), shown in panel B, turn 
out to be comparable to those between the SQA and the device, 
demonstrating that the observed deviations can be attributed to 
calibration errors.  

To minimise calibration errors, we then performed
annealing with multiple encodings of the same instance 
related by gauge transformations and averaged the success probabilities. The 
resulting correlations between the simulated quantum annealer and the 
gauge-averaged results from the D-Wave device, shown in panel C (identical to the plot shown in the main text), are 
significantly improved compared to a single gauge choice, with 
a substantial reduction of the number of extreme outliers. These 
correlations are even better than those between a single embedding 
and the gauge averaged results on the device (see panel D).

\begin{figure}[t]
  \centering
  \includegraphics[width=\columnwidth]{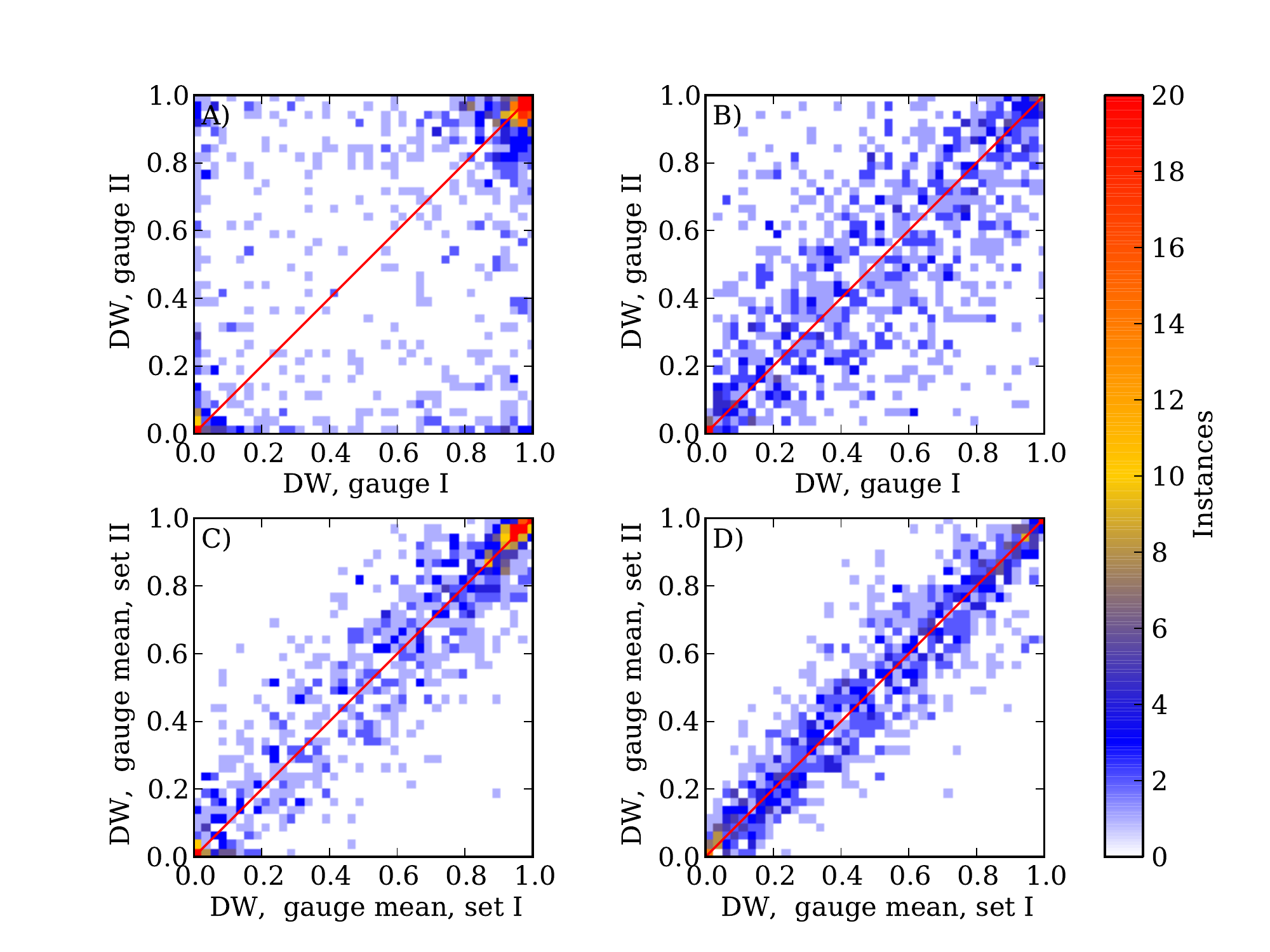}  
    \caption{{\bf Effect of gauge averaging on correlations}. Shown are in A) and B) are the correlations and copulae between two different gauge choices of 1000 instances with $N=108$ spins. In C) and D) we show correlations and copulae of results averaged over eight gauge choices against averages over eight different gauge choices. Gauge averaging significantly increases correlations. }
  \label{fig:withselfgauge}
\end{figure}

\begin{figure}[t]
  \centering
  \includegraphics[width=\columnwidth]{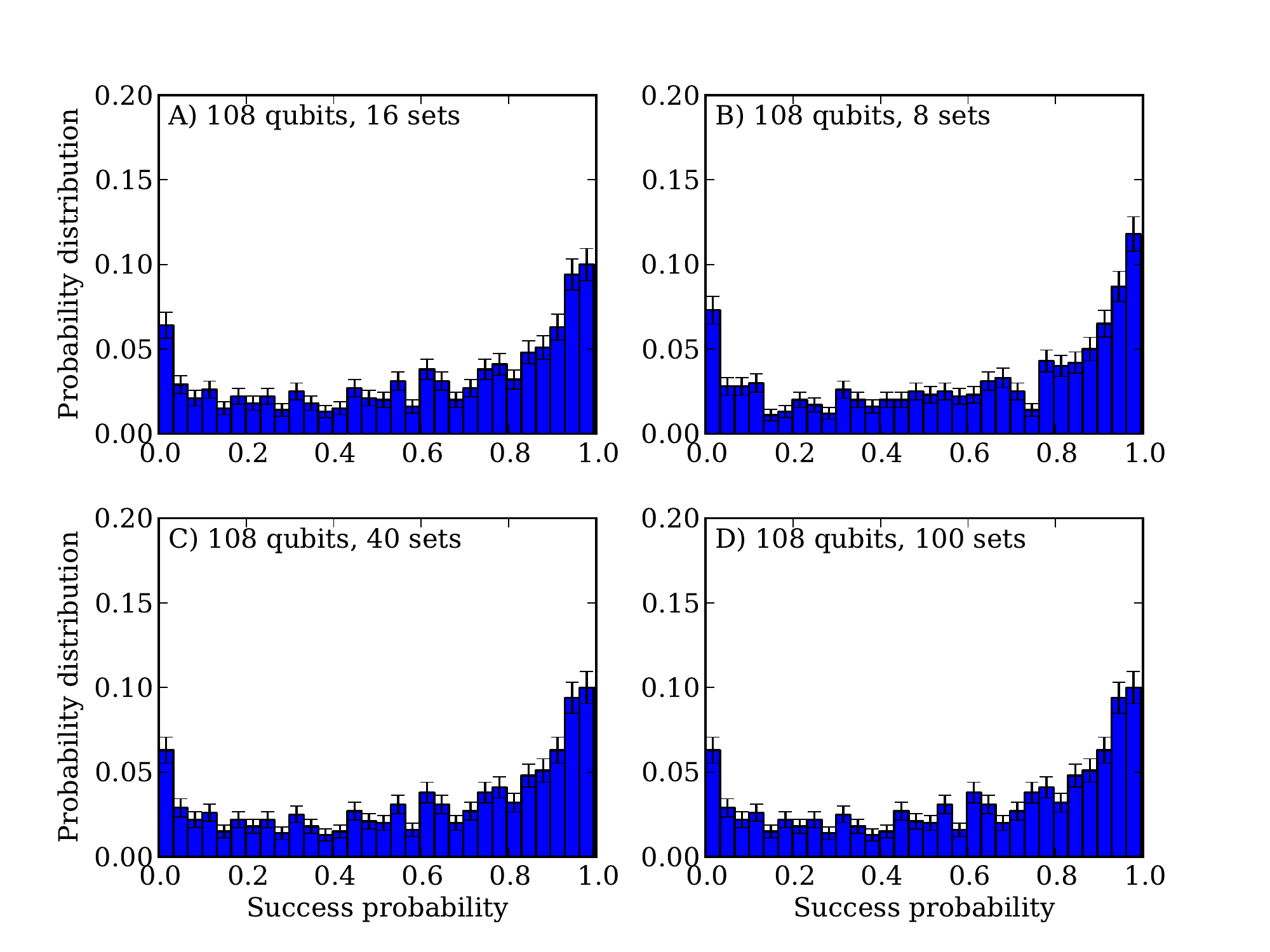}  
  \caption{{\bf Gauge averaged histograms.} Arithmetic averaging over success probabilities obtained using
    different gauge transformations, using A) 2 gauges, B) 8 gauges, C) 40
    gauges and D) 100 gauges. One sees that the marginal distribution
    converges after roughly 40 gauges. }
  \label{fig:gaugeaveraging}
\end{figure}

In \fig{withselfgauge} we show correlations between the arithmetic average over eight gauges compared against the
average over eight other gauges.  Gauge averaging significantly increases correlations compared to single gauge choices.
For the marginal distributions (see \fig{gaugeaveraging}),
the number of hard instances (weight of the peak close to zero) is reduced by gauge averaging, but the distribution remains bimodal even after averaging over many gauge choices.  While calibration errors  enhance the
bimodality, the convergence to a bimodal distribution after averaging many gauges shows that the bimodality is intrinsic and not  solely due to calibration errors.

For scaling plots of the total effort we  use geometric means of failure rates instead of arithmetic
means of success probabilities. If the probability for finding a ground state for a specific percentile and gauge choice $g$ is denoted by $s$ then the probability of achieving a ground state at least once in $R$ repetitions of the annealing is
\begin{equation}
 P=1-(1-s)^R.
 \label{eq:oneg}
 \end{equation}
 Splitting the $R$ repetitions into $R/G$ repetitions for each of $G$ gauge choices and denoting the success probabilities for a specific gauge choice by $s_g$ the total success probability becomes
 \begin{equation}
 P^{(G)} = 1-\prod_{g=1}^G(1-s_g)^{R/G},
 \label{eq:manyq}
 \end{equation}
which can be written in a form similar to equation (\ref{eq:oneg}) 
\begin{equation}
 P=1-(1-\overline{s})^R,
 \label{eq:onegav}
 \end{equation}
by using the {\em geometric mean} of the failure rates to define $\overline{s}$ as
 \begin{equation}
\overline{s} = 1-\prod_{g=1}^G(1-s_g)^{1/G}.
 \end{equation}

\begin{figure}[bt]
  \centering
  \includegraphics[width=\columnwidth]{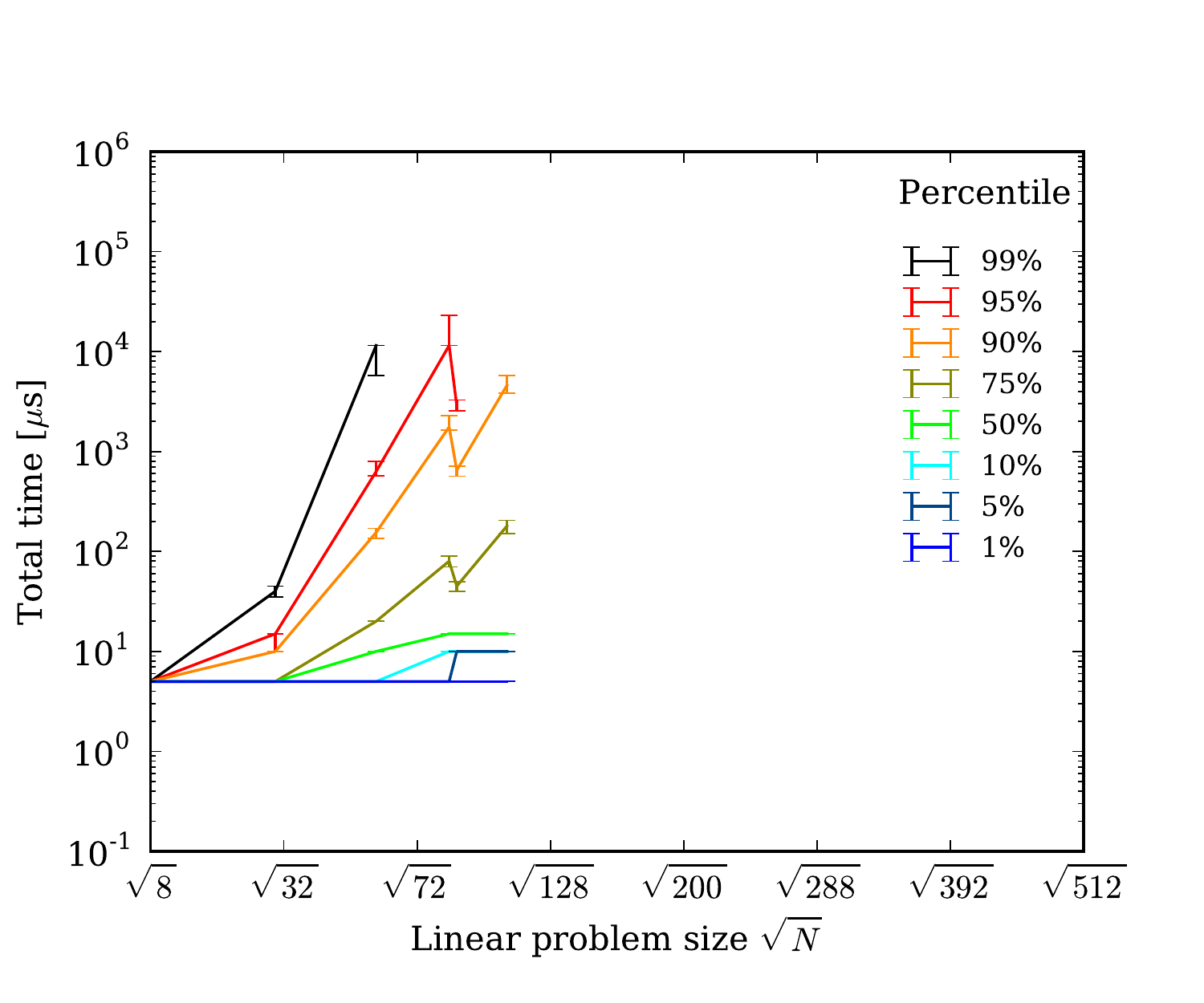}  
  \caption{{\bf Scaling without gauge averaging}. Shown is the total effort on the D-Wave device for a single gauge choice. The higher percentiles are not shown for large systems since 1000 repetitions of the annealing failed to find the ground states for these hardest instances.}
  \label{fig:scaling_sm}
\end{figure}

The higher percentiles grow much more rapidly in the absence of gauge averaging, as can be seen by comparing \fig{scaling_sm} to figure~4A) in the main text. Without gauge averaging the device actually fails to find the ground states for the 5\% hardest instances in 1000 repetitions and those percentiles are thus not shown.

\subsection{Correlations for the simulated annealer} 
\label{sec:simannealercorr}

\begin{figure}[b]
  \centering
  \includegraphics[width=\columnwidth]{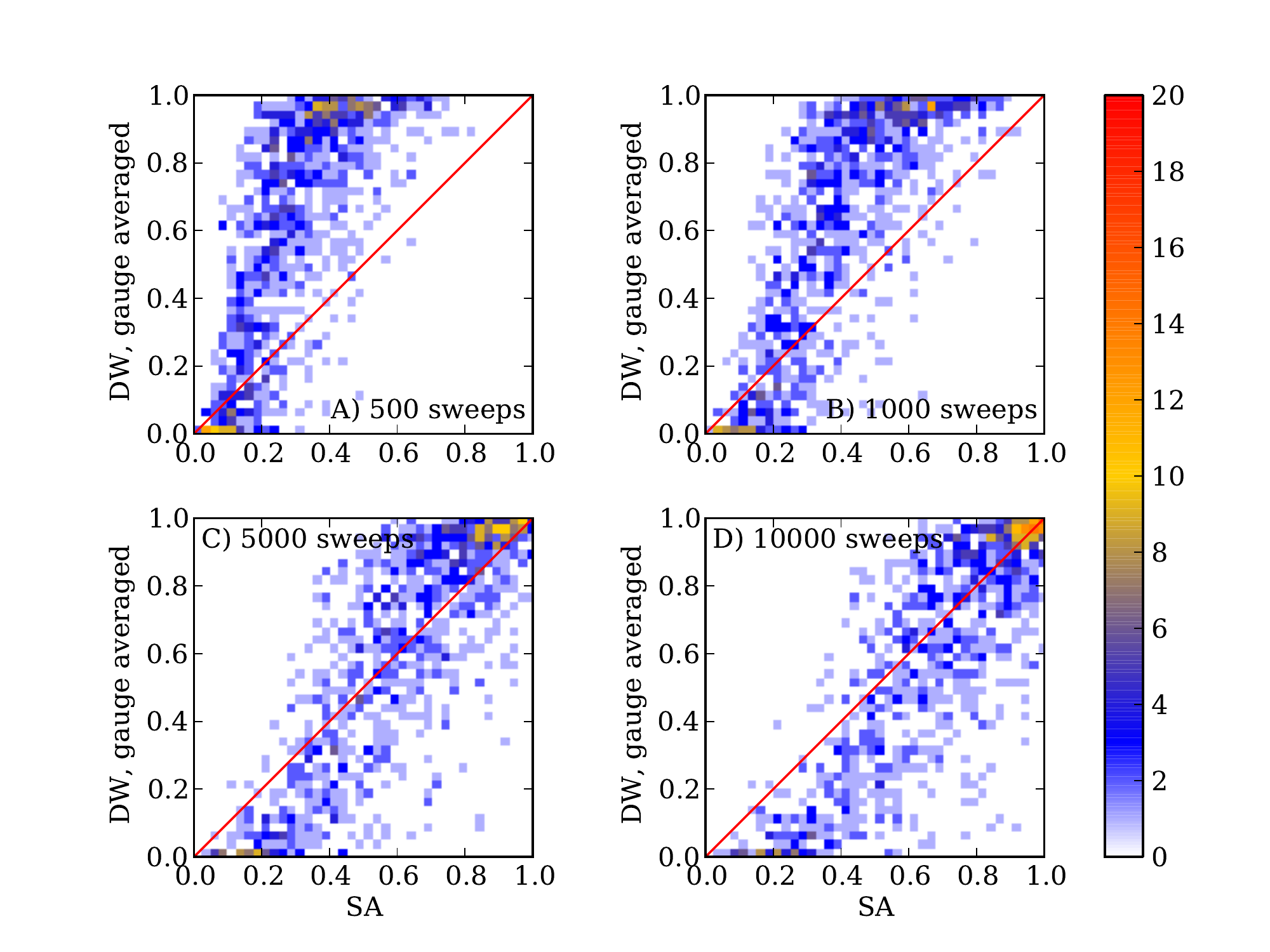}  
  \caption{{\bf Correlations between the D-Wave device and a simulated classical annealer}. 
  Shown are correlations at A) 500 sweeps, B) 1000 sweeps, C) 5000 sweeps and D) 10000 sweeps. The correlations are worse than for the simulated quantum annealer.}
  \label{fig:scaling_correlations_sa}
\end{figure}

\begin{figure}[t]
  \centering
  \includegraphics[width=\columnwidth]{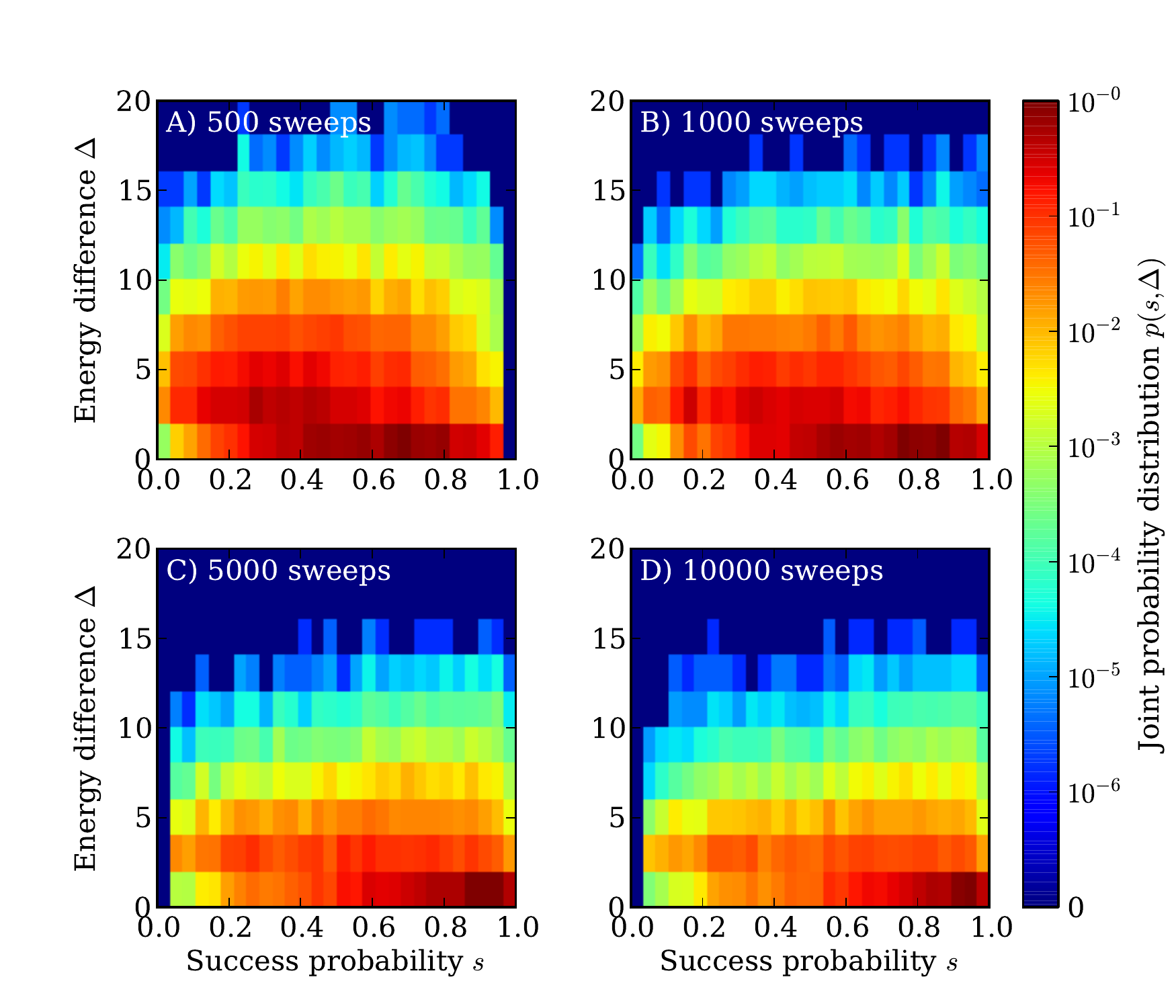}  
  \caption{{\bf Energy-success distributions for the simulated classical annealer.} Shown is the joint probability distribution $p(s,\Delta)$ (colour scale) of success probability $s$ and the final state energy  $\Delta$ measured relative to the ground state.  
  Shown are results for the simulated annealer at A) 500 sweeps, B) 1000 sweeps, C) 5000 sweeps and D) 10000 sweeps. Increasing the annealing time increases the  success probability and decreases the energy of states found. The distribution is always significantly different from  that of the D-Wave device.}
  \label{fig:E-success-SA}
\end{figure}

To complement figures 2 and 3 in the main text we show correlations between the D-Wave device and a simulated classical annealer in figure~\ref{fig:scaling_correlations_sa} and the energy-success distribution $p(s,\Delta)$ of the simulated classical annealer in figure~\ref{fig:E-success-SA}. As can already be expected from the different success distributions (figure 1 of the main text), the classical annealer does not correlate well with the D-Wave device and  $p(s,\Delta)$   is significantly different.

\begin{figure}[b]
  \centering 
  \includegraphics[width=\columnwidth]{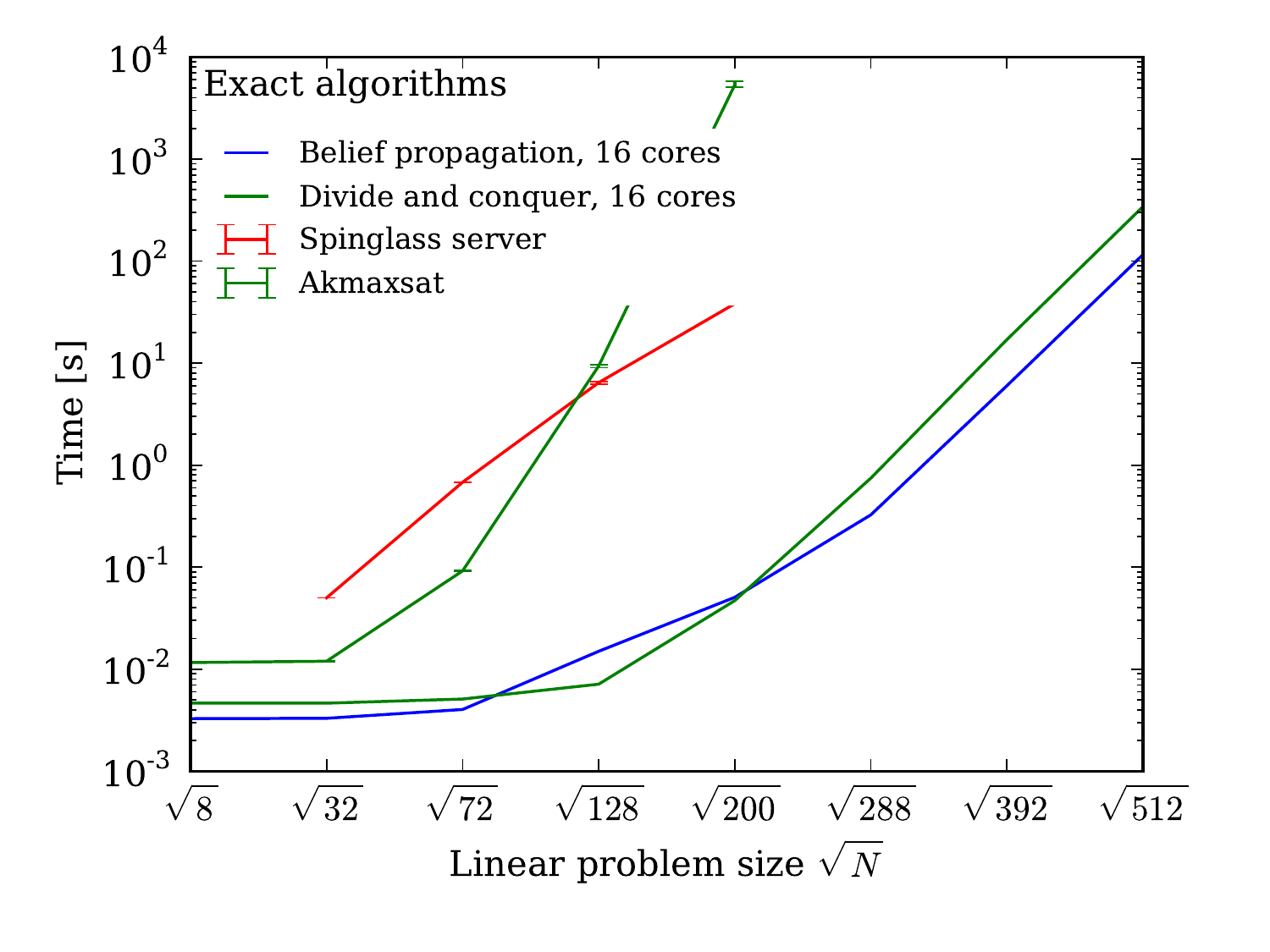}  
  \caption{{\bf Scaling with problem size for the exact solvers.}}
  \label{fig:scalingexact}
\end{figure}

\begin{figure}[t]
  \centering
  \includegraphics[width=\columnwidth]{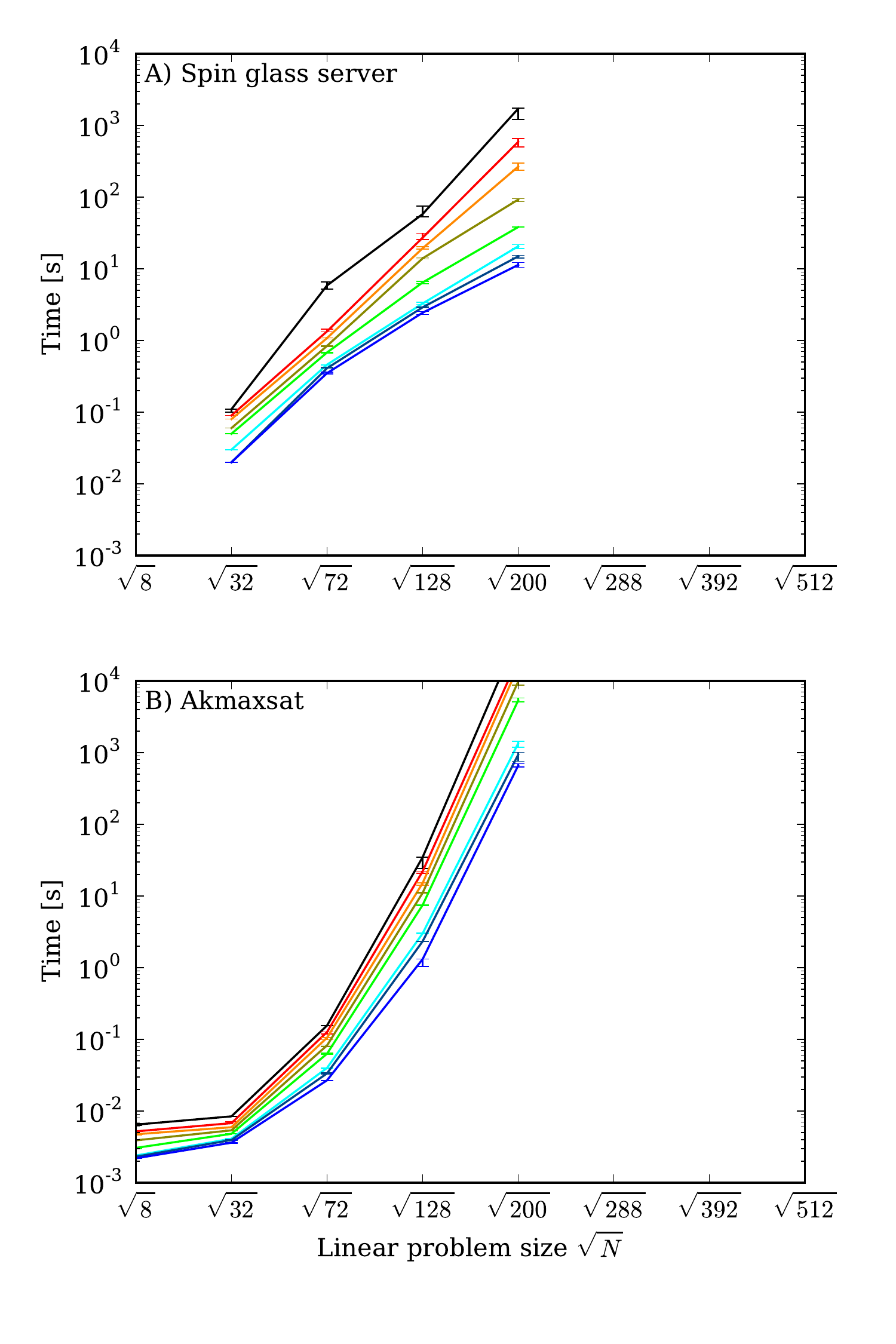}  
  \caption{{\bf Scaling plot for exact solvers and instances without fields} for A) the spin glass
      server and B) akmaxsat.}
  \label{fig:scaling_with_field2}
\end{figure}
\begin{figure}[t]
  \centering
  \includegraphics[width=\columnwidth]{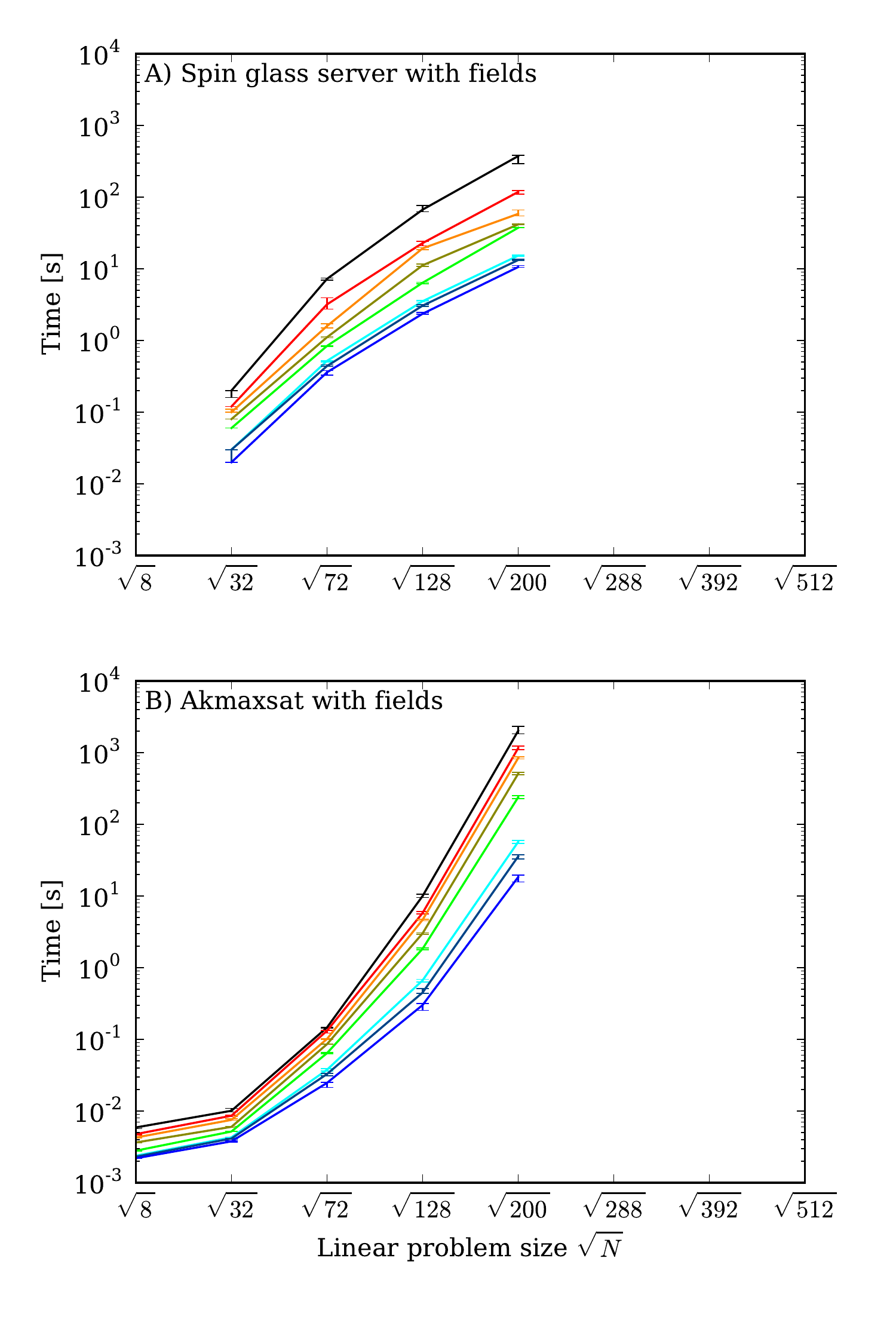}  
  \caption{{\bf Scaling plot for exact solvers and instances with fields} for A) the spin glass
      server and B) akmaxsat. }
        \label{fig:scaling_with_field3}
\end{figure}

\section{Scaling}

\label{sec:scaling}

\subsection{Scaling of the exact solvers}
\label{sec:exactscaling}

In this section we present the scaling of the time to solution for the various exact solvers used. Timings are from our reference CPU, except for the biqmac algorithm where the timings data from the spin glass server was used.

Figure~\ref{fig:scalingexact} shows the time scaling of the exact solvers for instances without fields. Two of the algorithms, an exact belief propagation algorithm using bucket sort \cite{dechter1999bucket} and the divide and conquer algorithm presented in section \ref{sec:bucket} do not depend on the specific instance. The time to find a solution is the same for all instances with and without field. These algorithms make use of the structure of the chimera graph with a tree width of $\sqrt{N}$ \cite{Choi1,Choi2} and thus show the expected scaling ${\cal O}(\exp(c\sqrt{N}))$.

The tree width of a tree decomposition of a graph is the size of the largest vertex set of the tree (minus one according to some definitions \cite{Choi1,Choi2}). A tree decomposition of a graph $G$ is a tree whose nodes are sets of vertices of $G$, and that satisfies the following properties:
\begin{enumerate}
\item Every vertex of $G$ is in at least one node of the tree.
\item For every edge $(v,w)$ in $G$ there is at least one node of the tree which includes both $v$ and $w$.
\item The nodes of the tree that contain any given vertex $v$ of $G$ form a connected subtree.
\end{enumerate}
The cost of doing exact belief propagation (or dynamical programming) on a tree decomposition of a graph $G$ is exponential in the tree width. Belief propagation proceeds from the leaves down. A tree node $t$ with a set of vertices $\{v_1,\ldots,v_l\}$ will calculate the minimum of the subgraph corresponding to the tree above $t$, conditional on all possible assignments to $\{v_1,\ldots,v_l\}$. The cost of this calculation is exponential in $l$, and the tree width is the largest $l$. A tree decomposition for chimera graphs with tree width ${\cal O}(\sqrt N)$ can be seen in Ref.~\onlinecite{klymko_adiabatic_2012}. Because a complete graph with ${\cal O}(\sqrt N)$ vertices can be minor-embedded in the same chimera graph~\cite{Choi2}, this scaling is optimal for exact belief propagation.

The time to solution of the other two solvers, akmaxsat \cite{akmaxsat}, and the biqmac algorithm \cite{biqmac} used in the spin glass server \cite{sgserver} depends on the specific instance. The scaling for the various percentiles from the easiest (1\% percentile) to the hardest (99\% percentile) is shown in figures \ref{fig:scaling_with_field2} and  \ref{fig:scaling_with_field3} for instances with and without local fields.

While the biqmac algorithm scales  as ${\cal O}(\exp(c\sqrt{N}))$, akmaxsat scales worse and -- unlike the other solvers -- does not benefit from the limited tree width of the chimera graph. Unfortunately, the
spin glass server constrained us to instances of up to $200$ spins. It would be
interesting to see if a crossover is present, i.e., to check whether the
best exact code scales better than ${\cal O}(\exp(c\sqrt{N}))$.

\subsection{Optimising the total annealing time}
For the purpose of scaling comparisons we consider the total annealing time needed to find a ground state with a probability of 99\%. Using Eq.~\eqref{eq:oneg} we find that the number of repetitions to find the ground state at least once with probability $p=0.99$ is
\begin{equation}
  R = \left\lceil
    \frac{\log (1-p)}{\log(1 - s)} 
    \right\rceil ,
    \label{eq:repetitions}
\end{equation}
where $s$ is the probability for a given percentile. The total annealing time is $Rt_f$. Note that this expression assumes uncorrelated repetitions. On the D-Wave device we have seen statistically significant correlations between repetitions in 15\% of the instances. In those instances the observed positive autocorrelations increase the length of ``runs'' of consecutive failures or successes on average by about a factor of two. This will result in an increase of the number of repetitions $R$ required on the device, as given by equation (\ref{eq:repetitions}) above. The simulated annealers, on the other hand, show no detectable correlations.

\begin{figure}
  \centering
  \includegraphics[width=\columnwidth]{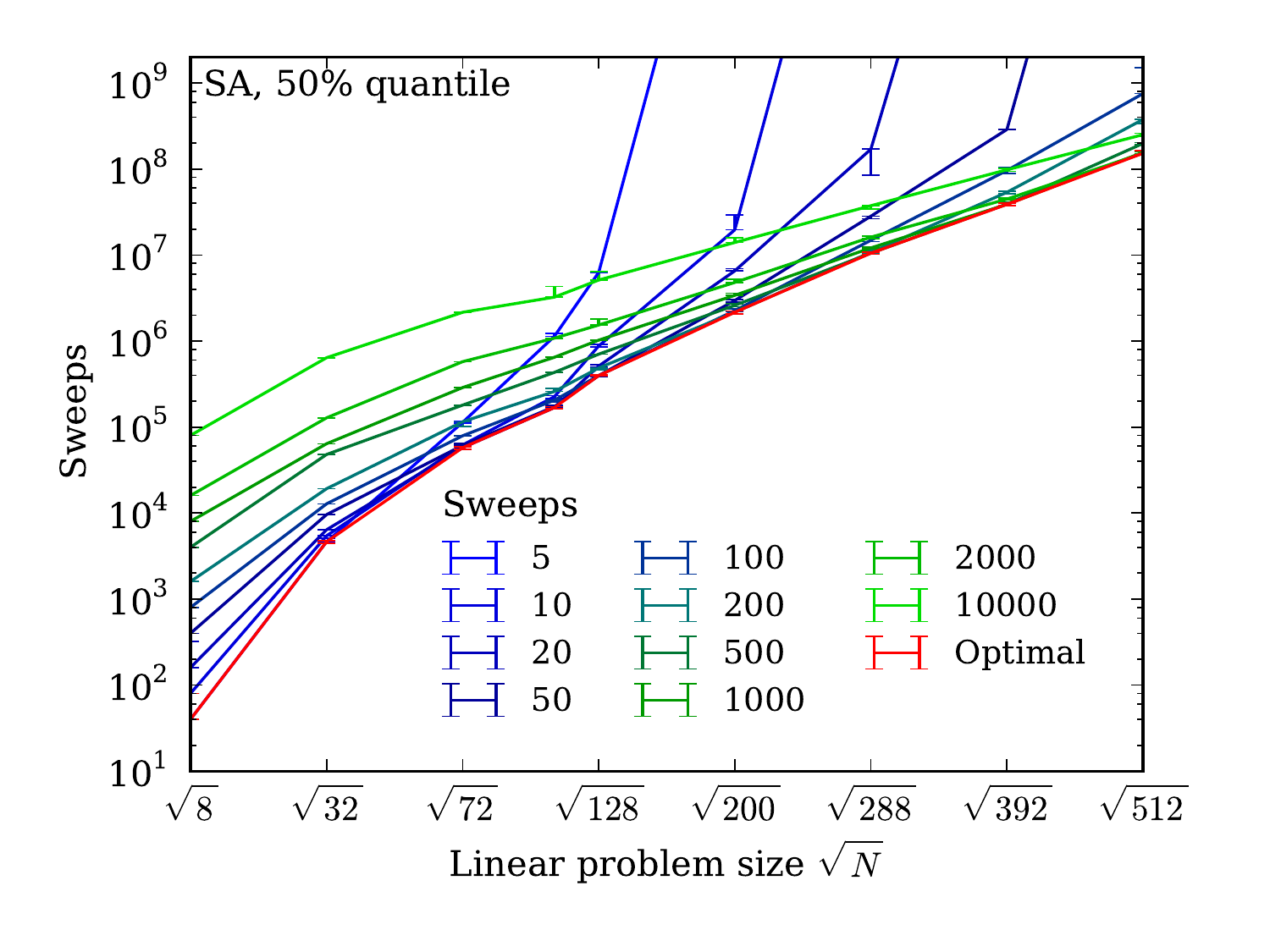}  
  \caption{{\bf Scaling with problem size at fixed annealing time.}  Shown is the scaling of the median total effort of the simulated annealer at constant annealing times measured in sweeps (one sweep is one attempted update per spin). This graph demonstrates that  extrapolating results for fixed annealing times does not give the true asymptotic behaviour.}
  \label{fig:scalingfixed}
\end{figure}

In the main text we argued that extrapolating the total annealing time on the D-Wave device at a fixed annealing $t_f$ is tempting but misleading. It can only be used to estimate the performance for slightly larger problem sizes, but will not give the true asymptotic scaling. To see this, consider  figure~\ref{fig:scalingfixed} where we show the scaling of the median time for a simulated annealer for various fixed annealing times (similar behaviour is observed for a simulated quantum annealer). The scaling at {\em fixed} annealing time $t_f$ increases only modestly for small system sizes, and then suddenly shoots upwards once $t_f$  becomes too short for the problem size.

\begin{figure}
  \centering
  \includegraphics[width=\columnwidth]{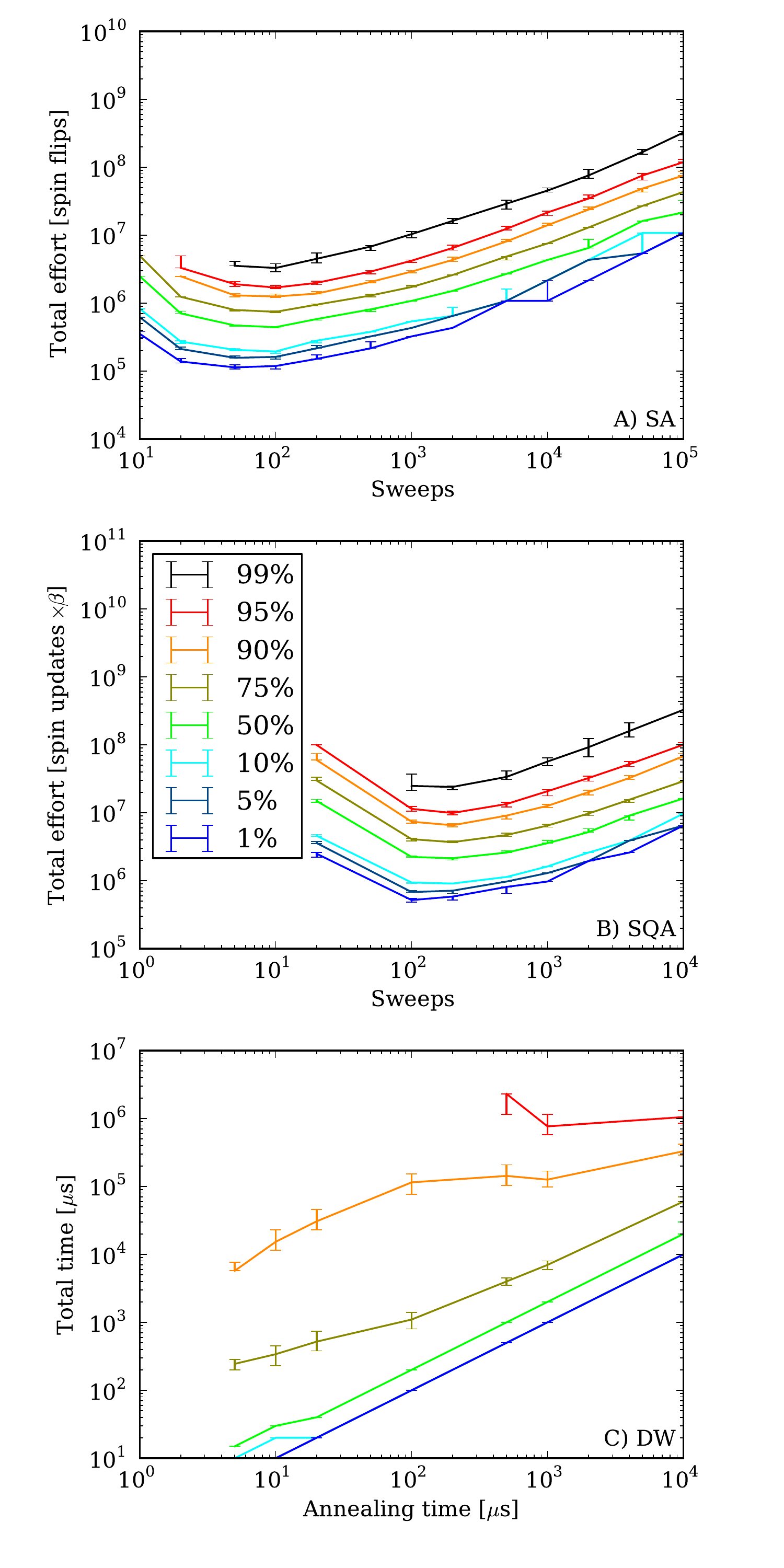}  
  \caption{{\bf Optimisation of the total annealing time for $N=108$ spin instances without local fields}. Shown is the total annealing time run time $Rt_f$ time as function of the annealing time $t_f$ for a single annealing run. }
  \label{fig:scaling_optimisation}
\end{figure}

What needs to be done instead is to optimise the total annealing time for each problem size $N$. To do this we vary $t_f$, measure the success probability $s$, calculate the required number of repetitions $R$, and plot the total annealing time $Rt_f$ as a function of $t_f$. The required number of repetitions $R$ diverges when the annealing time $t_f$ is too short, causing diabatic transitions. When the annealing time is too long it becomes disadvantageous to perform repetitions since a single run already optimises the success probability; however, Eq.~(6) always yields $R\geq 1$. We thus pick the optimal $t_f$ as the time that minimises $Rt_f$ and use it in the scaling plots.

In figure~\ref{fig:scaling_optimisation} we show typical data for simulated classical and quantum annealing (where time is measured in number of spin flips for SA and spin updates for SQA) and for the D-Wave device (measured in $\mu$s). For the simulated classical and quantum annealer we observe that both the optimal time and the required number of repetitions $R$ increase exponentially.
Note that a minimum is not found for the D-Wave device but instead a monotonic increase, indicating that even the fastest annealing time of the hardware of $t_f=5\mu$s is slower than the optimal time. All timings reported for the device are thus only {\em upper bounds} on the optimal time.

\begin{figure}[t]
  \centering
  \includegraphics[width=\columnwidth]{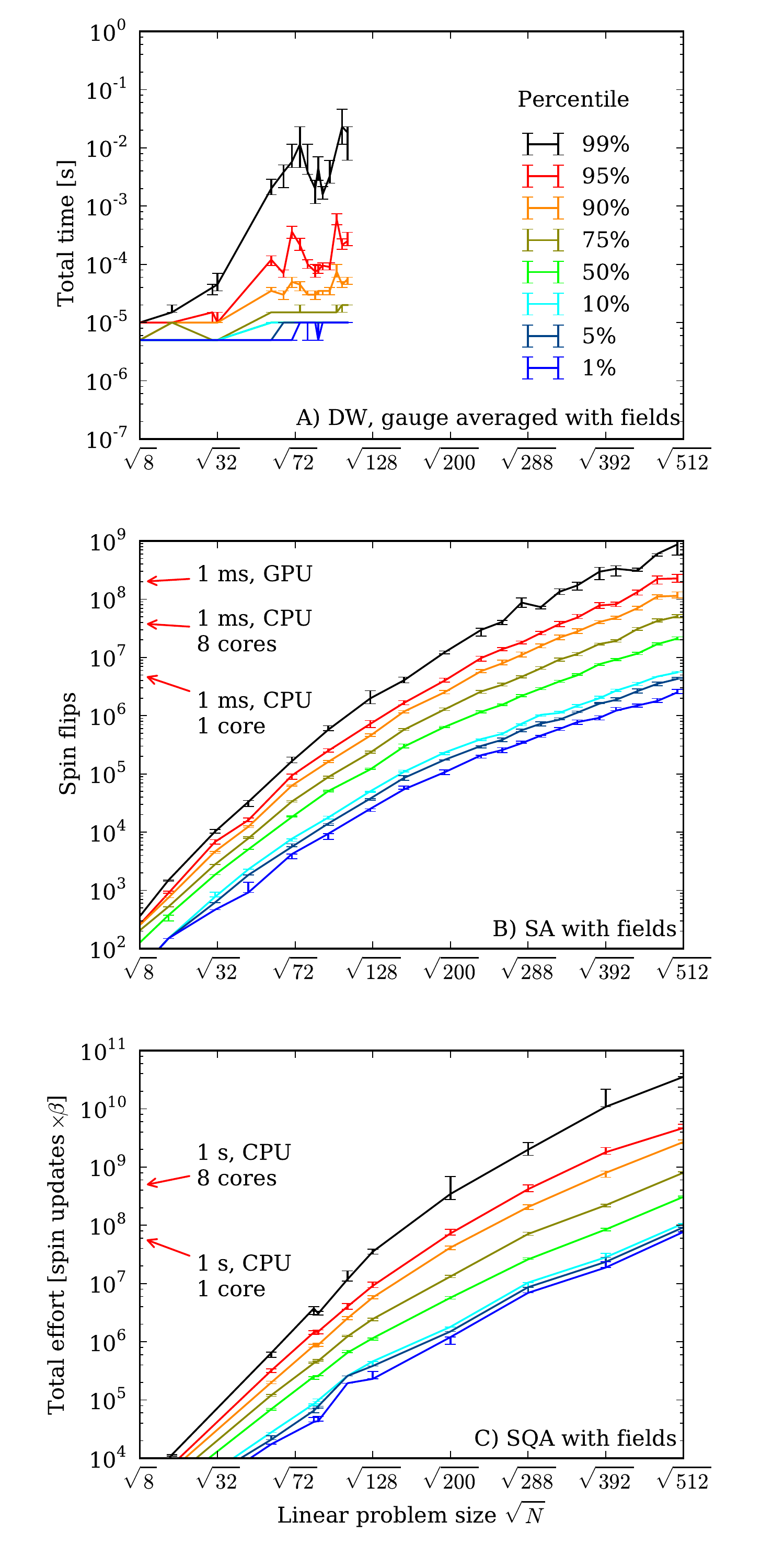}  
  \caption{{\bf Scaling plot with fields} for A) QA, B) SA and C) SQA. Comparing with figure~6, B), D) and E) in the main text, which shows the same plots for problem instances without fields, we see that problems with fields are easier for SA and SQA.}
  \label{fig:scaling_with_field}
\end{figure}

While the total annealing time for the D-Wave device can simply be given in microseconds, the annealing times of the simulated classical and quantum annealers depend on compiler options and the specific machine used. We thus give the total effort in terms of the number of attempted spin flips for the simulated classical annealer. For the simulated quantum annealer we specify the total effort as the number of spins updated multiplied by the inverse temperature $\beta$, to account for the complexity of updating the imaginary time world lines of the spins of length $\beta$.

Scaling plots of the total effort for instances with fields, complementing the results for instances without fields shown in the main text, are shown in
figure~\ref{fig:scaling_with_field}. In panel A, we show the scaling of the D-Wave device. Again simulated classical annealing scales slightly better than simulated quantum annealing. Comparing the scaling for instances without fields (figure~4 in the main text) and instances with local random fields (figure~\ref{fig:scaling_with_field}), we see that problems with fields are not only easier for small problem sizes but the total annealing time also scales better when going to larger problem sizes.

\subsection{Detecting quantum speed up}
Quantum speedup of a hardware quantum annealer can be detected by comparing the scaling of the total annealing times to that of the simulated classical or quantum annealer. To draw valid conclusions about a speedup one must ensure that the experimental annealing times are optimal. As we pointed out in the previous subsection, this is not the case for the current device: the annealing time of $5\mu$s is suboptimal, as demonstrated in figure~\ref{fig:scaling_optimisation}. It follows that the inferred total annealing times are only upper bounds, and those bounds are worse for smaller problem sizes, which leads to scaling plots with an underestimated scaling. To see the pitfall it suffices to consider extremely long fixed annealing times where a single repetition $R=1$ might be enough. We would then see a {\em constant} time needed to find the ground state.

In other words, since the fastest possible annealing time $t_f=5\mu$s on the D-Wave device is longer than the optimal time for the considered problem sizes, our experimental data is in the initial transient regime of modest increase, and thus cannot be used for reliable extrapolation or determination of quantum speedup.

\begin{figure}
  \centering
  \includegraphics[width=\columnwidth]{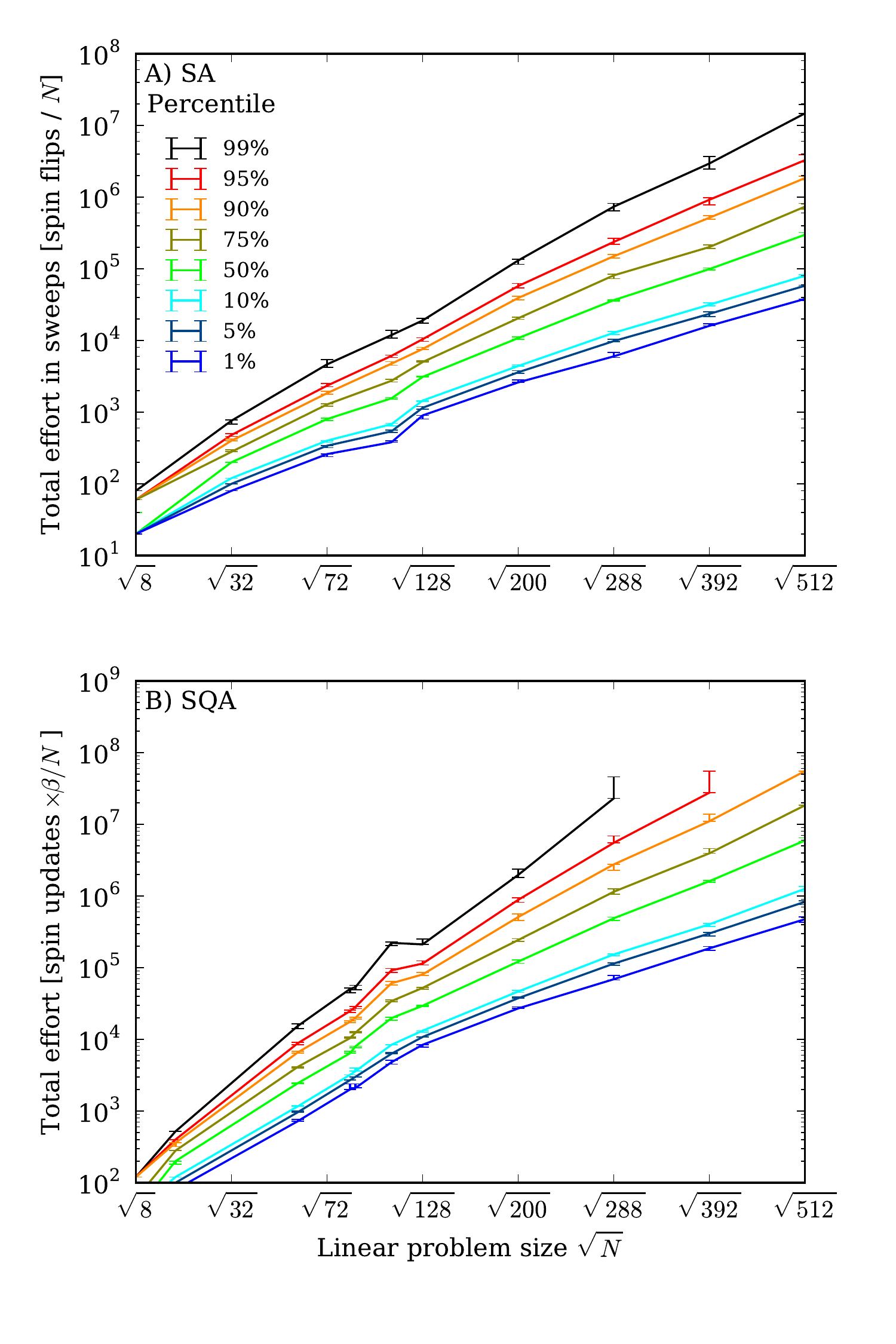}  
  \caption{{\bf Scaling of total annealing effort assuming parallel updates of spins for instances without fields}. Since experimental annealers (quantum or classical) have intrinsic parallelism of updating all spins simultaneously, we need to divide the total effort by the number of spins $N$ to obtain scaling plots against which quantum speedup can be detected.}
  \label{fig:scaling_pp}
\end{figure}

\begin{figure}
  \centering
  \includegraphics[width=\columnwidth]{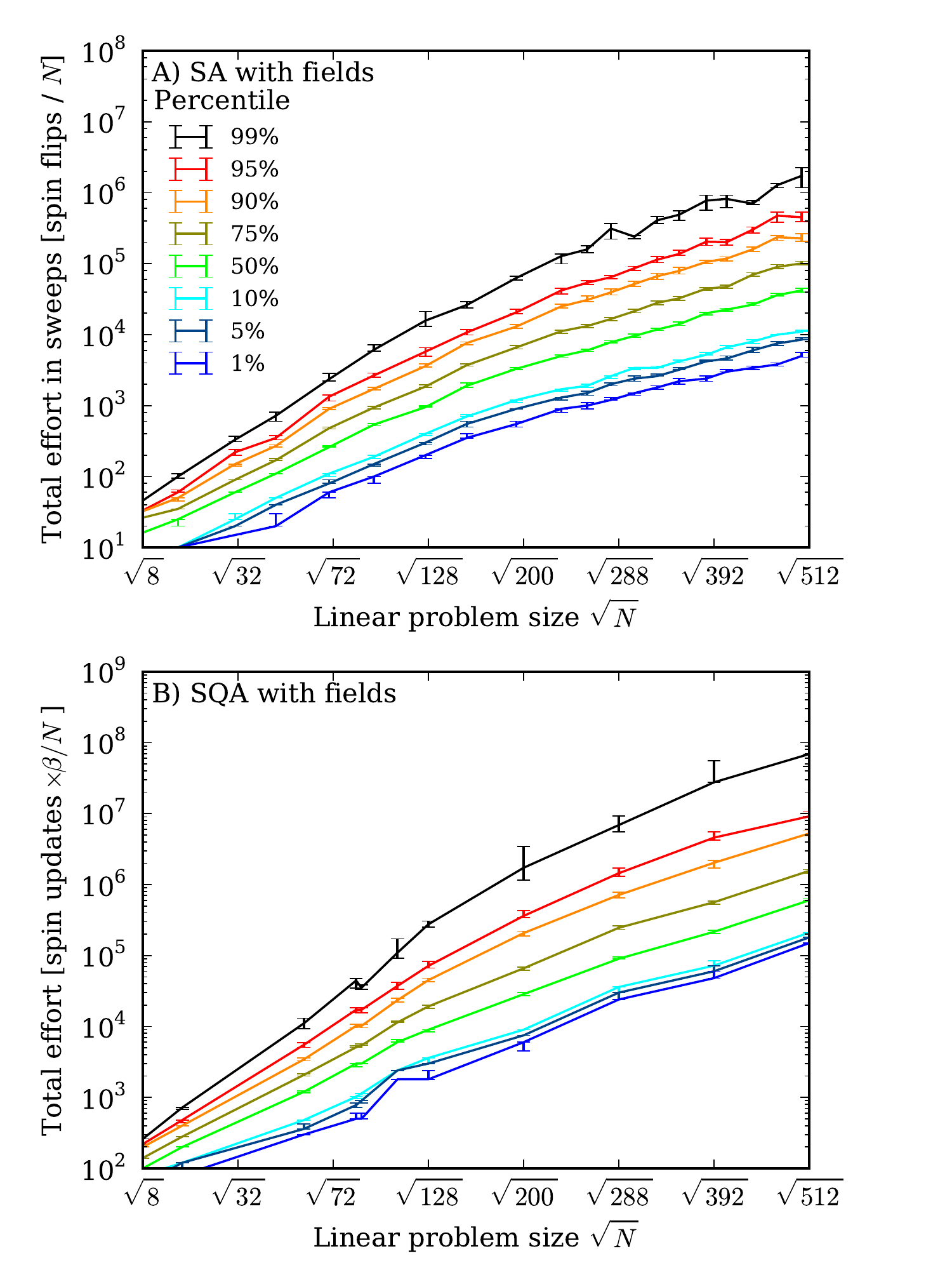}  
  \caption{{\bf Scaling of total annealing effort assuming parallel
      updates of spins for instances with fields}. As in figure
    \ref{fig:scaling_pp} the total effort is divided by the system size
    due to the intrinsic parallelism of an experimental quantum annealer.}
  \label{fig:scaling_pp_field}
\end{figure}

The optimal annealing time defined in the previous section (see figure~\ref{fig:scalingfixed}) increases exponentially with problem size for SA. It is expected to increase exponentially with problem size also for QA.  Therefore, assuming that the minimum programmable annealing time on future D-Wave devices does not increase too rapidly, we expect that perhaps already for a device with $N=512$ spins, or possibly $N=2048$ spins, the optimal annealing time $t_f$ for a single run will be within reach. This will allow us to determine the optimal total annealing times $Rt_f$ for QA. To detect quantum speedup one should then compare to the total effort of the simulated annealers {\em divided by the number of spins} $N$. This division is necessary to compensate for the trivial parallelism of the hardware annealer, which updates $N$ spins in parallel, since an analog classical annealing device would have the same parallelism. For completeness we provide the scaling plots of total effort in units of sweeps (spin flips divided by $N$) in figure~\ref{fig:scaling_pp}.

\section{Gap calculation}

\label{sec:gap}

We finally describe how the excitation gaps are obtained using a method similar to that of Refs.~\cite{Kashurnikov1999,Young2008}. For simplicity we consider the transverse field Ising spin glass Hamiltonian without fields,  
\begin{equation}
  H = - \sum_{i<j} J_{ij} \sigma_i^z \sigma_j^z - \Gamma \sum_i \sigma_i^x,
\end{equation}
where $\Gamma$ is the strength of a transverse magnetic field and for our instances $J_{ij}=\pm1$.

The excitation gap can be obtained from the connected correlation function
in imaginary time
\begin{equation}
  C(\tau) = \langle \hat{O}(\tau) \hat{O}(0)\rangle - \langle \hat{O} \rangle^2,
\end{equation}
where $\hat{O}$ is an observable with non-vanishing matrix element between the ground state and first excited state. The correlation function is given by a sum:
\begin{equation}
  C(\tau) = \sum_{i} c_i \exp(-\Delta_i \tau),
\end{equation}
where $\Delta_i$ is a gap to the $i$th eigenvalue above the ground state;
$i=0$ corresponds to the lowest gap. Only one term survives when $\tau$ is
large enough:
\begin{equation}
  C(\tau) = c_0 \exp(-\Delta_0 \tau).
\label{eq:ctau}
\end{equation}
Thus $\Delta \equiv \Delta_0$ can be obtained by fitting $C(\tau)$ at large
values of $\tau$. We use periodic boundary conditions in the imaginary time
direction. In this case, $C(\tau)$ can be efficiently calculated at discrete
points using the Fast Fourier Transformation.
\begin{figure}[t]
\centering
\includegraphics[width=\columnwidth]{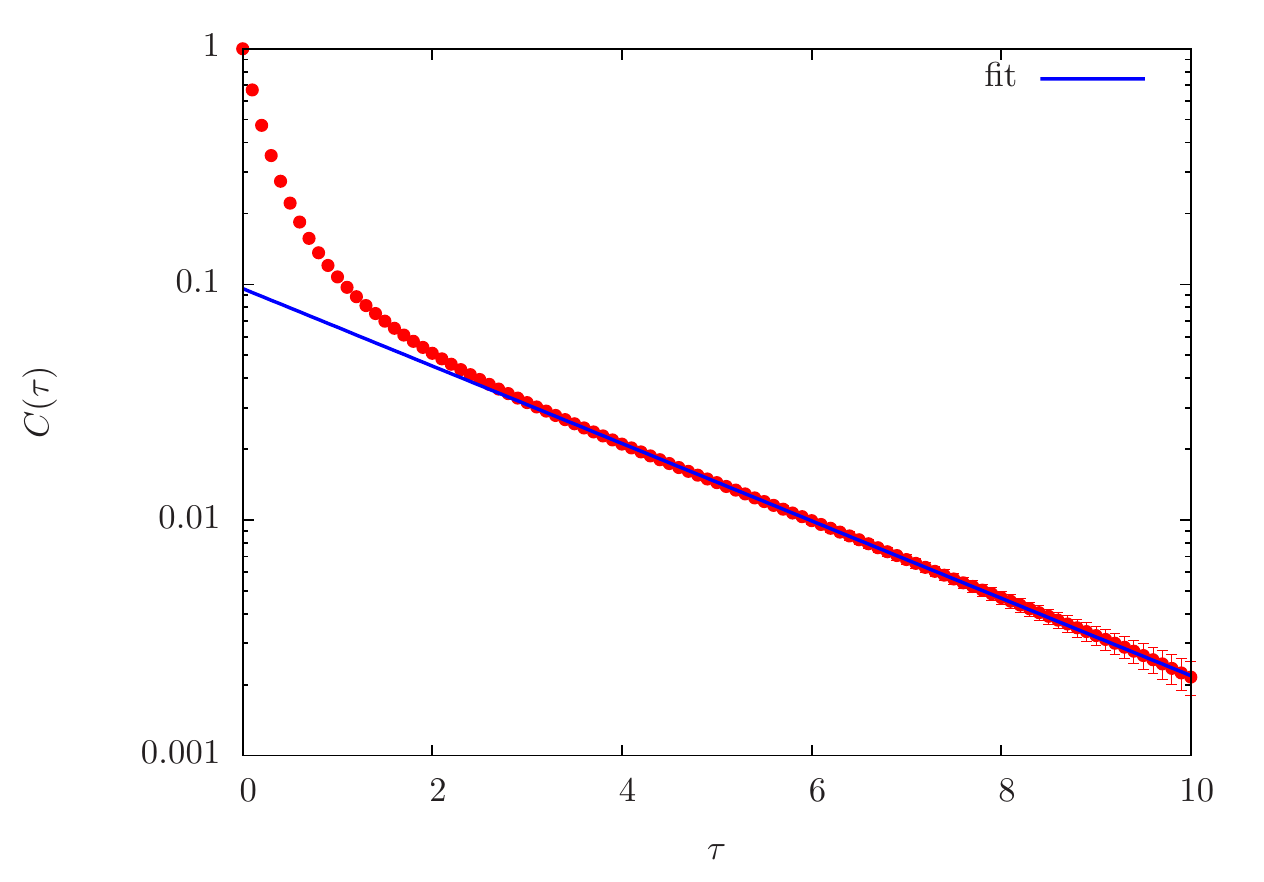}  
\caption{{\bf Correlation function $C(\tau)$}. The line shows an exponential fit.
The gap can be obtained from the slope of the line.}
\label{fig:corr}
\end{figure}

The observable $\hat{O}$ must be chosen carefully because for a poorly chosen
observable, $c_0$ can be much smaller than $c_1,c_2,\ldots$ leading to very
small values of $C(\tau)$ (comparable to statistical noise) at large $\tau$
and making the gap extraction very difficult. This issue is discussed in
Ref.~\onlinecite{PhysRevE.85.036705}. We use a simple observable,
the local magnetisation $\sigma_j^z(\tau)$ and its correlation function,
given by
\begin{equation}
  C(\tau) = \frac{1}{N} \sum_{j=0}^N
    \langle \sigma_j^z(\tau) \sigma_j^z(0)\rangle - \langle \sigma_j^z(0) \rangle^2,
\end{equation}
where the sum runs over all the spins $N$.

We use the continuous time algorithm described in \sect{qannealing} and
 ``anneal'' the system from  $\Gamma=3$ to small values of $\Gamma$ in steps of $0.1$. 
The simulation at a transverse field $\Gamma_i$ is started from the final configuration
obtained in the previous simulation  $\Gamma_{i-1}=\Gamma_i-0.1$, except
for the initial easy simulation at a strong transverse field $\Gamma=3$, where the simulation is started from a random configuration.
$200000/\Gamma$ Monte Carlo sweeps (one Monte Carlo sweep consists of $2N$ site updates) are performed for equilibration before measurements.
Simulations are done at  inverse temperatures $\beta=100$ and $\beta=200$.

\begin{figure}[b]
\centering
\includegraphics[width=0.48\columnwidth]{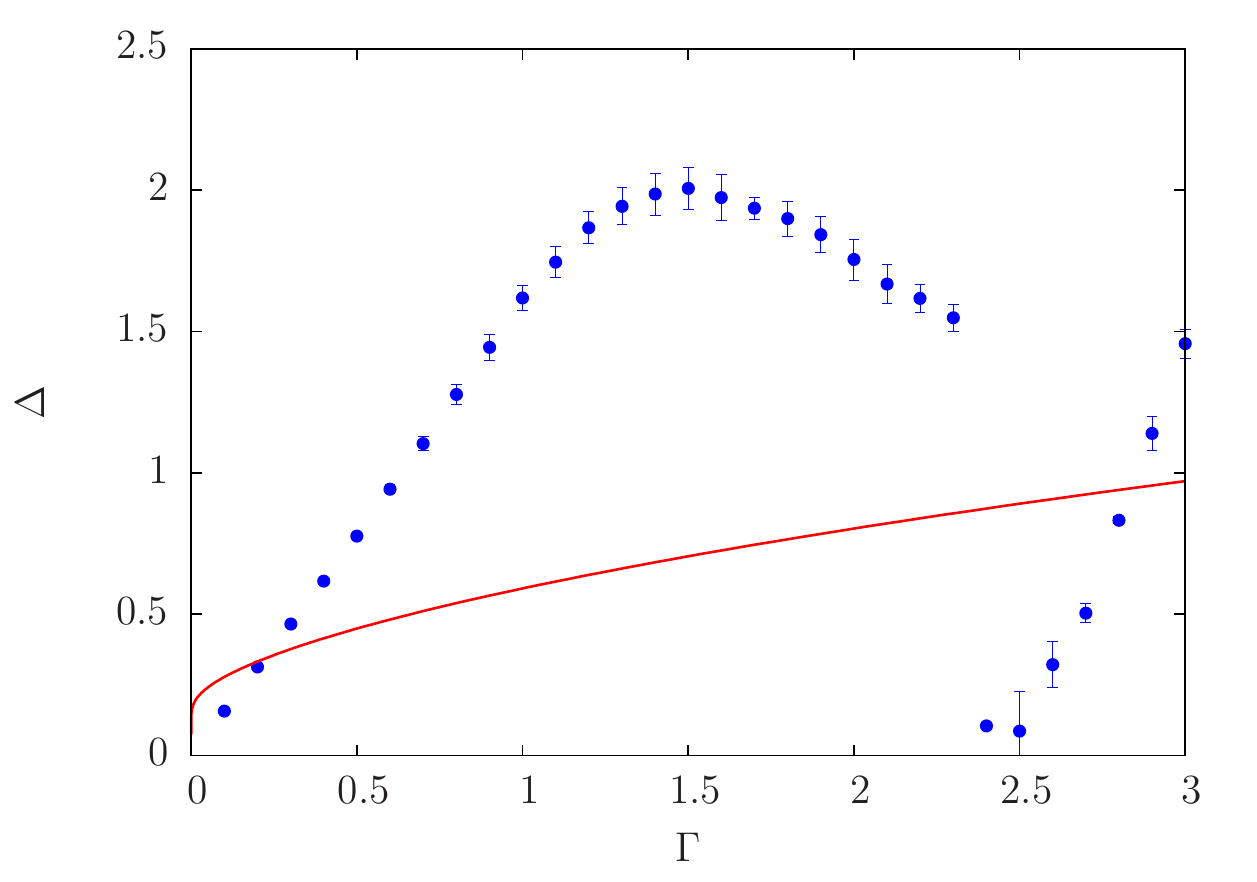}
\includegraphics[width=0.48\columnwidth]{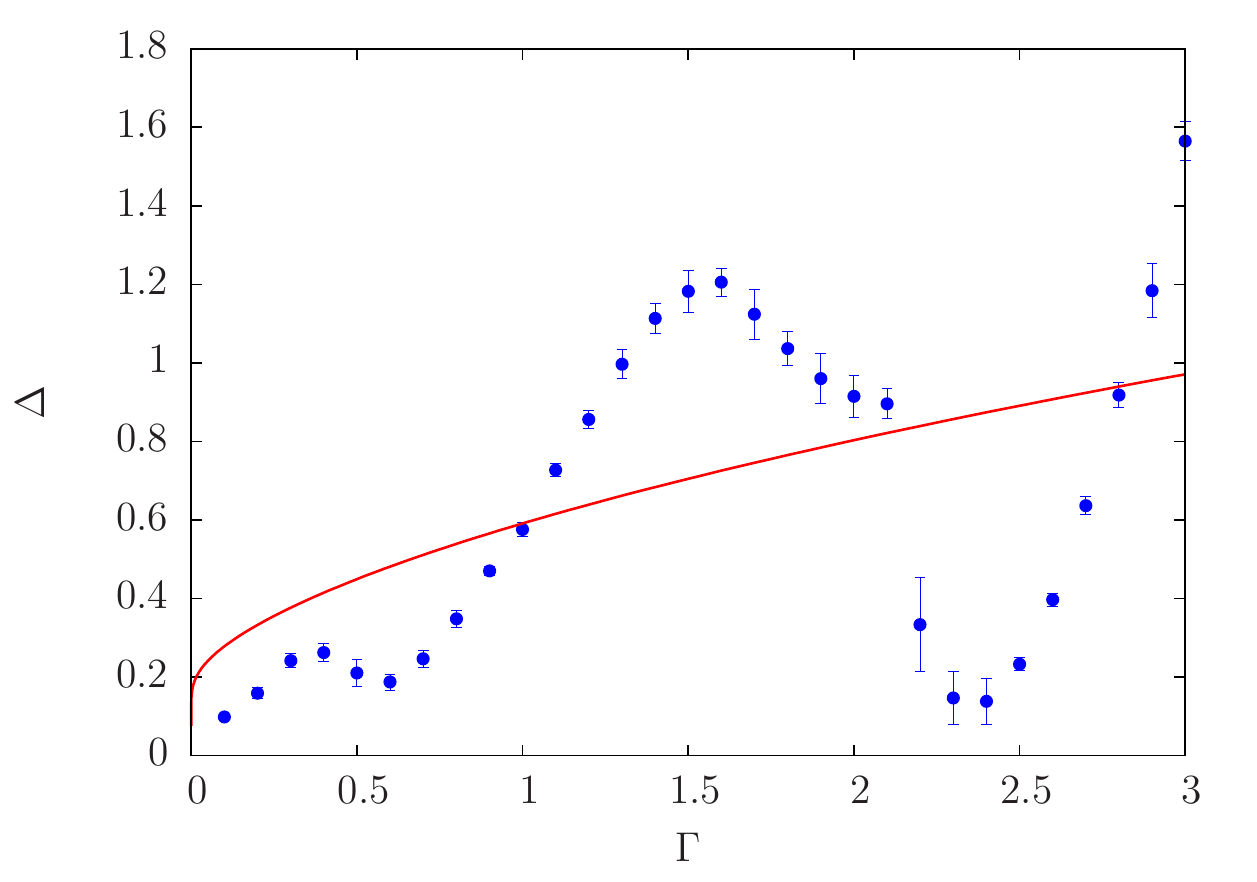}
\includegraphics[width=0.48\columnwidth]{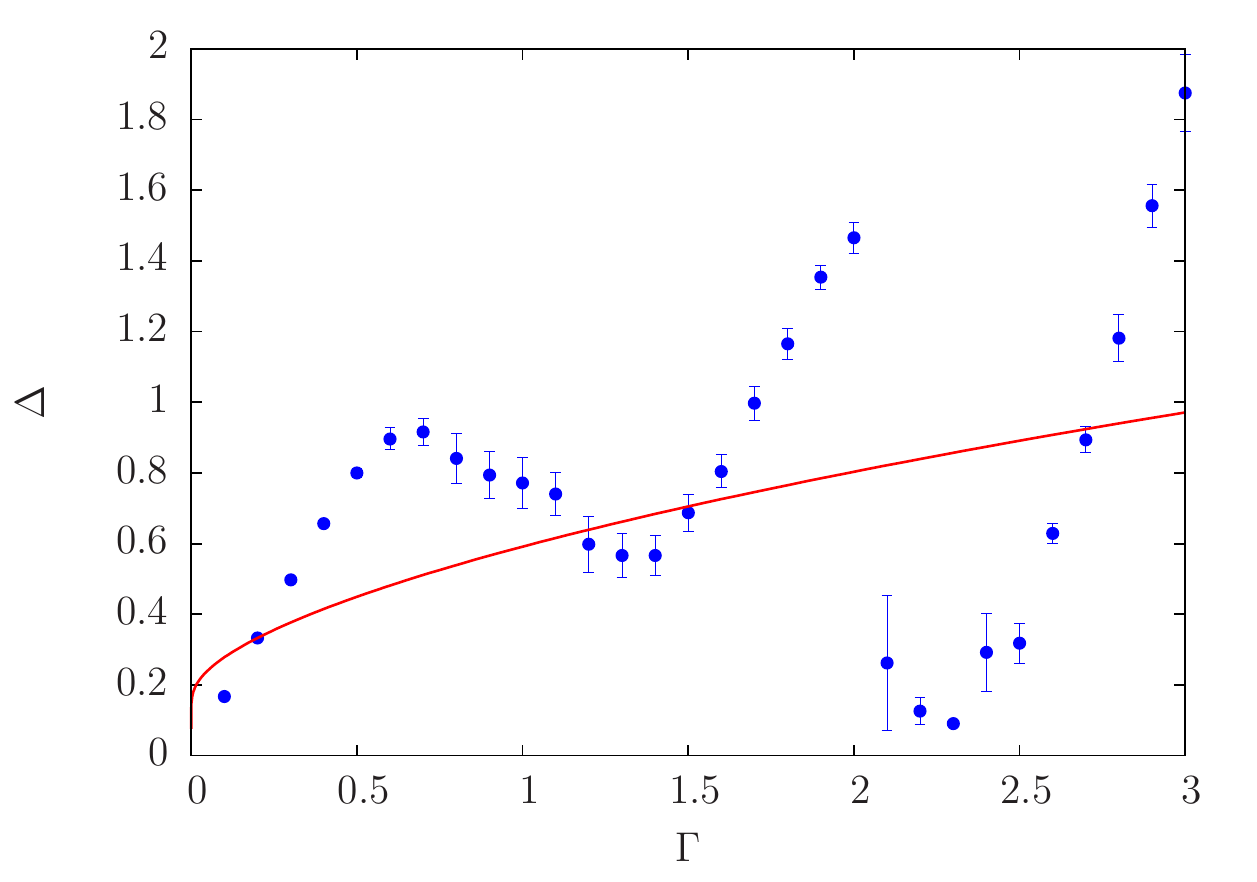}
\includegraphics[width=0.48\columnwidth]{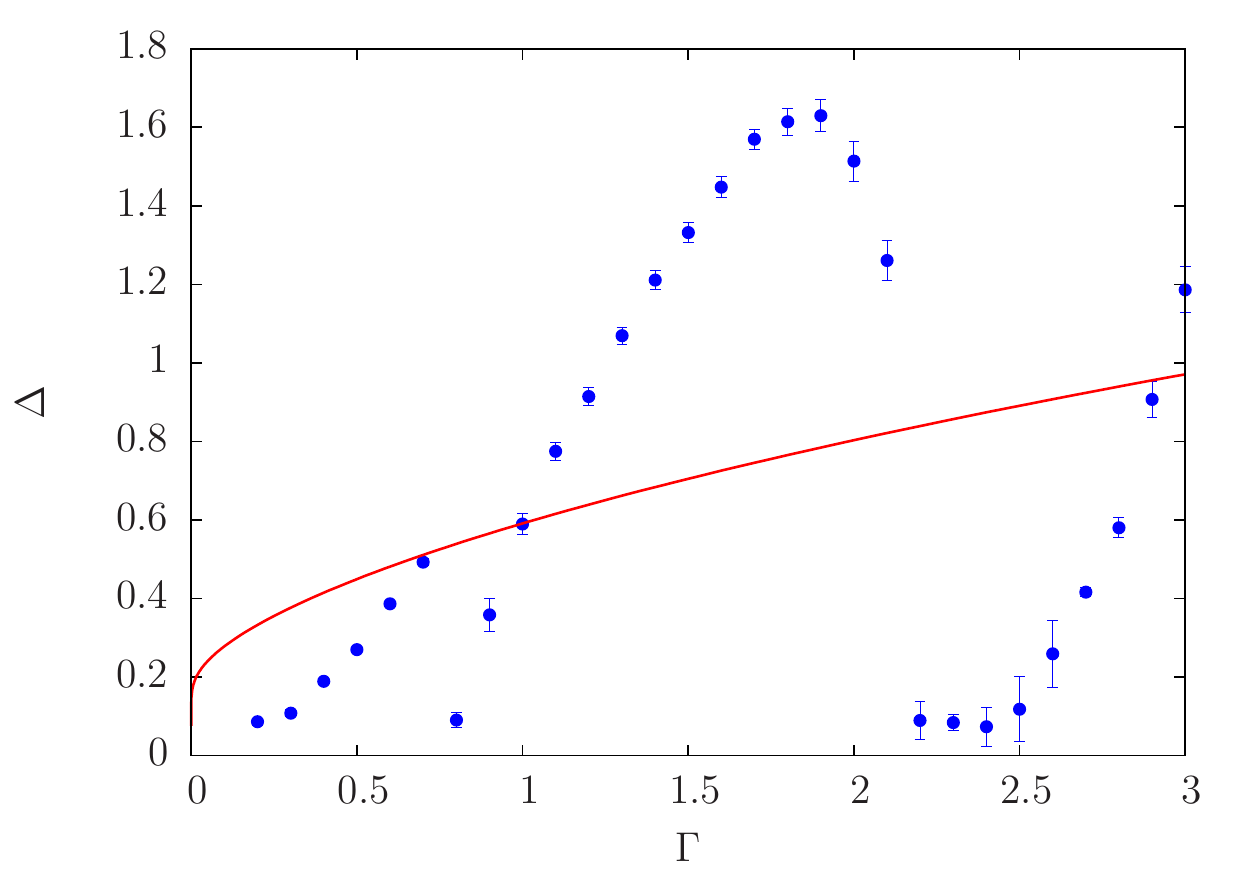}
\includegraphics[width=0.48\columnwidth]{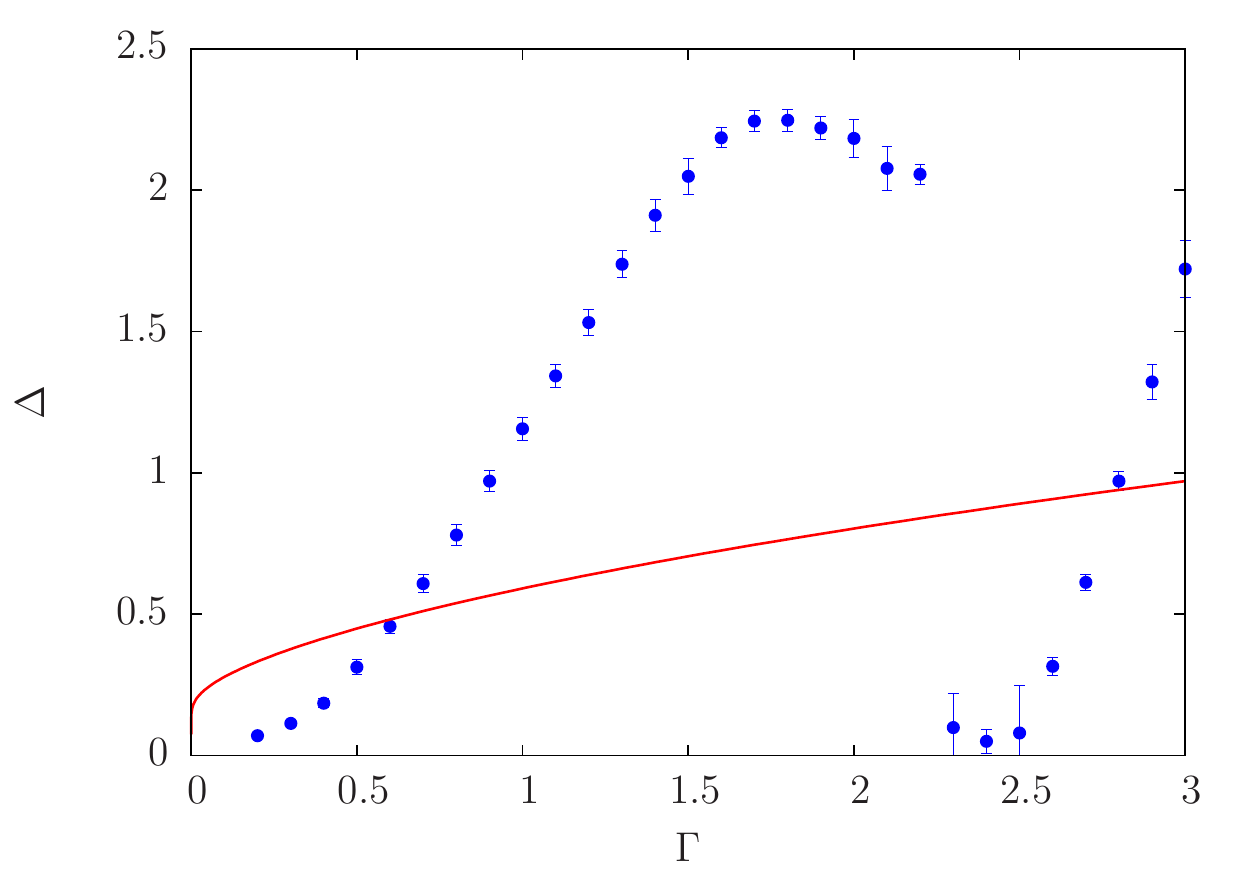}
\includegraphics[width=0.48\columnwidth]{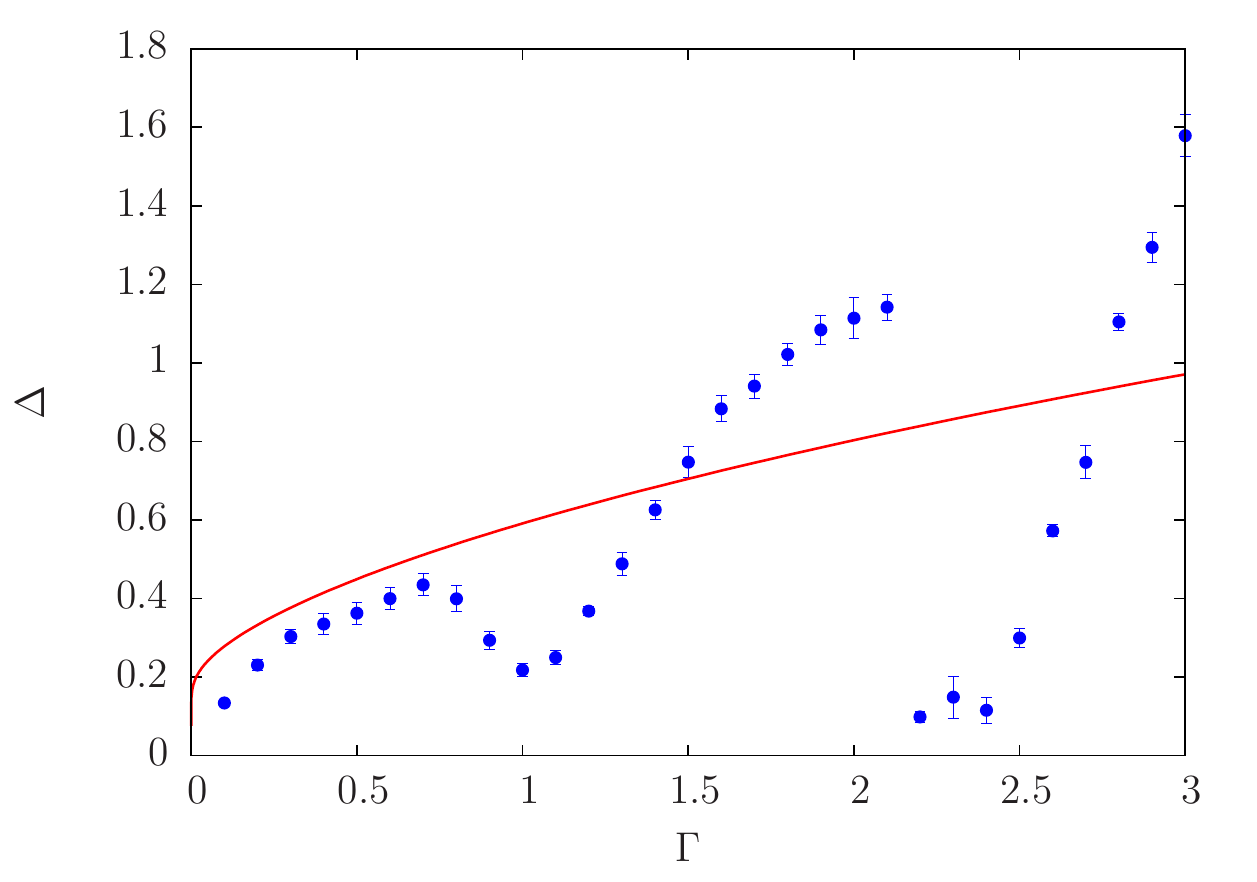}
\includegraphics[width=0.48\columnwidth]{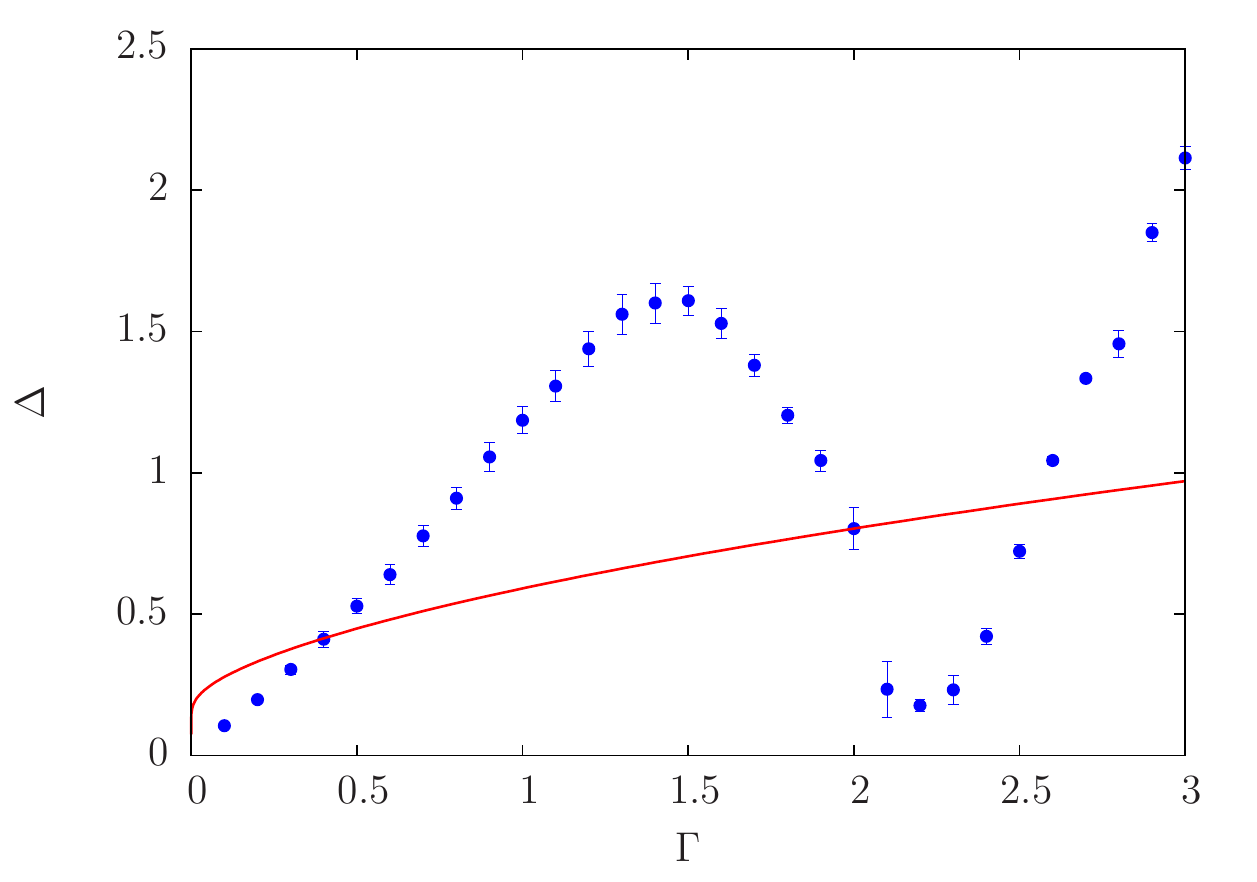}
\includegraphics[width=0.48\columnwidth]{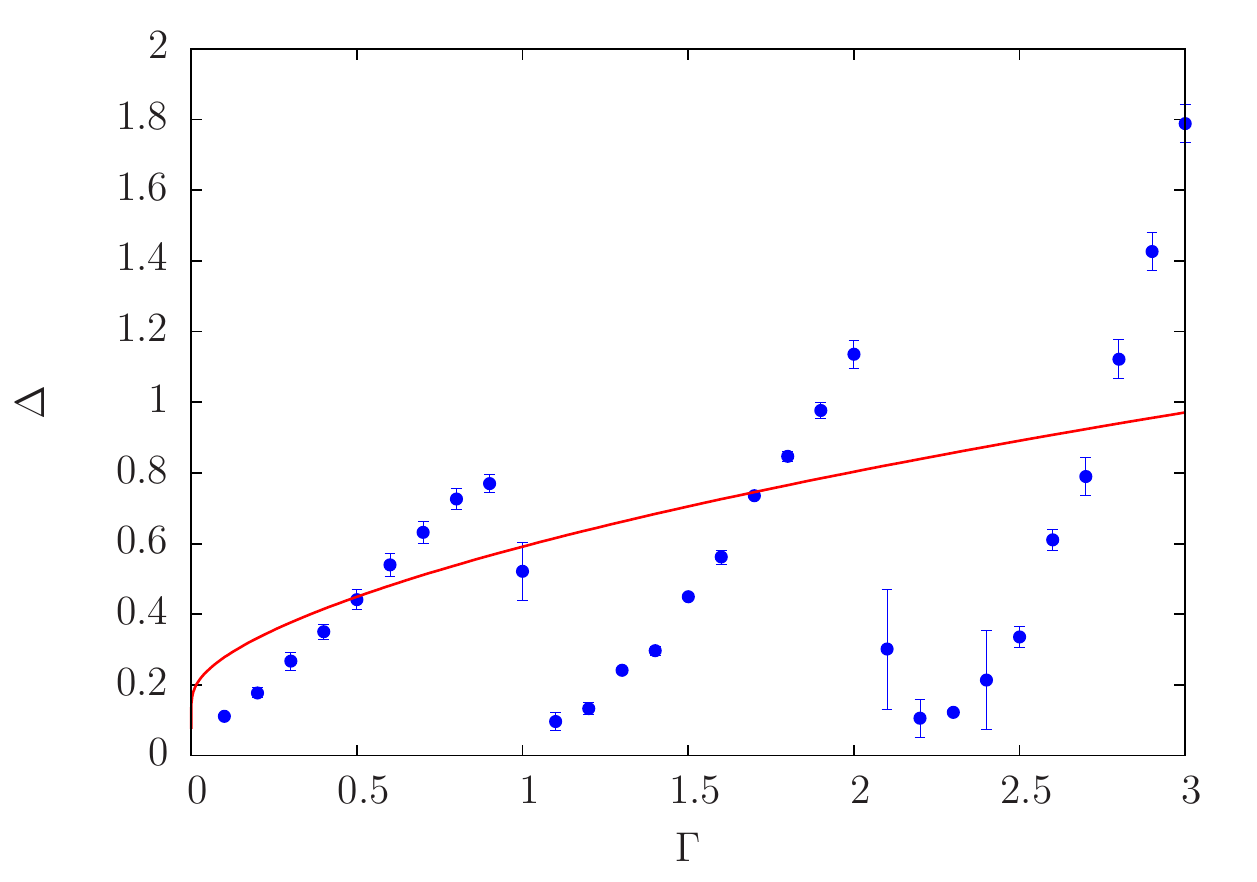}
\includegraphics[width=0.48\columnwidth]{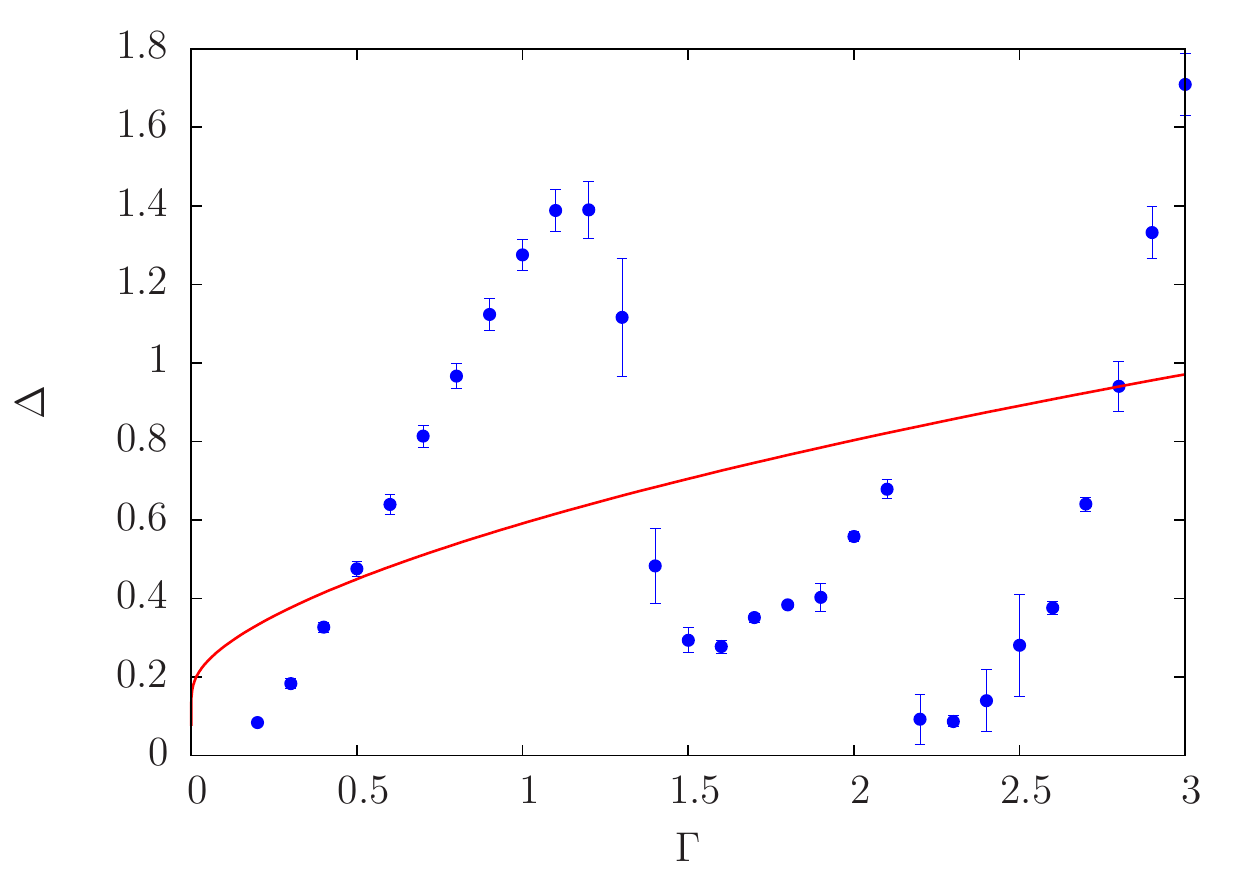}
\includegraphics[width=0.48\columnwidth]{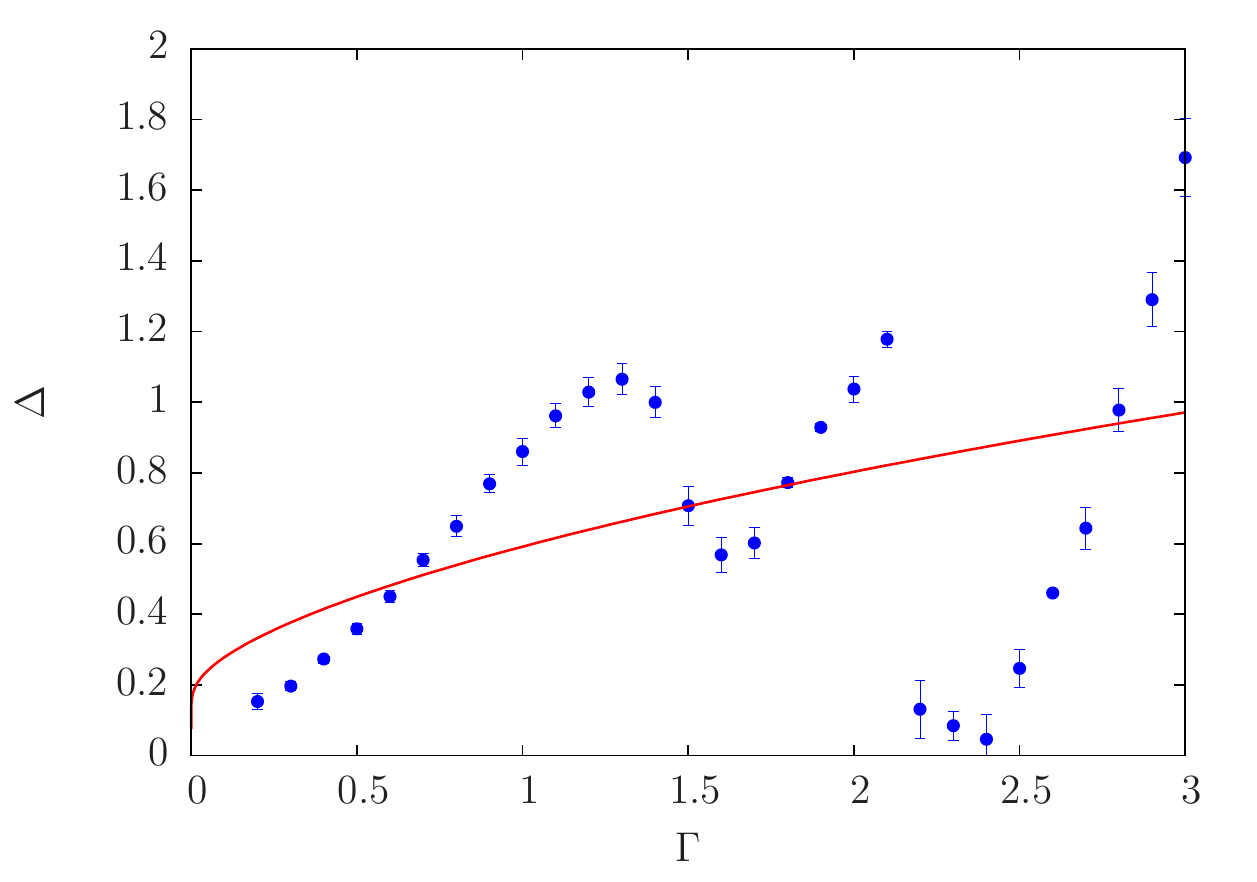}
\caption{{\bf Evolution of the excitation gap}. Shown is the lowest excitation gap for the two instances presented in the main text, along with eight additional, complementary instances. The gap is given in units of $B(t)$ so that it is a property of the model and not of the annealing schedule.
``Easy'' instances are shown on the left (success probabilities from top to
bottom: 0.98, 0.986, 0.995, 0.99, 0.932) and `hard` instances are shown on the right
(success probabilities from top to bottom: 0.08,0.001, 0.063, 0.000625, 0.0000625). The red line indicates the temperature in units of $B(t)$ for schedule II.}
\label{fig:gap}
\end{figure}

Extraction of the gap for many instances and many values of $\Gamma$ can be
a cumbersome task. To automate the process, we use the following approach.
We obtain the gap by fitting the correlation function to the exponential
function given by Eq.~\eqref{eq:ctau} in the range from $\tau_0$ to
$\tau_0+5/J$, where $\tau_0$ is chosen empirically in such a way that
$C(\tau)-C(\beta/2)>s_0$ for all $\tau<\tau_0$, $s_0=0.008 f(\Gamma)$ and
$f(\Gamma)=0.2+0.3\Gamma$. Let us denote this gap as $\Delta_0$.
To obtain the error bars, we calculate another two gaps $\Delta_1$ and
$\Delta_2$ by performing two extra fits with $s_1=0.006 f(\Gamma)$ and with
$s_2=0.011 f(\Gamma)$. Then the error bar is
just $\max(\Delta_2-\Delta_0,\Delta_0-\Delta_1)$. This procedure can be
fully automated. We show an example of such a fit in figure~\ref{fig:corr}.

In figure~4 of the main text we showed the excitation gap in units of temperature. In figure~\ref{fig:gap} we instead show the gap in units of $B(t)$, so that the gap shown is a property of the model only and not of the annealing schedule. While the physical temperature is constant ($17\,\textrm{mK}$), it increases as a function of $\Gamma$ when shown in units of $B(t)$. The gap and temperature are both shown in figure~\ref{fig:gap} for the two instances presented in figure~4 of the main text, as well as
for four ``easy" (left column) and four ``hard" (right column) additional instances. Note that the gap closes trivially around $\Gamma=2.2$, related to a global $Z_2$ symmetry breaking. Once the gap becomes very small it can no longer be detected by our procedure since the decay of $C(\tau)$ becomes too slow and is indistinguishable from a constant. Our procedure then picks up the gap to the next excited state, which results in an apparent jump of the gap to a bigger value. Generally, all gaps shown here are upper bounds for the gap to the lowest excited state.

The gap always closes also in the limit of a weak transverse field $\Gamma\longrightarrow 0$, where multiple ground states become degenerate. Some of the instances, however, have an additional
small gap (relative to the temperature) that can be associated with an avoided level crossing. These instances
are ``hard'' for both the D-Wave device and SQA. Usually, the instances that do not have
such a small gap are ``easy'' for both the D-Wave device and SQA. However, note that the last ``easy" instance shown in figure~\ref{fig:gap} does have a small gap. This shows that a small gap is not a sufficient condition for making a problem ``hard". The fact that this last instance is ``easy" can have multiple reasons, for example a second avoided level crossing with a small gap that takes the annealing back to a ground state or an avoided level crossing with a small Hamming distance from the ground state, from where thermal relaxation is not difficult.

\bibliographystyle{naturemag}
\bibliography{qa}

\end{document}